\documentclass[12pt,preprint]{aastex}         
\def\be{\begin{equation}}
\def\ee{\end{equation}}

\def\lambar{\lambda\llap {--}}

\def\lsim{\lower 2pt \hbox{$\, \buildrel {\scriptstyle <}\over
         {\scriptstyle \sim}\,$}}
\newcommand\gsim{\buildrel > \over \sim}
\begin{document}
\newcommand{\figureout}[2]{ \figcaption[#1]{#2} }       

\title{High-Altitude Emission from Pulsar Slot Gaps:  The Crab Pulsar}

\author{Alice K. Harding\altaffilmark{1},
Julie V. Stern\altaffilmark{1,2}, Jaroslaw Dyks\altaffilmark{3} \& Michal Frackowiak\altaffilmark{3,4}}   

\altaffiltext{1}{Astrophysics Science Division, 
NASA/Goddard Space Flight Center, Greenbelt, MD 20771}
 
\altaffiltext{2}{Stony Brook University, Stony Brook, NY}

\altaffiltext{3}{Nicolaus Copernicus Astronomical Center, Torun, Poland}

\altaffiltext{4}{Current address: redbeard@openlabs.pl}


\begin{abstract}
We present results of a 3D model of optical to $\gamma$-ray emission from the slot gap accelerator of
a rotation-powered pulsar.  Primary electrons accelerating to high-altitudes in the unscreened electric 
field of the slot gap reach radiation-reaction limited Lorentz factors of $\sim 2 \times 10^7$, 
while electron-positron pairs from lower-altitude cascades flow along field lines interior to the slot gap.  
The curvature, synchrotron and inverse Compton radiation of both primary electrons and pairs produce a 
broad spectrum of emission from infra-red to GeV energies.  Both primaries and pairs undergo cyclotron
resonant absorption of radio photons, allowing them to maintain significant pitch angles.  
Synchrotron radiation from pairs with a
power-law energy spectrum from $\gamma = 10^2 - 10^5$, dominate the spectrum up to $\sim 10$ MeV.  
Synchrotron and curvature radiation of primaries dominates from 10 MeV up to a few GeV.  We examine
the energy-dependent pulse profiles and phase-resolved spectra for parameters of the Crab pulsar as a
function of magnetic inclination $\alpha$ and viewing angle $\zeta$, comparing to broad-band data.  
In most cases, the pulse
profiles are dominated by caustics on trailing field lines. We also explore the relation of the high-energy
and the radio profiles, as well as the possibility of caustic 
formation in the radio cone emission.  We find that the Crab pulsar profiles and 
spectrum can be reasonably well reproduced by a model with $\alpha = 45^{\circ}$ and $\zeta \sim 100^{\circ}$
or $80^{\circ}$. This model predicts that the slot gap emission below 200 MeV will exhibit correlations
in time and phase with the radio emission.

\end{abstract} 


\pagebreak
  
\section{INTRODUCTION}

The study of emission from rotation-powered pulsars will soon undergo a major advance with
observations by two new $\gamma$-ray telescopes.  The AGILE mission launched in April 2007
and carries a pair-production telescope with sensitivity at 100 MeV comparable to the EGRET telescope
on the Compton Observatory, but with much larger effective area at energies above 1 GeV.  AGILE
is expected to find $\gamma$-ray pulsar counterparts to some of the unidentified EGRET sources.
The Large Area Telescope (LAT) on the Gamma-Ray Large Area Space Telescope (GLAST), due to launch
in early 2008, will have unprecedented sensitivity (30 times better than EGRET) and energy resolution for $\gamma$-rays in the range of 20 MeV to more than 300 GeV.  GLAST is expected to significantly
advance our knowledge of particle acceleration and radiation from rotation-powered pulsars, by
discovering many more $\gamma$-ray pulsars and measuring pulse profiles and phase-resolved spectra to
much higher precision.  Specifically, the LAT may increase the population of $\gamma$-ray pulsars to
several hundred and obtain flux measurements with error bars five-times smaller than those of EGRET.
In order to make the best use of the power of LAT observations, one needs the most sophisticated and
diverse models available.  

This paper adds to this effort by presenting, for the first time, results of a full model of high-altitude 
radiation from a pulsar slot gap and its relation to the geometry of the radio emission.  
Even 40 years after the discovery of pulsars, the origin of the particles that radiate high-energy photons as
well as how and where they are accelerated is still not known.  A number of models have
been proposed and studies over the years, including polar cap models, putting the site near the neutron
star at the magnetic poles (Arons \& Scharlemann 1979, Daugherty \& Harding 1982), and outer gap models, 
placing the site of acceleration and emission in the
outermost magnetosphere near the light cylinder (Cheng et al. 1986, Romani 1996, Hirotani 2001).  Such
models have had varying degrees of success, with the polar cap models being able to explain very well
the phase-resolved spectra of Vela -like pulsars (Daugherty \& Harding 1996) and the outer gap models 
being able to explain more easily the double-peaked profiles of the Vela, Geminga and Crab-like $\gamma$-ray pulsars.  However, even multiwavelength observations have not been able to definitively rule out any
of the proposed models.  

Studies of the geometry of polar cap accelerators (Arons \& Scharlemann 
1979, Arons 1983) discovered the possibility of a slot gap, a narrow bundle of field lines bordering
the closed-field region where the parallel electric field is decreasing and accelerating particles cannot create
pairs.  The pair formation front across the polar cap thus forms a bowl shape as pairs are produced at higher and higher altitude approaching the last open field line.  Although the existence of a slot gap has been
known for some time, its potential as a high-energy emission site has only recently been explored.
Muslimov \& Harding (2003) derived the electric potential and field in the slot gap, and modeled the
acceleration and pair cascades at altitudes from the neutron star surface up to several stellar radii.
They found that cascades develop over a range of 3 - 4 stellar radii along the slot gap inner boundary.
These cascades have higher pair multiplicity (number of pairs per primary electron) 
than cascades nearer the magnetic axis because the primary
electrons accelerate and produce pairs over a larger distance.  The radiation from the slot gap cascades
naturally produces a wider hollow cone of emission than the cascades near the surface.  Muslimov \& Harding 
(2004) extended the electric field solution in the slot gap to high altitudes and modeled only the curvature
radiation of the primary particles.  They found that such emission formed caustic patterns, like those
of the ``two-pole caustic" model proposed by Dyks \& Rudak (2003) to naturally explain the double-peaked 
profiles of $\gamma$-ray pulsars.  However, the spectrum of curvature radiation from 
radiation-reaction-limited particles alone is too hard to match that of observed $\gamma$-ray pulsars.

The radiation model of Muslimov \& Harding (2004) was not complete since it did not include
all sources of primary particle radiation and all particles capable of contributing to high-altitude
radiation.  We have now included additional radiation components of the primary particles as well as
radiation from electron-positron pairs, produced in polar cap and slot gap cascades at lower altitude, that
continue to flow along open field lines with relativistic velocity.  We investigate both the synchrotron
radiation of both primaries and pairs, that undergo cyclotron resonant absorption of radio photons and 
significant increases in pitch angle, and non-resonant inverse-Compton scattering of radio photons.  
We also explore the
geometry of the radio emission in relation to that of the high energy emission.  In Section \ref{sec:geom}, 
we describe the geometry of the magnetic field and of the slot gap, Section \ref{sec:acc} discusses the
model of acceleration in the slot gap and Section \ref{sec:emission} details our treatment of the emission processes we model.  We describe the simulation of radiation in the slot gap in Section \ref{sec:sim} and present the results for the Crab pulsar in Section \ref{sec:results}.  Finally, in Section \ref{sec:dis} 
we discuss the results and how the characteristic features of the model could be measured by GLAST and 
AGILE.

\section{3D GEOMETRY OF THE MAGNETOSPHERE}
\label{sec:geom}

\subsection{Retarded Vacuum Dipole and Open Volume Coordinates}  \label{sec:ovc}

The modeling of the neutron star magnetospheric geometry is based on the calculation of Dyks, Harding \& Rudak
(2004, DHR04), who adopted the retarded vacuum dipole solution (e.g. Romani \& Yadigaroglu 1995, 
Cheng et al. 2000) to account for the rotational distortions and sweepback of the magnetic field lines.
While the vacuum dipole solution does not accurately apply to a real pulsar magnetosphere that is filled with 
charges, and more realistic ideal MHD solutions now exist (Spitkovsky 2006, Timokhin 2006), the MHD
solutions are numerical and time-consuming to simulate.  
The vacuum dipole solution exists in analytic form, is
thus much easier to implement and we believe it exhibits the features like sweepback of field lines 
that approximate the MHD solutions. 

Our code first performs Kunge-Rutta integrations along the distorted field lines to determine the 
location of the open field footpoints at the neutron star surface that form the distorted polar cap rim.  
Details of this 
calculation are given in DHR04 and Dyks \& Harding (2004).  We then define ``open volume coordinates" $(r_{\rm ovc},
l_{\rm ovc})$ at the stellar surface to identify footpoints of field lines on which particles 
are traced.  The ``radial" coordinate $r_{\rm ovc}$ ranges from 0 at the magnetic pole 
to 1 at the polar cap rim.  Lines of constant $r_{\rm ovc}$ thus form a set of distorted concentric 
rings that cover the polar cap surface (see Figure 2 of DHR04).  The ``azimuthal" coordinate $l_{\rm ovc}$
measures the arclength along each deformed ring of fixed $r_{\rm ovc}$.  The arclength is measured in
the direction of increasing magnetic azimuth $\phi_{pc}$, 
counterclockwise around the polar cap, starting from
$l_{\rm ovc} = 0$ at $\phi_{pc} = 0$, defined to be at the magnetic meridian (i.e. line between the
magnetic and spin axes).  One notable feature of the distorted polar cap rim is the appearance of 
a ``notch" at $\phi_{pc} \sim 30^{\circ}$, first discovered by Arendt \& Eilek (1998) and most prominent
between inclination angles of $40^{\circ}$ and $60^{\circ}$ (see Figure 2 of Dyks \& Harding 2004).  
The notches are
caused by field lines crossing near the light cylinder because of the sweepback, and produce bunching
of fields lines at lower altitudes.  This bunching of field lines will produce enhanced radiation at
certain azimuth angles and noticeable features in the radiation phase plots, as will be discussed in 
\S \ref{sec:results}. 

The open volume coordinates (`ovc') enable us to selectively model different particle distributions 
across the polar cap, such as electrons accelerating and radiating in the slot gap.  To model a 
distribution of particles that is uniform in magnetic azimuth, we define a set of rings between
$r_{\rm ovc}^{\rm min}$ and $r_{\rm ovc}^{\rm max}$ with equal spacing $d_{\rm ovc}$.  The footpoints
are then placed uniformly around each ring with $N_l = \Delta_{\rm azim}360$ equal divisions.  
Since we trace emission from
particles on the same number of magnetic field lines around each ring, and the rings have different 
circumferences, the contribution from different rings is weighted by $l_{\rm ring}/l_{\rm rim}$,
where $l_{\rm ring}$ is the ring length and $l_{\rm rim}$ is the polar cap rim length.

\subsection{Geometry of the Slot Gap}  \label{sec:SGgeom}

The slot gap, a narrow region bordering the last open magnetic field line where particles cannot
accelerate rapidly enough to produce pairs before the field drops, is an unavoidable feature 
of polar cap space-charge limited flow (SCLF) models (Arons \& Scharlemann 1979 [AS79], Arons 1983).  
SCLF models assume that charges are freely emitted from the neutron star polar cap surface and flow out 
along open field lines.  Since this charge flow is not sufficient to supply the
Goldreich-Julian charge above the surface, an $E_{\parallel}$ exists and the charges are accelerated.
Radiation from these charges forms electron-positron pairs in the strong magnetic field, which can screen
the $E_{\parallel}$ above a pair formation front (PFF) in a distance small compared to the acceleration distance,
which is typically less than a stellar radius
(AS79, Harding \& Muslimov 2001).  On field lines well inside the polar cap rim,
$E_{\parallel}$ is relatively strong and the PFF is very near the neutron star surface. 
These models assume a boundary condition that the accelerating electric field and potential vanish 
at the last open field line.  Near the boundary, the electric field is decreasing and a larger distance is 
required for the electrons to accelerate to the Lorentz factor needed to radiate photons energetic 
enough to produce pairs.  The pair front thus occurs at higher and higher altitudes near the boundary 
and curves upward, asymptotically approaching the 
last open field line (Harding \& Muslimov 1998), forming a narrow slot gap.  
Since $E_{\parallel}$ is unscreened in the slot gap, particles continue to
accelerate and radiate to high altitude along the last open field lines.  
The width of the slot gap, $\Delta \xi_{\rm SG}$, is a function of pulsar period and surface magnetic 
field, and can be expressed as a fraction in colatitude 
$\xi \equiv \theta/ \theta_0$ of the polar cap opening angle, $\theta_0 \simeq (\Omega R/c)^{1/2}$
(Muslimov \& Harding 2003),
\be  \label{SGwidth}
\Delta \xi _{_{\sc SG}} \approx 2~P(\lambda B_{12})^{-4/7}I_{45}^{-3/7}.
\label{deltaxi1}
\ee
where we will take $\lambda = 0.1$.  For the Crab parameters, rotation period $P = 0.033$, surface magnetic
field $B_{12} = B_0/10^{12} = 8$ and moment of inertia $I_{45} = I /10^{45}$ g$\cdot $cm$^2 = 4$, 
the slot gap width $\Delta \xi _{_{\sc SG}} \simeq 0.04$.  Here and in the rest of the paper, we will 
adopt these values, where $B_{12}$ comes from the measured $P$ and $\dot P$, and $\lambda$ is 
a free parameter within the range $0.1 - 0.5$ that sets the width of the slot gap.  Although the slot gap
width itself is somewhat weakly dependent on these parameters, the slot gap potential and electric field
are more strongly dependent on these parameter values (see \S \ref{sec:acc}).

MH03 found that pair cascades on the inner edge of the slot gap occur at altitudes of 3-4 stellar 
radii and have higher multiplicities $M_+ \sim 10^4 - 10^5$ than the polar cap cascades. 
Since the slot gap is very narrow for young
pulsars having short periods and high fields, the corresponding solid angle of the gap emission 
$\Omega _{SG}  \propto \theta _0^2 \eta \Delta \xi _{SG}$ is quite small at low altitudes, 
$\eta \ll \eta_{LC}$, where $\eta \equiv r/R$ is the dimensionless radius.  
However, this approximate expression does not describe the solid 
angle of radiation from the high-altitude slot gap, where relativistic effects strongly distort the
emission pattern (see \S \ref{sec:results}) into narrow caustics, and the solid angle must be computed by 3D mapping
on the sky.  So even though only a small fraction of the polar cap flux is accelerated 
in the slot gap, the radiated flux $\Phi_{\rm SG} = L_{SG}/\Omega _{SG}\,d^2$ can be substantial.  
The particles in the slot gap can achieve very high Lorentz factors which at altitudes of
several stellar radii are limited by curvature radiation losses, to $\gamma_{SG} \simeq 
3-4 \times 10^7$ (Muslimov \& Harding 2004).  In this paper, we model the emission from the electrons 
in the slot gap as they continue to accelerate to high altitude above
the region of pair formation.

\section{ACCELERATION OF PARTICLES IN THE SLOT GAP}  \label{sec:acc}

The solution for the electric potential and parallel electric field in the slot gap was presented
by Muslimov \& Harding (2003, MH03) in the low altitude limit and by Muslimov \& Harding (2004, MH04) 
in the high altitude limit.  Both solutions assume a space-charge limited flow of electrons from the 
neutron star surface, with the boundary condition $\rho = \rho_{GJ}$ at the stellar surface, where
$\rho$ is the true space charge density and $\rho_{GJ}$ is the Goldreich-Julian charge density.
The space charge density at the neutron star surface that includes relativistic frame-dragging, 
from Eqn (4) of MH03, is
\be
\rho_0 = -{\Omega B_0\over 2\pi c\alpha_0}\,[(1-\kappa)\cos\alpha + {3\over2}\theta_0 H(1)\sin\alpha\cos\phi_{pc}],
\ee
where $B_0$ is the surface magnetic field strength in Gauss, $\Omega = 2\pi/P$ and $R$ are the neutron
star rotation rate and radius, $\alpha $ is the  pulsar obliquity, $\phi _{\rm pc}$ is the magnetic 
azimuthal angle and $\theta_0 \approx [\Omega R/c f(1)]^{1/2}$ 
is the polar cap half-angle.  Here, $\kappa = (r_{\rm g}/R) (I/I_0) \approx 0.15~I_{45}/R_6^3$ is the 
general relativistic parameter from the frame-dragging effect, $r_{\rm g}$ is the NS gravitational radius, 
 $R_6 = R/10^6$ cm and $\alpha_0 = (1 - r_{\rm g}/R)^{1/2}$.  
In addition, $f(1) \sim 1.4$ and $H(1) \simeq 0.8$ are general relativistic parameters.
Since the $\rho (r)$ decreases faster than $\rho_{GJ}(r)$ above the surface (for the explicit expressions 
see MH03), the deficit
$(\rho - \rho_{GJ})$ increases with distance, causing a parallel electric field, 
$\nabla \cdot E_{\parallel} = 4\pi(\rho - \rho_{GJ})$ to develop.  The treatment of the electric field
in the slot gap by MH03 and MH04 differs from that of Arons \& Scharlemann (1979) and Arons (1983) in that
MH03 take into account the screening of the electric field by pairs on field lines interior to the slot gap (see \S \ref{sec:SGgeom}).
This forms fully conducting boundaries on both the inner and outer edges of the slot gap, causing a lower
electric field in the slot gap.
From Eqn (58) of MH04, the low-altitude solution for $E_{\parallel}$ is
\be  
E_{\parallel, {\rm low}}  \simeq  -3\left({{\Omega R}\over {c}}\right) ^2 {{B_0}\over {f(1)}} \nu _{_{\rm SG}}
\left[{\kappa\over \eta^4}  \cos \alpha + 
 {1\over 2} \theta _0 H(1) \delta(\eta) \sin \alpha \cos \phi _{pc} \right], 
\label{eq:Epar - lo}
\ee
where  
$\delta (\eta )$ (Muslimov \& Harding 1997) varies between $\sim 0.5$ and 1, and $\nu _{_{\sc SG}} = {1\over 4} \Delta \xi _{_{\sc SG}}^2$ is a parameter related to the width of 
the slot gap, $\Delta \xi _{_{\sc SG}}$, given by Eqn (\ref{SGwidth}) above.
The solution in Eqn (\ref{eq:Epar - lo}) is
valid for radii $\eta \lsim \eta_c$, where $\eta_c$ is a free parameter to be determined by matching 
smoothly to the high-altitude solution.  The high-altitude solution for $E_{\parallel}$, valid for
$\eta \gsim \eta _c$, is given by Eqn (53) of MH04,
\begin{eqnarray} 
E_{\parallel, {\rm high }} &\approx &- {3\over 8}\left({{\Omega R}\over {c}}\right) ^3 {{B_0}\over {f(1)}} 
\nu _{_{\rm SG}} \left\{ \left[ 1+ {1\over 3}\kappa \left( 5 - {8\over {\eta _c^3}}\right) +2{\eta \over \eta _{lc}} \right] \cos \alpha + \right. \nonumber \\
&& \left. {3\over 2} \theta _0 H(1) \sin \alpha \cos \phi _{pc} \right\}.
\label{eq:Epar-high}
\end{eqnarray}
It is interesting that frame-dragging's dominant effect on the accelerating field persists even at large 
distances from the neutron star surface, since the high-altitude SCLF solution depends on surface 
boundary conditions.   Since a pure dipole field was assumed to derive the above expression for $E_{\parallel}$,
it is probably not accurate for very high altitudes approaching the light cylinder ($r \gsim 0.7 R_{\rm LC}$), where the field lines
will be distorted by effects of rotation and particle inertia.
We combine the low and high-altitude solutions by use of the general
expression
\be  \label{eq:Epar - comb}
E_{\parallel } \simeq E_{\parallel, {\rm low}}\exp{[-(\eta-1)/(\eta_c-1)]} 
\label{Epar - tot} + E_{\parallel, {\rm high }} 
\ee
and determine $\eta_c$ to give a smooth transition between $E_{\parallel, {\rm high}}$ and $E_{\parallel, {\rm low}}$.  For the Crab pulsar, we find that $\eta_c = 1.2$ gives a smooth transition at most
inclination angles.

An estimate of the full potential drop of the slot gap can be obtained by summing that of the low
altitude potential (Eqn [12] of MH03) from the surface up to $\eta_c = 1.2$ and that of the high altitude 
potential (Eqn [52] of MH04) at the $\eta _{lc}$:
\be  \label{Phi}
\Phi^{\rm SG}_{\rm Tot} = \Phi^{\rm SG}_{\rm low}(\eta_c) + \Phi^{\rm SG}_{\rm high}(\eta _{lc}) = 
\Phi_0\theta_0^2\nu _{_{\rm SG}}\,0.5(1+\kappa)\cos\alpha,
\ee
where $\Phi_0 \equiv (\Omega R/c)B_0 R$.
In the case of the Crab pulsar, taking $B_0 = 8 \times 10^{12}$ G, $\Omega = 190$, $\nu _{_{\rm SG}}= 
\Delta \xi _{_{\sc SG}}^2/4 = 4 \times 10^{-4}$ and $\kappa=0.14$, $\Phi^{\rm SG}_{\rm Tot} \simeq 1.3 
\times 10^{13}$ eV.

The particle flux from the polar cap and slot gap is
\be  \label{ndot}
\dot n = {\rho\over e}c A_{\rm sur},
\ee
where 
\be \label{Asur}
A_{\rm sur} = \pi R\theta_0^2 [(r_{\rm ovc}^{\rm max})^2 - (r_{\rm ovc}^{\rm min})^2] \simeq \pi R\theta_0^2 
\Delta \xi _{_{\sc SG}}
\ee
is the surface area either of the slot gap or of the field line surface footpoints of the pairs, with
$\rho = \rho_0$ for the primary electrons and $\rho = {\cal M}_{\rm pairs}(r_{\rm ovc})\rho_0$ for the pairs.
Again for the case of the Crab, $\dot n_{\rm SG} \sim 7 \times 10^{32}$ for the slot gap.

\section{EMISSION PROCESSES} \label{sec:emission}

We model radiation over the entire spectrum from radio to $\gamma$-ray wavelengths.
The radio emission is necessarily treated phenomenologically, since no consensus physical model has yet
emerged.  We therefore adopt a geometrical description of the radio emission beam as consisting of a core component and single conal component, described below.  For the emission from optical through high-energy $\gamma$-ray, we simulate the radiation from primary electrons accelerating in the slot gap and also the radiation from secondary electron-positron pairs, that are produced in cascades near the neutron star surface and are flowing on field lines interior to the 
slot gap.  
  
\subsection{Radio beam model} \label{sec:radio}

We model the radio emission beam using an empirical core and cone model that has developed over the years 
through detailed study of pulse morphology and polarization characteristics.  The average-pulse
profiles are quite stable over long timescales and typically show a variety of shapes, ranging from a single 
peak to as many as five separate peaks.  The emission is also highly polarized, and displays changes in polarization position angle across the profile that often matches the position angle swing expected for a sweep across 
the open field lines near the magnetic poles in the Rotating Vector Model (Radhakrishnan \& Cooke 1969).  
Rankin's (Rankin 1993) study of pulse morphology concluded that pulsar radio emission can be 
characterized as having a core beam centered on the magnetic axis and one or more hollow cone beams also centered on the magnetic axis surrounding the core.  Although Rankin's model assumes that emission fills the core and cone beams, other studies
(Lyne \& Manchester 1988)
conclude that emission is patchy and only partially fills the core
and cone beam patterns.  

The particular description we adopt is from Gonthier et al (2004) and is based on work of Arzoumanian et al. (Arzoumanian, Chernoff \& Cordes 2002, ACC), who fit average-pulse 
profiles of a small collection of pulsars at 400 MHz to a core
and single cone beam model based on the work of Rankin.  The summed flux from the two components seen at 
angle $\theta$ to the magnetic field axis (modified by Gonthier, Van Guilder \& Harding (2004) to include frequency dependence $\varepsilon_R$) is
\be  \label{eq:Stheta}
S(\theta, \varepsilon_R ) = F_{\rm core} e^{ - \theta ^2 /\rho _{\rm core}^2 }  + 
F_{\rm cone} e^{ - (\theta  - \bar \theta )^2 /\omega _e^2 } 
\ee
where 
\be \label{eq:Fi}
F_i(\varepsilon_R) = {-(1+\nu_i) \over \varepsilon_R}\left({\varepsilon_R\over 50 {\rm MHz}}\right)^{\nu_i+1}{L_i\over \Omega_i D^2}
\ee
and the index $i$ refers to the core or cone, $\nu_i$ is the spectral index of the total angle-integrated flux, $L_i$ is the luminosity of
component $i$ and $d$ is the distance to the pulsar.  The width of the Gaussian describing the core beam is
\be  \label{eq:rhocone}
\rho _{\rm core}  = 1.5^{\circ} P^{ - 0.5} 
\ee
where $P$ is the pulsar period in seconds.  The annulus and width of the cone beam are
\be \label{eq:thetabar}
\bar \theta  =  (1.-2.63\,\delta_w) \rho_{\rm cone}
\ee
\be  \label{eq:widann}
w_e = \delta_w \rho_{\rm cone}
\ee
where $\delta_w = 0.18$ (Gonthier et al. 2006), and
\be  \label{rhocone}
\rho_{\rm cone} = 1.24^{\circ}\,  r_{\rm KG}^{0.5}\, P^{- 0.5}
\ee
where
\be  \label{eq:rKG}
r_{\rm KG}  \approx 40\, \left({\dot P\over 10^{ - 15}{\rm s\,s^{-1}}}\right)^{0.07} P^{0.3} \varepsilon_{GHz}^{ - 0.26} 
\ee
is the radio emission altitude in units of stellar radius (Kijak \& Gil 2003).

The luminosities of the core and cone components are then
\be  \label{eq:Lcc}
L_{\rm cone} = {L_{\rm radio}\over 1+(r/r_0)},   L_{\rm core} = {L_{\rm radio}\over 1+(r_0/r)},
\ee
where
\be   \label{eq:r0}
r_0 = {\Omega_{\rm cone}\over \Omega_{\rm core}}{(\nu_{\rm core}+1)\over (\nu_{\rm cone}+1)}
{1\over r}\left({\varepsilon_R\over \varepsilon_0}\right)^{\nu_{\rm core}-\nu_{\rm cone}}
\ee
where $\nu_{\rm core} = -1.96$ and $\nu_{\rm cone} = -1.32$, and
\be   \label{eq:Lradio}
L_{\rm radio}  = 2.87 \times 10^{10}\,P^{ - 1.3} \dot P^{0.4}\, {\rm mJy\cdot kpc^2\cdot MHz} 
\ee
where $\dot P$ is in units of $1\rm s\,s^{-1}$.
The core-to-cone peak flux ratio is (Gonthier et al. 2006)
\be  \label{eq:Gratio}
r = \frac{{F_{\rm core} }}{{F_{\rm cone} }} = \left\{ {\begin{array}{*{20}c}
   {25\,P^{1.3} \varepsilon_{\rm GHz}^{ -0.9}, ~~~~~~~~~~~~ P < 0.7s}  \\
   {4\,P^{ - 1.8} \varepsilon_{\rm GHz}^{ -0.9}, ~~~~~~~~~~~~ P > 0.7s}  \\
\end{array}} \right.
\ee

According to Eqn (\ref{eq:rKG}), the altitude of the conal radio emission is a weak function of the pulsar period but
the emission occurs increasingly close to the light cylinder as the period decreases.  For Crab-like periods, the
conal emission occurs at altitudes of $10\% - 20\%$ of the light cylinder radius.

We will use this radio emission beam model both to calculate the radio emission profile and to compute the 
intensity and angles of radio photons for the cyclotron absorption/synchrotron emission component described in 
Section \ref{sec:CycAbs} below.  To incorporate this radio emission geometry in the retarded dipole magnetic field that 
we are using to simulate the high energy emission, we modulate the field lines of the ovc with the flux 
$S(\theta,\varepsilon_R)$ given by Eqn (\ref{eq:Stheta}).  The differential flux radiated from a bundle of field lines
centered at ovc coordinates ($r_{\rm ovc}$, $l_{\rm ovc}$) is
\be  \label{eq:L_r}
dS(\theta, \varepsilon_{_R} ) = S_i(\theta, \varepsilon_{_R} )\sin\theta \, d_{\rm ovc}\, \theta_0 \,r_{\rm ovc}^{\rm max}\, {2\pi\over N_l} \,d\varepsilon_{_R} 
\ee
where $N_l$ is the number of azimuthal divisions of each ring.  The flux is assumed to be
emitted at altitude $1.8 R$ for the core component and at altitude given by Eq (\ref{eq:rKG}) for the conal
component.

\subsection{Curvature Radiation of Primary Electrons}

Primary electrons accelerating in the slot gap will produce curvature radiation up to $\gamma$-ray energies.  The
curvature radiation losses will be balanced by the acceleration due to $E_{\parallel}$ so that the particles 
reach steady-state Lorentz factors.  The curvature radiation energy spectrum from a single electron with Lorentz
factor $\gamma$ is
\be
N_{CR}(\varepsilon) = \sqrt{3}\,{e^2\over c}\,\gamma\,\kappa\left({\varepsilon\over 
\varepsilon_{cr}}\right)
\ee
where $\varepsilon$ is the radiated photon energy in units of $mc^2$ and
\be
\varepsilon_{cr} = {3\over 2} {c\over \rho_c}\,\gamma^3,
\ee
and the function $\kappa(x)$ is defined as
\be
\kappa(x) \equiv 2x \int_{2x}^{\infty} K_{5/3}(x')dx'.
\ee
The field line radius of curvature $\rho_c$ is not the standard one for a pure dipole field, but is 
determined in the ininertial observer's frame by interpolating from tables computed from variations of the vacuum retarded 
field lines (see \S \ref{sec:ovc}).  The approximate form of the photon spectrum is a power law with an exponential 
cutoff at $\varepsilon_{cr}$,
\be
{N_{CR}(\varepsilon)\over d\varepsilon} = {\alpha_f \over (\lambar mc)^{1/3}} \left({c\over \rho_c}\right)^{2/3}\,
\varepsilon^{-2/3}\exp(-\varepsilon/\varepsilon_{\rm cr}).
\label{insCR} 
\ee

\subsection{Cyclotron Resonant Absorption and Synchrotron Radiation} \label{sec:CycAbs}

Relativistic particles moving parallel to magnetic field lines can gain perpendicular momentum by undergoing 
cyclotron resonant absorption of low-energy photons.  The cyclotron resonant absorption of radio emission by 
relativistic particles in pulsar magnetospheres, followed by spontaneous synchrotron emission, 
was first proposed some years ago by Shklovsky (1970) as a mechanism for generating the optical
radiation from the Crab pulsar.  The process involves the absorption of photons at the cyclotron 
resonant frequency in the rest frame of the particle, resulting in an increase in the particle pitch angle.
The particle then spontaneously emits cyclotron or synchrotron radiation, depending on whether its
momentum perdendicular to the magnetic field, in the frame in which the parallel momentum vanishes, is
non-relativistic or relativistic.  Blandford \& Scharlemann (1976) computed the cross section for 
cyclotron resonant absorption, but their application of the process to the Crab pulsar resulted in
a re-radiated cyclotron radiation flux that was too small to explain the Crab optical emission.  However, they assumed
that the perpendicular momentum remained non-relativistic, in which case the applicable rate is that of 
cyclotron emission from the first excited Landau state, which is small relative to the rate from 
highly excited states.  Lyubarski \& Petrova (1998, LP98) performed
a more detailed analysis of the distribution functions of particles undergoing synchrotron resonant absorption 
of radio photons and found that the particles can increase their pitch angles rapidly enough in the outer magnetosphere 
to attain relativistic perpendicular momentum.  They found that the pitch angle excitation rate due to
resonant absorption is much higher than the de-excitation rate due to synchrotron radiation until the particles 
reach very large pitch angles.  
The resulting synchrotron radiation 
explain the optical and X-ray emission spectrum of the Crab and other young pulsars (Petrova 2003).  Harding et al.
(2005, HVM05) showed that cyclotron resonant absorption of radio emission can work very efficiently 
for millisecond pulsars, 
especially for those pulsars where the accelerating electric field is unscreened.  

We apply the formulation of HUM05 to model the synchrotron radiation components from both primary electrons
accelerating in the slot gap and from non-accelerating pairs on field lines just inside the slot gap.  In the case 
of the primary electrons, the calculation closely parallels that of HUM05, who found that accelerating particles
undergoing cyclotron resonant absorption will reach a steady-state where the synchrotron radiation losses are balanced
by the acceleration gain.  The resonant absorption condition is 
\be  \label{rescond}
B' = \gamma\varepsilon_0\,(1-\beta\mu)
\ee
where $\gamma$ is the particle Lorentz factor, $\varepsilon_0$ is the energy of the radio photon in the lab frame
(in units of $mc^2$), $\beta = (1-1/\gamma^2)^{1/2}$, $B' = B/B_{\rm cr}$ is the local magnetic field strength in
units of the critical field $B_{\rm cr} = 4.4 \times 10^{13}$ G, 
$\mu =\cos \theta$, and $\theta$  is the angle in the lab frame between the photon direction 
and the particle momentum (to good approximation the direction of particle momentum is the same as the magnetic 
field direction).  If this
condition is met, the radio photon energy is at the local cyclotron energy in the particle rest frame.  
The resonant condition is not satisfied near the
neutron star surface, where the magnetic field is too strong, but may be achieved when the particle reaches the lower
magnetic fields at high altitudes.  When a particle is undergoing absorption initially in low Landau states, the
rate of cyclotron emission is much lower than the rate of absorption.  The Landau state and pitch angle of the 
particle will therefore increase continuously until an equilibrium is reached between gain in pitch angle through
resonant absorption and the loss in pitch angle through synchrotron emission.  Since this equilibrium is achieved at high
Landau states, the emission is synchrotron rather than cyclotron.

Lyubarski \& Petrova 1998 identified two regimes of resonant absorption as the pitch angle of a particle
increases.  When $\psi \ll \theta$ (i.e. the pitch angle of the particle, $\psi$ is less than the incident angle of the 
radio photons, $\theta$), the particle pitch angle increases but the total momentum stays roughly constant.  When 
$(\theta - \psi) \ll \theta$, the pitch angle stays constant while the total momentum increases.
Petrova (2002) has derived the
solution for the distribution function of electrons undergoing resonant absorption of radio emission in a
pulsar magnetosphere and the corresponding mean-square value of the pitch angle for both of these regimes.
In the case where $\psi \ll \theta$ (equation (2.17) of Petrova (2002)), the mean square of the pitch angle is
\be
\langle {\psi ^2} \rangle = 4\,R\int _{\eta _R}^{\eta } {\it a}(\eta ')d\eta ' ,
\label{psi-msq}
\ee
where $\eta = r/R$ and 
\be
{\it a_0}(\eta ) = {{2\pi ^2 {\it e}^2 (1-\beta\mu_0) I_0}\over {\gamma ^2 m^2c^4}} 
\left( {{\varepsilon _0 \gamma (1-\beta\mu_0)}\over {B'}} \right) ^{\nu },~~~~  \eta > \eta_{_R}.
\label{a0}
\ee 
Here $I_0$ is the intensity of observed radio emission measured in $erg\cdot cm^{-2}\cdot s^{-1}\cdot Hz^{-1}$
and $\nu$ is the radio spectra index. 
Thus, for the perpendicular momentum change due to cyclotron resonant absorption we can write 
\be
\left({{dp_{\perp }}\over {dt}}\right)^{abs} = 2~{\it a_0(\eta)}~c {{\gamma ^2}\over {p_{\perp}}} + 
{{p_{\perp}}\over p} \left({{dp}\over {dt}}\right)^{abs} 
\label{dp_dt_eval}
\ee
where we used the relationship $p_{\perp } = p {\langle {\psi ^2} \rangle}^{1/2}$.  Thus, we 
assume that $p_{\perp }$ is proportional to the root mean-square value of the pitch angle.  
We also make the further approximation of computing the evolution of the root mean-square value
of $p_{\perp }$ rather than the evolution of the particle distribution function.  Since the 
primary and secondary electrons are continuously accelerating, $\gamma$ remains very high and 
$p_{\perp }/p = \sin\psi \ll 1$.  According to Petrova (2003), the width of the $p_{\perp }$ 
distribution is of order $p_{\perp }$, so that the large variations in $\gamma$ and $p$ in r 
along the field lines is much more important in the formation of the spectrum than the spread in the
$p_{\perp }$ distribution.

Combining Eqn (\ref{a0}) and (\ref{dp_dt_eval}), the rate of resonant absorption can be written
\be
\left({{dp_{\perp}}\over {dt}}\right)^{abs} = D{{\gamma^{\nu}}\over {p_{\perp }}} + 
{{p_{\perp } \gamma }\over {\gamma^2 -1}} \left({{d\gamma }\over {dt}}\right)^{abs},~~~~~~~~~\gamma < \gamma _R
\label{dp_perp_abs}
\ee
where
\be  \label{D}
D = 5.7 \times 10^{9}\,\rm s^{-1}\,\gamma_R^{-\nu}\,\left({d_{kpc}\over \eta}\right)^2\,
\Phi_0[{\rm mJy}]\,(1-\beta\mu_0),
\ee 
and we also can neglect the $({d\gamma }/{dt})^{abs}$ term in (\ref{dp_perp_abs}) since it
is small compared to $({d\gamma }/{dt})$ from acceleration and from curvature and synchrotron losses.
In the above expression, we have assumed that $I _0 = \Phi _0\Omega_{rad} d^2/A$, 
where $\Phi _0$ is the measured radio flux (in mJy), $d$ is the source distance, $\Omega_{rad} \sim A/r^2$ is 
the radio emission solid angle, with $A$ and $r$ the cross-sectional area and radius at the 
absorption radius. Also, from the resonant condition (see equation [\ref{rescond}]), $\gamma < \gamma _R$,  
$\gamma _R$ is defined as
\be
\gamma _R = {{B'}\over {\varepsilon _0(1-\beta \mu _0)}} 
= 2.8\times 10^5 {{B_8}\over {\varepsilon _{0, {\rm GHz}} (1-\beta\mu_0)}}. 
\label{gamma_R}
\ee
The resonant terms will switch on only when the resonant condition is satisfied.

In the case where $\theta - \psi \ll \theta$, we assume that the pitch angle remains at the value $\psi = \theta/2$
and the total mean particle momentum evolves as
\be
\bar p = {\Gamma({3\over 3-\nu})\over \Gamma({2\over 3-\nu})}
\left[{(3-\nu)a_1\over b_1\theta^2}\right]^{1/(3-\nu)}
\ee
(Petrova 2003), where   
\be
{\it a_1}(\eta ) = {4\pi ^2 {\it e}^2 {J'}_1^2 I_0\over c^2} 
\left( {{\varepsilon _0 \theta}\over {B'}mc} \right) ^{\nu },~~~~  \eta > \eta_{_R}.
\label{a1}
\ee 
and $b_1 = 2 e^2 B'^2/3 \hbar^2 c$.

To evaluate the local intensity of radio photons $\Phi_0$ and the incident absorption angle $\mu_0$ 
needed in Eqns (\ref{D}) and (\ref{a1}) 
we use the radio core/cone beam model described in Section \ref{sec:radio} above.  
To compute $\Phi_0$, we represent the emission at each point of the core and cone beam in ovc coordinates as a ``beamlet" with opening angle $\mu_{bm}$ centered on the tangent to 
the field line and sum the contributions from all the separate beamlets.  The beamlets that
are modulated by the core beam are located at $1.8 R$ and the beamlets that are modulated by
the cone beam are located at $r_{KG}$ (see Eqn (\ref{eq:rKG})). To evaluate $\mu_o$ for each beamlet that will
encounter a particle at position ($x_p$,$y_p$,$z_p$), we must take into account the aberration of the radio photons 
at their emission point ($x_{bm}$,$y_{bm}$,$z_{bm}$) as well as the light travel time from the radio photon emission point to the particle and the rotation of the field line (and of the particle on that field line).  
Because of the time delays and the rotation of the
magnetic field, not all beamlets will encounter a particle at a given position in the outer magnetosphere.  

At each step along a field line (see \S \ref{sec:sim}), the particle radiates an instantaneous synchrotron spectrum,
given by (Tademaru 1973)
\be
\dot N_{SR} (\varepsilon) = {2^{2/3}\over \Gamma({1\over 3})}\,\alpha_f B' \sin\psi\, \varepsilon^{-2/3}\, 
\varepsilon_{_{SR}}^{-1/3}\exp(-\varepsilon/\varepsilon_{_{SR}}),
\label{nSR}
\ee
where $\sin\psi = p_\perp/p$, $p^2 = \gamma^2 - 1$ and $\varepsilon_{_{SR}} = (3/2)\gamma^2 \, B'\sin\psi$ 
is the synchrotron critical frequency. It is important to point out that the main radiation power produced 
by the particles undergoing resonant absorption of radio photons comes mostly from the parallel energy of the 
relativistic particles, not from the power of the observed radio emission (which is relatively much smaller).
The absorption of radio photons increases the pitch angles of the particles which already have very high $\gamma$.
For high $\gamma$, the pitch angle remains approximately constant while the particle radiates synchrotron radiation,  
because the radiated photons are emitted nearly perpendicular to the magnetic field in the frame of circular motion, and therefore nearly along (within angle $\sim 1/\gamma$) the particle momentum in the lab frame.  The particle then experiences recoil in a direction opposite to its motion, decreasing $\gamma_\perp$ and $\gamma_\parallel$ in proportion to each other, 
allowing the parallel component of energy to be tapped.

\subsection{Non-resonant inverse Compton}  \label{sec:ICS}

The primary electrons will also scatter radio photons that are not in the cyclotron resonance in their rest frames.
We treat this component as a non-resonant Compton scattering, which will be in the Thomson limit since 
$\gamma \varepsilon_0 \ll 1$.  The spectrum of scattered photons can be written
\be  \label{eq:dIph}
{dN_{ICS}\over d\varepsilon_s d\mu_s} = {c\over \gamma(1 -\beta\mu_s)}\,\int d\phi \int d\varepsilon \int_{\mu_{\rm min}}
^{\mu_{\rm max}} d\mu n_{ph}(\varepsilon, \mu)\,{d\sigma'(\varepsilon', \mu')\over d\varepsilon_s' d\mu_s'}\,(1-\beta\mu)
\ee
where $\varepsilon$ and $\varepsilon_s$ are the incident and scattered photon energies in the lab frame, $\mu$ and $\mu_s$ 
are the cosines of the incident and scattered photon angles in the lab frame and 
$n_{ph}(\varepsilon, \mu)$ is the number density of incident radio photons.  The primes denote the corresponding 
quantities in the electron rest frame and are related to those in the lab frame by the Lorentz transformations
\begin{eqnarray}
\varepsilon' - \gamma\varepsilon(1-\beta\mu) \label{eq:LT1} \\
\mu' = {\mu-\beta\over (1-\beta\mu)}
\end{eqnarray}
with same type of expression for the scattered quantities.  The differential cross section for scattering
of an electron in a magnetic field, averaged over photon polarization, in the Thomson limit, is written 
(e.g. Dermer 1990)
\be
{d\sigma'(\varepsilon', \mu')\over d\varepsilon_s' d\mu_s'} = {3\over 8}\,\sigma_T\,\delta(\varepsilon_s'-\varepsilon_s)
\left\{(1-\mu_s'^2)(1-\mu'^2) + {1\over 4}(1+\mu_s'^2)(1+\mu'^2)\left[{u^2\over (u+1)^2} + {u^2\over (u-1)^2}
\right]\right\}
\ee
where $u = \varepsilon'/B'$.  Since the primary electrons have high Lorentz factors, the incident photons will
be beamed into a narrow cone with $\theta' \sim 1/\gamma$ so that $\mu' \approx 1$.  The radio photons will also 
lie at energies near or above the cyclotron resonance in the electron rest frame (see \S \ref{sec:CycAbs}), and 
since we are treating only the non resonant part of the scattering here, 
can approximate the cross section as :
\be
{d\sigma'(\varepsilon', \mu')\over d\varepsilon_s' d\mu_s'} = {3\over 16}\,\sigma_T\,\delta(\varepsilon_s'-\varepsilon_s)
(1+\mu_s'^2)(1+\mu'^2)
\ee
For the incident radio photon distribution local to the primary electrons, we use the form
\be
n_{ph}(\varepsilon, \mu) = n_R \delta(\varepsilon - \varepsilon_R) \Theta(\mu)
\ee
where $n_R$ is the local radio photon density, we approximate the energy distribution as a $\delta$-
function at energy $\varepsilon_R$.  We estimate the radio photon density as
\be
n_R = {L_{\rm radio}\over \varepsilon_0 A(r)c} 
\ee
where comes $L_{\rm radio}$ from Eqn (\ref{eq:Lradio}) and 
$A(r) \simeq \pi r^2 [(\bar\theta+w_e)^2 - (\bar\theta-w_e)^2 ]$, assuming the electron interact
mostly with photons from the conal component.  For the angular distribution we adopt
\be  \label{Theta}
\Theta(\mu) = {1\over (1-\beta_R\mu)^3}\left[1 - {(1-\mu^2)\cos^2\phi\over \gamma_R^2(1-\beta_R\mu)^2}\right]
\ee  
which is that expected for emission from relativistic particles with Lorentz factor $\gamma_R$
(Jackson 1965).  Equation (\ref{Theta}) is meant to describe the distribution of coherent radio 
emission from pairs with Lorentz factors in the range $\gamma_R \sim 10^2$.  In order to calculate
the spectrum of scattered photons, we change variables from $\mu$ to $\varepsilon'$ in Eqn (\ref{eq:dIph})
by means of the Lorentz transform in Eqn (\ref{eq:LT1}).  The $\phi$ integration can be done easily.
We then make use of the $\delta$-functions to perform the $\varepsilon$ and $\varepsilon'$ integrations
and perform the integration over the scattered photon angle $\mu_s$ numerically.

\section{SIMULATION OF MULTIWAVELENGTH SLOT GAP RADIATION}  \label{sec:sim}

In order to simulate the radiation on the sky for a non-rotating distant observer, we have adopted two coordinate 
systems. The acceleration and emission takes place in the corotating frame
(CF).  The frame of the distant, non-rotating observer we will refer to as the inertial observer frame (IOF) 
where the radiation is observed.
We start the simulation by computing the distorted polar cap rim at the neutron star surface in 
the CF.  The foot-point of the field lines on which the 
primary electrons are accelerated are positioned along concentric rings in the ovc and the $(x_0,y_0,z_0)$
coordinates along each field line are computed.  The primary electrons are
confined to the slot gap, which we assume is bounded by the ovc coordinate rings $r_{\rm ovc}^{\rm min} = 0.95$ 
and $r_{\rm ovc}^{\rm max} = 1.0$ with equal spacing $d_{\rm ovc} = 0.004$ giving 11 rings.  Each ring
is divided into 180 equal divisions.  Each electron starts at the stellar 
surface on a trajectory that advances along the field line in increments determined by either
its rate of acceleration, radiation loss rate, perpendicular momentum gain or distance gain. 
The electron trajectory extends to a radius $r_{\rm max} = 0.8\,R_{\rm lc}$, since the field line structure
of the retarded vacuum dipole solution is not accurate very near the light cylinder.  However, the resulting
emission pattern and profiles do not depend very sensitively on emission patterns in the outermost parts of the
magnetosphere, since the bulk of emission originates at lower altitudes (see \S \ref{sec:results}).  
At each step along the field line in the CF, 
the equations of motion for the electron (Eqns [\ref{dgamma}] and [\ref{dp_perp}]) are solved to give the 
new Lorentz factor and perdendicular momentum.
The spectrum of curvature, synchrotron and ICS photons radiated in that step are computed and the radiation
direction is assumed to be tangent to the local field direction in the CF.  The emission direction 
is then transformed to the IOF (aberration), appropriate time delays are added and the radiation from 
curvature, synchrotron and ICS are
accumulated in separate arrays $P(\varepsilon, \zeta, \phi)$, where $\zeta$ and $\phi$ are the 
viewing angle and phase with respect to the pulsar rotation axis for a distant, non-rotating observer.

The electron-positron pairs are also treated in a similar way as the primary electrons, but with the 
acceleration gain rate set to zero, since they are created above the pair formation front which forms 
the inside boundary of the slot gap where the parallel electric field is screened. The pairs are
assumed to flow along open field lines bounded by the ovc coordinate rings $r_{\rm ovc}^{\rm min} = 0.25$ 
and $r_{\rm ovc}^{\rm max} = 0.99$ with equal spacing $d_{\rm ovc} = 0.014$ giving 54 rings. 
On rings $r_{\rm ovc} < 0.25$, the field line radius of curvature increases near the magnetic axis 
and the pair 
multiplicity rapidly decreases (Daugherty \& Harding 1996), so we ignore pair radiation on these
very inner rings.  We ran our code on a parallel processor, where each ring was assigned to a different processor.

\subsection{Particle dynamics and radiation}

The Lorentz factors, $\gamma$, and perpendicular momentum, $p_{\perp}$ (in units of $mc$), 
of each particle will evolve along the field lines according to its equation of motion, which
may be written
\be \label{dgamma}
{d\gamma\over dt}={eE_\parallel\over mc}-{2e^4\over 3m^3c^5}
B^2\,p_\perp^2 - {{2e^2\gamma ^4}\over {3\rho _c^2}}
+ \left({d \gamma\over dt}\right)^{abs} \,
\ee

\be  \label{dp_perp}
{d p_\perp\over dt}=
-{3\over 2}{c\over r}{p_\perp}
-{2e^4\over 3m^3c^5}B^2\,{p_\perp^3\over \gamma} + \left({d p_\perp(\gamma)\over dt}\right)^{abs}.
\ee
The terms of the right hand side of equation (\ref{dgamma}) are acceleration, synchrotron losses,
curvature radiation losses and cyclotron/synchrotron absorption.  The terms of the right hand side 
of equation (\ref{dp_perp}) are adiabatic changes along the dipole field line, synchrotron losses and 
cyclotron/synchrotron resonant absorption.  A derivation of the above equations (minus the CR loss and 
resonant absorption terms) is given in the Appendix of Harding, Usov \& Muslimov (2005).
The ICS losses may be neglected for the primary particles and may also be neglected
for the pairs since the acceleration and synchrotron loss rates are much larger. 

By substituting equation (\ref{dp_dt_eval}) into the right hand sides of equations 
(\ref{dgamma}) and (\ref{dp_perp}), we get 
\be  \label{gamma}
{d\gamma\over dt}=A_1 E_{\parallel}- B_1B_8^2\,p_\perp^2
-C_1\gamma^4 
\ee 
\be  \label{p_perp}
{dp_\perp\over dt}=-A_2 \eta^{-1}\,p_\perp - B_1 B_8^2\,p_\perp^3
{1\over \gamma} + \left({{dp_{\perp }}\over {dt}}\right)^{abs},
\ee
where $A_1 = 1.76 \times 10^{7}\,\rm s^{-1}$, $B_1 = 1.93 \times 10^7\,\rm s^{-1}$, 
$C_1 = 5.6 \times 10^{-3}\,\rm s^{-1}$, $A_2 = 4.5 \times 10^4\,\rm s^{-1}$, 
$E_{\parallel}$ is in e.s.u. and $B_8 \equiv B/10^8$ G. Since Petrova (2002) has assumed 
that $p$ and $\gamma$ are constant to compute the change in pitch 
angle due to resonant absorption, we have neglected the change in $\gamma$ due to 
absorption in equation (\ref{gamma}).  Both $E_{\parallel}$ and
$B_8$ are functions of $\eta$.  For the primary electrons, $E_{\parallel}$ is computed using
Eqns (\ref{eq:Epar - lo}), (\ref{eq:Epar-high}) and (\ref{eq:Epar - comb}).  For the pairs,
we assume $E_{\parallel} = 0$.

\section{RESULTS FOR THE CRAB PULSAR}  \label{sec:results}

The Crab, with a rotation period $P = 33$ ms, and period derivative 
$\dot P = 4.22 \times 10^{-13}\, s/s$, implying a surface
magnetic field of $B_0 = 8 \times 10^{12}$ G, is the youngest pulsar having detected pulsed $\gamma$-rays.  
The $\gamma$-ray profile shows
two sharp peaks with a phase separation of about $0.4$ or $144^{\circ}$.  Very similar profiles appear at
all wavelengths, including radio, where an additional pre-cursor peak leading the first or main peak 
is present at frequencies below about 450 MHz.  We model the Crab spectrum and pulse profile using our 
model of emission from primary electrons accelerating in the slot gap and non-accelerating electron-positrons 
pairs flowing along field lines inside the slot gap (i.e. field lines at smaller colatitude).  Polar cap 
pair cascade models (Daugherty \& Harding 1996; Muslimov \& Harding 2003) find that pair multiplicity 
(number of pairs per primary particle) for the Crab is in the range $10^4 - 10^5$, where the lower number
applies to colatitudes nearer the magnetic axis while the latter number applies to cascades extending
to higher altitudes of $3 -4$ stellar radii, initiated by electrons accelerating in the slot gap.
In simulating the radiation from pairs, we trace particles along all open field lines with an
assumed broken power law spectrum of energies
\be
{\cal N}_{\rm pairs}(\gamma_p) = 
\left\{\begin{array}{lr} 

A_1 \gamma_p^{-\delta_1}, & \gamma_p^{\rm min} < \gamma_p < \gamma_p^{\rm br} \\
A_2 \gamma_p^{-\delta_2}, &  \gamma_p^{\rm br} < \gamma_p < \gamma_p^{\rm max}
\end{array}
\right.
\ee
where the normalization constants are set to preserve the total pair multiplicity 
${\cal M}_{\rm pairs}(r_{\rm ovc})$ in each ring.   We explore two geometries for the radio
emission beam: the `standard' beam model described in \S \ref{sec:radio} with the core emission at
a fixed altitude and cone emission at a single frequency-dependent altitude, and a non-standard
model with core emission at the same fixed altitude but the cone emission is extended along the 
last open field line to produce caustic peaks.  

Figure 1 plots several examples of the evolution of particle Lorentz factor and perpendicular momentum
that are solutions to the equations of motion (\ref{dgamma}) and (\ref{dp_perp}) along magnetic field lines at 
leading and trailing edges of the open volume, for the extended radio cone emission case.  
Also plotted are the acceleration gain rate, 
$d\gamma_{\rm acc}/dt$, curvature loss rate, $d\gamma_{\rm cr}/dt$, 
synchrotron loss rate, $d\gamma_{\rm sr}/dt$ , cyclotron absorption rate, $d\gamma_{\rm abs}/dt$, 
and critical synchrotron energy $\varepsilon_{_{\rm SR}}$.  In the top plots, showing these quantities for
primary electrons, the electric field acceleration falls off rapidly from its initially high value near the 
neutron star surface but maintains a nearly constant lower value out to high altitudes.  The electron
Lorentz factor is limited by curvature radiation losses at fairly low altitude, with the gain rate from
acceleration balancing the curvature loss rate along nearly its entire path (thus the two curves lie on
top of each other).  The absorption begins at the lowest altitude of conal radio emission, 
$r_{\rm min} \sim 0.2 R_{lc}$ at 1 GHz, producing a sudden increase in $p_\perp$, which then increases 
steadily with altitude.  As a result, the synchrotron losses turn on and reach a level nearly that of 
curvature losses on the leading field lines, but somewhat lower on trailing field lines.  The fluctuations
in absorption rate are due to the electron encountering radio emission from different beamlets 
(see \S \ref{sec:CycAbs}) and so reflects the numerical resolution of these sub-elements of the 
radio cone beam.  At higher altitude,
the radio cone emission turns off, the electrons on trailing field lines no longer encounter radio photons
and $p_\perp$ begins to drop along with synchrotron losses.  The synchrotron radiation reaches energies 
of several hundred MeV.  The bottom plots show the evolution of these quantities, except for acceleration
gain (which is zero) and curvature losses (which are negligible), for an electron or positron of Lorentz
factor $\gamma_p = 10^5$ near the upper end of the pair spectrum (see below).  The evolution of $p_\perp$
and $d\gamma_{\rm abs}/dt$ show similar behavior to that of the primary electrons, except that the pairs
`see' the radio photons out to higher altitude since they are on more interior field lines.  The synchrotron 
emission of these highest-energy pairs reaches energies of a few MeV, while that of the lowest energy
pairs with $\gamma_p = 10^2$ peaks at optical frequency.

We adopt a model where the slot gap high-altitude emission best reproduces the Crab phase-averaged spectrum.  
This model has $M = 1.4 M_{\sun}$, $R = 14$ km, $I = 4 \times 10^{45}\,\rm g\,cm^2$ and $\lambda = 0.1$ for the 
parameter values determining the primary electron acceleration.  For the parameters of the pairs spectrum, we
assume $\gamma_p^{\rm min} = 10^2$, $\gamma_p^{\rm max} = 2 \times 10^5$,
$\gamma_p^{\rm br} = 5 \times 10^3$, $\delta_1 = 2.0$, $\delta_2 = 2.8$, which matches the pair distribution derived
for the Crab parameters in Fig. 7 of Daugherty \& Harding (1982).  The pair multiplicity was 
varied in rings over the polar cap, such that 
\be
{\cal M}_{\rm pairs}(r_{\rm ovc}) = 
\left\{\begin{array}{lr} 
4 \times 10^3 & 0.25 < r_{\rm ovc} < 0.5 \\
2 \times 10^4 & 0.5 < r_{\rm ovc} < 0.9  \\
4 \times 10^5 & 0.9 < r_{\rm ovc} < 0.99
\end{array}
\right.
\ee
The multiplicity of cascades on the field lines near the magnetic axis is lower than that for cascades near the slot gap.  
Since the $E_{\parallel}$ near the slot gap has a lower magnitude, primary electrons
accelerate and produce extended cascades over much longer distances.  The primary electrons also produce
most of the cascades during their radiation-reaction limited phase, further increasing the multiplicity of
pairs near the slot gap.

\subsection{Standard radio beam}

Using the output arrays $P(\varepsilon, \zeta, \phi)$ from the simulations, we can display plots of
radiation intensity on the sky, or phaseplots, in different energy bands.  Figure 2 shows phaseplots of 
high-energy emission assuming the standard radio emission model for
three energy bands: $1-20$ keV, $0.1 - 10$ MeV, and $> 100$ MeV.  The radiation distribution shown in the 
high-energy phaseplots exhibits caustics,
extended bright lines of emission from particles on the trailing field lines from each pole.  The near 
cancellation of phase shifts due to retardation, aberration and field line curvature cause emission at
a wide range of altitudes to arrive in phase, while emission on the leading edge of the open volume is 
spread-out in phase (Morini 1983, Dyks \& Rudak 2003).  Pulse profiles for these energy 
bands are obtained by displaying the intensity as a function of phase at a particular viewing 
angle $\zeta$.  At viewing angles that cut across caustics, the profiles show two peaks with phase
separation less than $180^{\circ}$.  We also show the phaseplot at 400 MHz
of the radio emission, which is dominated by the cone beam.  The cone beam is shifted earlier in phase
relative to the core beam due to the difference in aberration and retardation of the higher-altitude cone
emission.
We find that inclination angles in the range $\alpha = 40^{\circ} - 55^{\circ}$ give the
best combination of spectrum and profile to match the Crab profile and spectra.  For large inclination angles, the 
parallel electric field reverses direction on some field lines (see MH03), and for 
small inclination angles, radiation from the low-altitude pair cascades dominates the emission and
produces double-peaked profiles for small viewing angles (MH03).  
For inclination angle $\alpha = 45^{\circ}$, viewing angles in the range 
$\zeta = 78^{\circ} - 82^{\circ}$ and $\zeta = 98^{\circ} - 102^{\circ}$ produce profiles having two peaks
with phase separation near 0.4. The phase-averaged emission spectra are very similar for the different angles in this range.  
The profiles for energy bands $1-20$ keV, $0.1 - 10$ MeV look identical
because both are due entirely to pair synchrotron radiation whose geometry is not energy
dependent.  The profile for energies $> 100$ MeV are significantly different, with the first peak now
larger than the second peak, and the phases somewhat shifted from those at lower energy.  
While the high-energy profiles can reasonably reproduce the observed
Crab profiles, the standard model radio profile does not reproduce that observed.
The high energy phaseplots and profiles are in units of $\rm ph \,s^{-1}/ster$ and 
$\rm ph \,s^{-1}/ster/N_{\phi}$, 
respectively, where $N_{\phi} = 180$ are the number of phase bins.  The phase-averaged flux for an
observer at viewing angle $\zeta$ is then the sum over the emission in the profile divided by source
distance squared.  For the viewing angle $\zeta = 100^{\circ}$ shown in Figures 2 and distance of 2 kpc, the 
phase-averaged flux is $\langle \Phi(> 100 {\rm MeV})\rangle = 4.3 \times 10^{-6}\,\rm ph \,s^{-1}\,cm^{-1}$ 
and $\langle \Phi(0.1-10 {\rm MeV})\rangle = 0.05\,\rm ph \,s^{-1}\,cm^{-1}$, which are in good
agreement with observed values.

In Figure 3, we show the our model phase-averaged spectrum for the same case 
$\alpha = 45^{\circ}, \zeta = 100^{\circ}$ shown in Figure 1.  Four components are visible and three 
components make significant contributions to the total spectrum.  Synchrotron radiation from pairs 
contributes at the lowest energies, from infra-red and optical through hard X-rays, turning over at
around $20$ MeV.  The range of this component reflects the pair spectrum, with the low energy turnover 
determined by $\gamma_p^{\rm min}$, the high-energy turnover set by $\gamma_p^{\rm max}$, the break
dividing lower and upper slopes determined by $\gamma_p^{\rm br}$, $\delta_1$ and $\delta_2$.  
Synchrotron radiation from primary electrons contributes at the mid-range of $20 - 300$ MeV and has 
a smaller energy range, reflecting the smaller energy range of the primaries.  Finally, curvature 
radiation from primaries contributes at the highest energies, from $\sim 100$ MeV to the turnover
at around 5 GeV that is determined by the parameters of $E_{\parallel}$.  The component due to non-resonant
ICS from primaries and from pairs (which appears below the scale of the plot) make negligible contributions. 
The non-resonant Compton scattering contribution is much lower than that of resonant absorption because
the cyclotron absorption cross section is orders of magnitude higher, being a first order process, than the 
Thomson cross section for non-resonant ICS, being a second-order process.  Furthermore, the particles
undergo absorptions of many photons at the resonant before radiating the most significant synchrotron emission, 
reaching
high Landau levels.  Thus the ratio of resonant absorption to ICS is even larger than the simple ratio 
of the cross sections.  Curvature radiation from pairs is also negligible since their Lorentz factors
are much lower than that of the primary electrons.  
Our model spectrum plotted in Figure 3 has not been adjusted arbitrarily to fit the 
data.  Rather, we have adjusted the model parameters within a reasonable range.  
The pair spectrum can be tuned to match the optical-to-X-ray spectrum very well by
adjusting its energy range and shape, as well as the variation of multiplicity across the polar cap. 
However, the values that best fit the data are not far from those that come out of pair cascade
calculations.  The model spectrum does not match the high-energy spectrum quite as well, with the
peak of the curvature spectrum that gives a high enough cutoff energy being somewhat above the data
points, although the EGRET data points have large errors.  The EGRET sensitivity above 1 GeV has recently 
been re-examined by Stecker et al. (2007), and they concluded that the Crab phase-averaged spectral points (and 
indeed those of all EGRET sources) 
should be systematically lowered by as much as a factor of 2 above 1 GeV.  We also plot these corrected points
in Figure 3 and they provide an improved match to our model spectrum.  The multicomponent nature of the model 
spectrum produces dips around 20 and 200 MeV.  
The observed spectrum seems to show the dip at 20 MeV but a dip at 200 MeV is not clear, given the large errors.
Measurements with the GLAST LAT should define such details much better.

The phase-resolved spectra are shown in Figure 4, for the phase ranges adopted by Kuiper et al. (2001)
to study the Crab pulsed emission.  Our model spectra are obtained by summing the 
emission for the equivalent phase intervals (our Peak 1 occurs at phase 0.41 and Kuiper et al.'s 
Peak 1 occurs at phase 0.0) in the phaseplots at the chosen viewing angle.  The flux levels of the 
resulting model spectra were then plotted with the data from the corresponding phase intervals without
any renormalization.  Generally, the phase-resolved spectra do not fit the data as well as the 
phase-averaged spectrum.  This may be due to the fact that the low-energy profile peaks (due to 
pair synchrotron emission) are not exactly phase-aligned with the peaks of the high-energy profile (due mostly
to primary curvature emission).  However, the data and model spectra match fairly well for the P1, P2 and
bridge intervals.  The primary synchrotron component makes a larger contribution in the Peaks and the 
Bridge, which is expected since the primary electrons radiate only along a narrow set of field lines. 

\subsection{Extended radio cone beam}

The phase alignment of the first and second peaks in the profile across frequency bands from radio 
to high-energy $\gamma$-ray is a hallmark of the Crab pulsar and difficult to explain unless the 
emission originates from the same location in the magnetosphere.  As we have seen in the last section,
the caustics that form in emission along the trailing field lines at the edge of the open volume naturally 
produce phase-aligned peaks, even if the emission does not originate at the same location along the 
field lines.  A requirement for caustic peak formation is emission that is extended along these
field lines above altitudes of a few tenths of $R_{lc}$.  One possibility to explain the phase alignment
of radio and high-energy peaks in the Crab profile is that the radio cone emission, which is already 
expected in the standard model to come from altitudes above $0.1R_{lc}$, is extended by several tenths
of $R_{lc}$ in altitude.  To explore this possibility, we modify the standard cone beam model by
spreading the emission from $r_{\rm min} = r_{\rm KG}$ to $r_{\rm max} = N r_{\rm KG}$ and decreasing its  
width $w_e$, preserving the same luminosity.  To implement this extended radio beam in the cyclotron
resonant absorption/synchrotron emission component, we divide the beamlets of incident radio photons 
among a number of discrete altitudes between $r_{\rm min}$ and $r_{\rm max}$.

The resulting phaseplots and profiles for the extended radio emission model are shown in Figure 5.  The radio
emission indeed forms caustics similar to those seen in the high-energy emission phaseplots.  To produce 
radio caustics we need to extend the radio conal emission from $r_{\rm min} = r_{\rm KG}$ to 
$r_{\rm max} = 5 r_{\rm KG}$ and take its width as $0.2 w_e$.  The resulting radio profile at the same 
$\zeta = 100^{\circ}$ displays two peaks separated by 0.4 in phase that are at the same phases as the
high-energy peaks.  The high-energy phaseplots and profile calculations use the same parameters for the 
primary electrons and pairs as the model described in the previous section, the only difference being the
radio beam geometry.  There are only minor differences in the high-energy profiles, with a higher and 
more evenly distributed emission level in the bridge region and a less extended trailing wing on Peak 2
above 100 MeV. 

The resulting high-energy phase-averaged spectra, shown in Figure 6, are also very similar to those of 
the standard radio beam case in Figure 3.  The level of the pair synchrotron spectrum is slightly lower 
but the shape is the same.  The shape of the primary synchrotron spectrum is somewhat different, but the
peak is still around 100 MeV.  The primary curvature radiation is of course unchanged since it does not
depend of the radio emission.  There are somewhat larger differences in the phase-resolved spectra, 
shown in Figure 7, since they are more sensitive to the details of the radio emission geometry.  
The pair synchrotron components are lower
in the leading wings, LW1 and LW2, but higher in the trailing wings, TW1 and TW2, and about the same in 
the peaks, P1 and P2, and bridge intervals.   Similarly, the primary synchrotron components are lower
in the leading wings and higher in the trailing wings, with the exception of TW2, but higher in the
Bridge.

\section{DISCUSSION}   \label{sec:dis}

A model of radiation from young, rotation-powered pulsars has been investigated, assuming that the emission
we observe occurs at high altitudes in the magnetosphere.  We include two distributions of particles
radiating along open field lines at high altitude: primary electrons accelerating in the slot gap from the
stellar surface to near the light cylinder, and 
non-accelerating electron-positron pairs on field lines bordering and interior to the slot gap.  Simulating
spectral formation in 3D, we are able to study pulse profiles in different energy bands as well as 
phase-resolved spectra.  We choose to first model the Crab pulsar, since it is bright in all wavebands from
optical to high-energy $\gamma$-rays, with good phase-resolved spectral measurements.  
In our model, three different components dominate the Crab phase-averaged spectrum, curvature and synchrotron
radiation of primary electrons and synchrotron radiation of pairs.  The synchrotron radiation results
from cyclotron resonant absorption of radio photons in the low magnetic field at high altitude.  
Cyclotron resonant absorption of radio photons using standard core and cone emission models can reproduce
high-energy pulse profiles but not the observed radio profiles or the phase alignment of high-energy and radio peaks.  
Extension of conal radio emission along the last open field lines results in caustic formation that can both 
produce a double-peaked radio profile and the phase-alignment with high-energy peaks.
However, we find that the peaks in the model profiles at energies below 100 MeV are not exactly phase-aligned 
with the peaks in the profile above 100 MeV.  Also, our model phase-averaged spectrum does not reproduce exactly the 
observed spectrum above 100 MeV, because the curvature radiation component appears high relative to the that of
the primary synchrotron radiation.  The observed spectrum in the peak intervals is matched better by the 
model spectra where the curvature radiation component is relatively lower.  We predict that, due to the 
multicomponent nature of the model spectra, transitions between different components may be observed as dips 
in the peak and phase-averaged spectrum around 20 and 200 MeV which may be verified by GLAST LAT observations. 

Since absorption of radio photons is critical to the production of the synchrotron emission components, this model 
predicts that the radio and high-energy emission in the synchrotron dominated parts of the spectrum, 
should be correlated in time as well as in phase.  
In our phase-averaged model spectrum of the Crab, the synchrotron radiationn dominates below about
200 MeV, so we predict that correlated variability should be seen in optical, X-ray and $\gamma$ rays below
200 MeV.
Although Lundgren et al. (1995) detected no observable correlation between radio
giant pulses and EGRET flux, Shearer et al. (2003) found small but significant (3 \%) enhancement 
of optical flux during Crab giant radio pulses.  Lommen et al. (2007) found a correlation between the 
arrival time of pulses from Vela with X-ray profile shape as measured by RXTE.  
Such a correlation suggests a strong connection between radio and X-ray emisson.  All of these results
are consistent with our prediction if the EGRET photons that Lundgren et al. (1995) considered were dominated by the
curvature radiation component ($\gsim 200$ MeV), which should not exhibit any correlations with radio emission.

Although we have found from purely geometrical consideration that radio emission extended along the outer edge of
the open field lines can produce a double-peaked profile whose peaks are in phase with those of the high-energy
profiles, we must ask why the radio emission geometry of the Crab pulsar should be so different from that of
other pulsars, including apparently the other known $\gamma$-ray pulsars.  But if one looks at known X-ray
pulsars, the radio and X-ray profiles of some millisecond pulsars such as PSR B1957+20 and PSR B1821-24 also 
have phase-aligned peaks.  In fact, the standard core and cone radio geometry is based primarily on emission
morphology of pulsars having much longer periods in the range 0.5 - 1 s.  It is possible that the radio emission
geometry in faster pulsars could be quite different.  One could speculate that the slot gaps, that are predicted
to exist only in the young and/or fast pulsars, sustain a beam of high energy primary particles near the last open
field line that can interact with a flow of lower energy pairs, or an opposite flow of particles in a return
current, to produce instabilities that lead to an extended radio component.  These pulsars are also the ones that have 
giant radio pulses, a phenomenon that may be related to the non-standard emission geometry (Hankins \& Eilek 2007).

Takata \& Chang (2007) and Tang et al (2007) recently presented results of a modified outer gap model for energy 
dependent pulse profiles and phase-resolved spectra of the Crab pulsar.  In that work the classic outer gap, which 
extended from the null charge surface to the light cylinder along the last open field line, is assume to be extended
below the null chanrge surface, so that the LW1 and TW2 parts of the pulse can be produced by emission from the
second magnetic pole.  The geometry of the modified outer gap model and that of the slot gap model presented in this
paper are now quite similar.  All the different parts of the profile come from the same regions of the magnetosphere in 
both models, with the exception of the first peak, which in the outer gap primarily comes from near the light cylinder.
In the slot gap model, the first peak comes mostly from the caustic of the pole opposite to the one producing the second 
peak, although our $> 100$ MeV profile also has substantial components from both poles.  The radiation mechanisms
of the outer gap are very different from those of our slot gap model, with the major difference being the link to radio
emission through resonant absorption of the radio emission in the slot gap model.  While the optical to hard X-ray 
component of the Crab spectrum is produced by synchrotron radiation of pairs in both models, resonant absorption is 
responsisble for the pair pitch angles in the slot gap model but in the outer gap the pair 
pitch angles come from cascades outside the gap.  Also, the 10 MeV to 500 MeV spectrum comes from inverse Compton
emission in Tang et al (2007), while in our model this component comes from synchrotron radiation of primaries undergoing
resonant absorption.  The highest energy part of the Crab spectrum, above a few hundred MeV, comes from primary
curvature radiation in both models, although this component originates only from very high altitude in the
outer gap model.

Although we have not presented results on the polarization here, this will be explored in future work.  The
expected properties of polarization in a two-pole caustic geometry have been investigated by Dyks et al. (2004).
They found that the position angle and polarization degree as a function of phase produce a reasonable
match to the optical polarization of the Crab pulsar (Kanbach et al. 2005).  Since the radiation modeled 
in this paper includes
emission on interior field lines as well as in the slot gap, we expect that the polarization properties could
be somewhat different and perhaps also energy dependent.

Other Crab-like pulsars, such as PSR B0540-60 and PSR B1509-58, have very similar X-ray to low-energy $\gamma$-ray  
spectra to that of the Crab, but notably no EGRET detections.   However, their single peaked high-energy profiles 
are quite different from the Crab profile, suggesting a different viewing angle.  Both of these pulsars are at
larger distances, so that the EGRET upper limits are above the level of emission they would have with an exact
match of the Crab spectrum.  A GLAST LAT detection of the high-energy emission component(s) would thus be able to
probe the geometry of the slot gap or of other radiation models.

These pulsars and also the Vela-like pulsars will be investigated in future work.  Vela is predicted to have 
a slot gap so the spectrum above 100 MeV should be similar to that of the Crab.  But since the pair multiplicity 
should be smaller than for the Crab, one expects a lower pair synchrotron component which may also peak at
a different energy.  The relation of the radio to high-energy peaks in Vela-like pulsars is quite different,
with a single radio peak leading double-peaks at high energy.   One possibility is that the radio cone is at too
low an altitude to form caustic peaks (Harding 2005).

In this work we have assumed steady-state acceleration in a space-charge limited flow model.  However, since the
time-dependence of such accelerators has not been investigated, this may not be correct.  In fact, the  
magnitude of the current in SCLF models has not been determined self-consistently but is assumed to be  
constant at the Goldreich-Julian value across the polar cap.  This current distribution is not consistent with that 
found in global force-free models of the pulsar magnetosphere (Spitkovsky 2006, Timokhin 2006), that require a
current that departs from the Goldreich-Julian value in the outer parts of the polar cap.  Matching the current
distribution required by the global models with the average boundary condition on the charge flow of the SCLF 
accelerator implies either temporal or spatial fluctuations in the current and voltage.  A lower primary charge
flow in the slot gap or a smaller slot gap width may in fact produce a better match of the model spectrum to the observed spectrum of the Crab.
Investigation of the temporal stability and boundary conditions of acceleration models will ultimately be needed in 
order to accurately model and understand pulsar high-energy emission.

We thank Alex Muslimov and Kouichi Hirotani for comments on the manuscript and an anonymous referee for
very helpful suggestions.  We
would like to acknowledge support from the NASA Astrophysics Theory Program (AKH), the NASA 
Undergraduate Student Researchers Program (JVS) and KBN grant N203 017 31/2872 (JD).  
We also thank the NASA Center for Computational
Science for allotment of time on parallel processors Halem and Discover, and for their help with parallel
coding.

\newpage

\hskip -0.5cm\includegraphics[width=10cm]{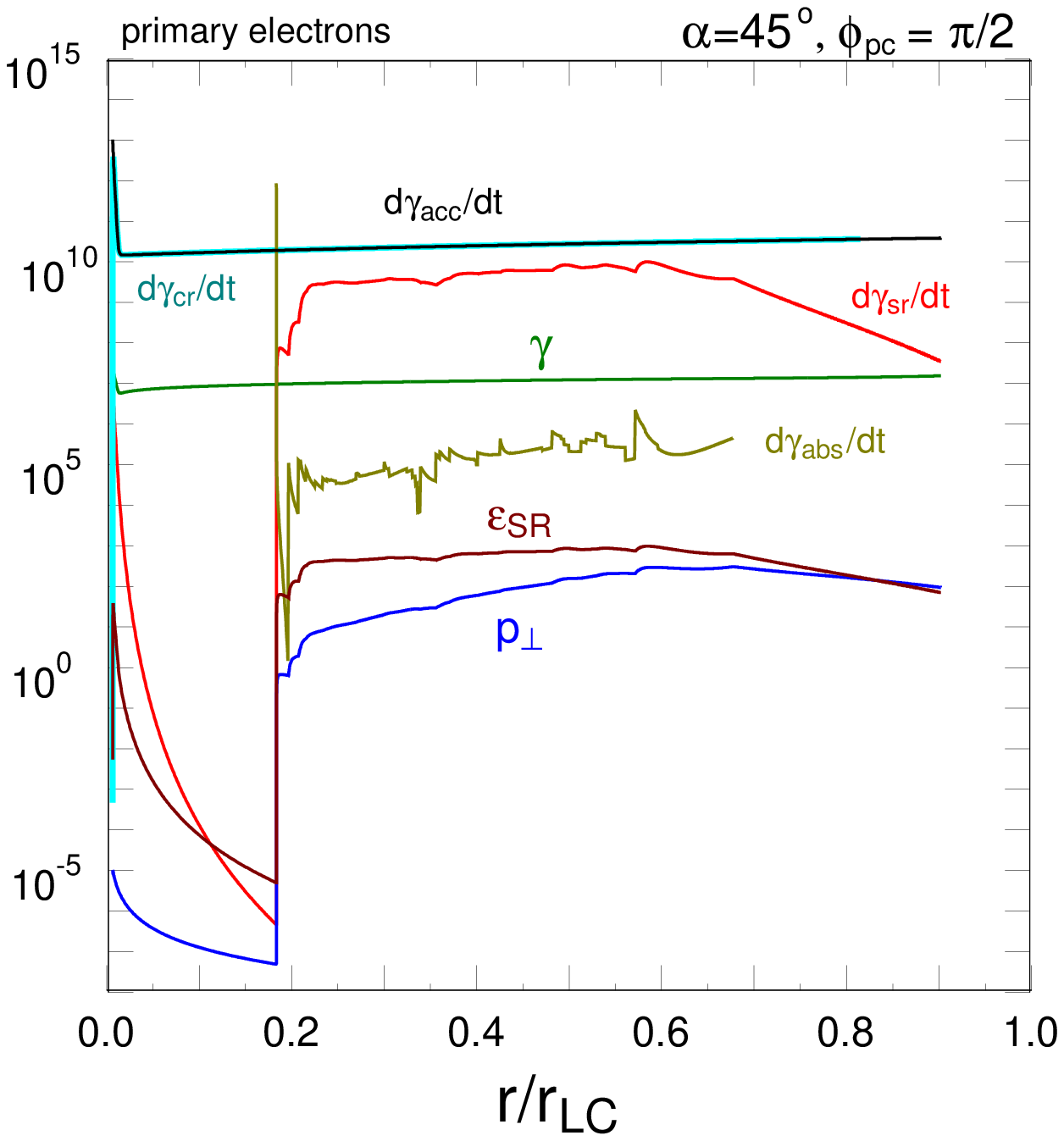}\hskip -2cm\includegraphics[width=10cm]{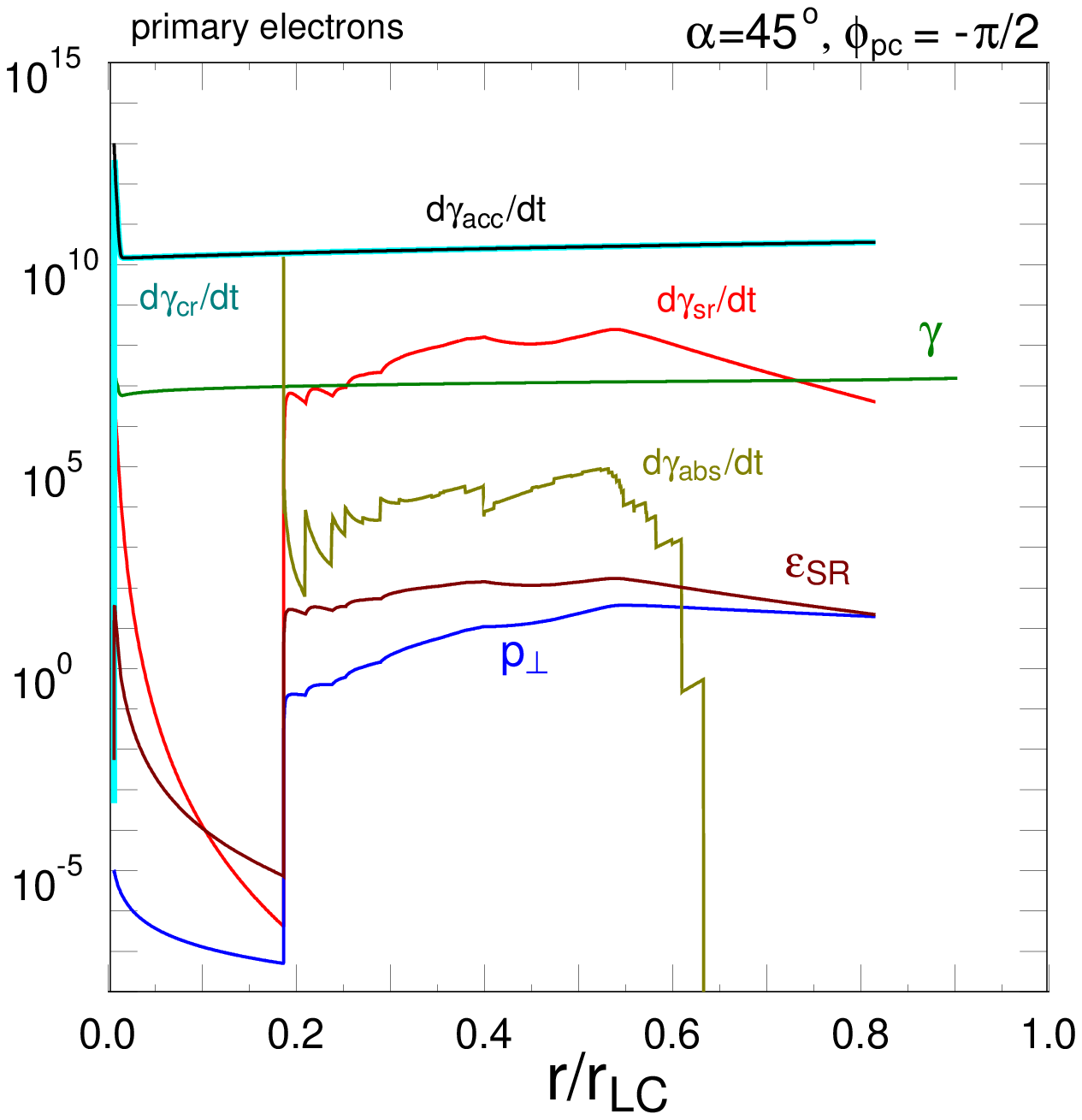}

\hskip -0.5cm\includegraphics[width=10cm]{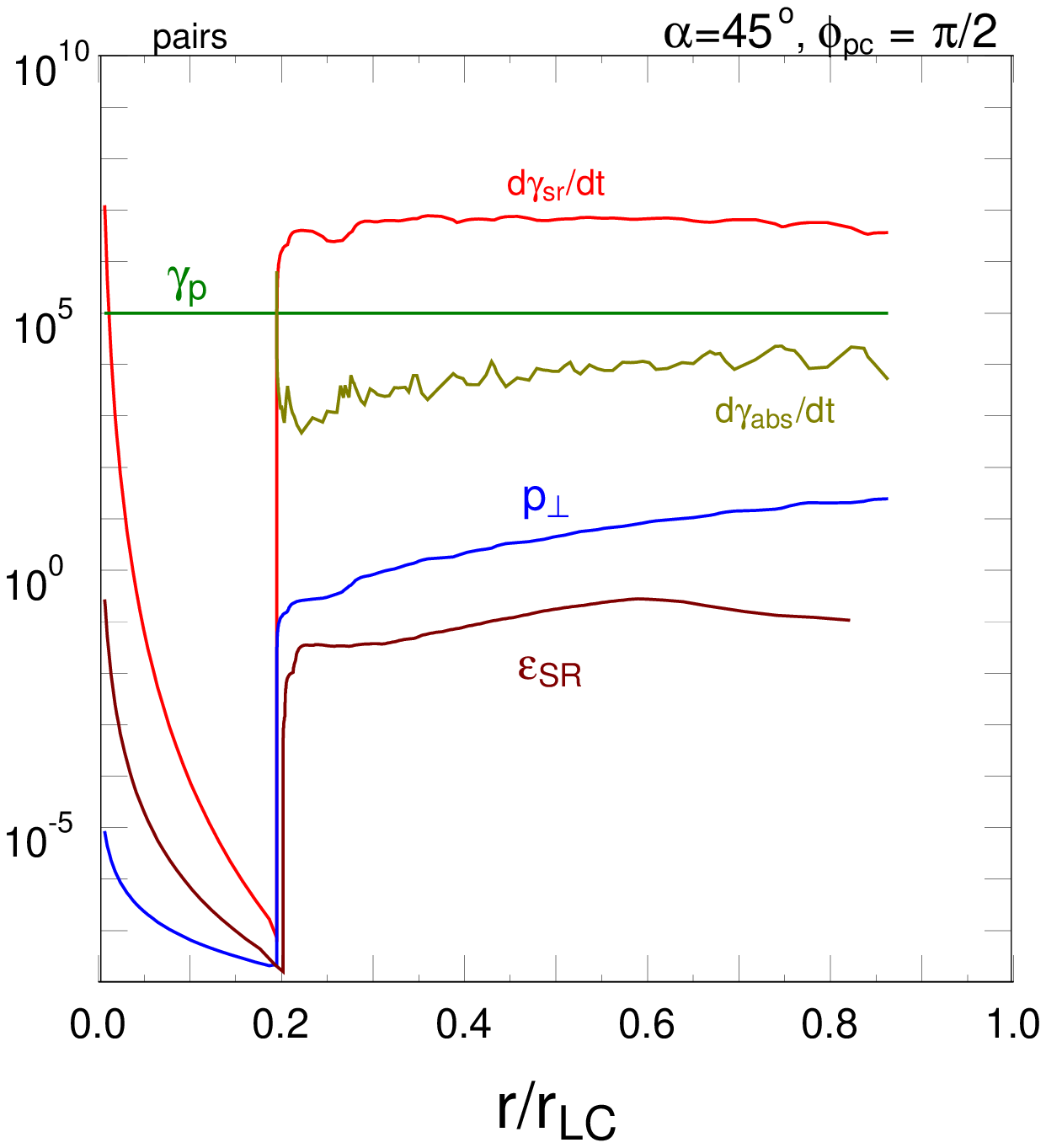}\hskip -2cm\includegraphics[width=10cm]{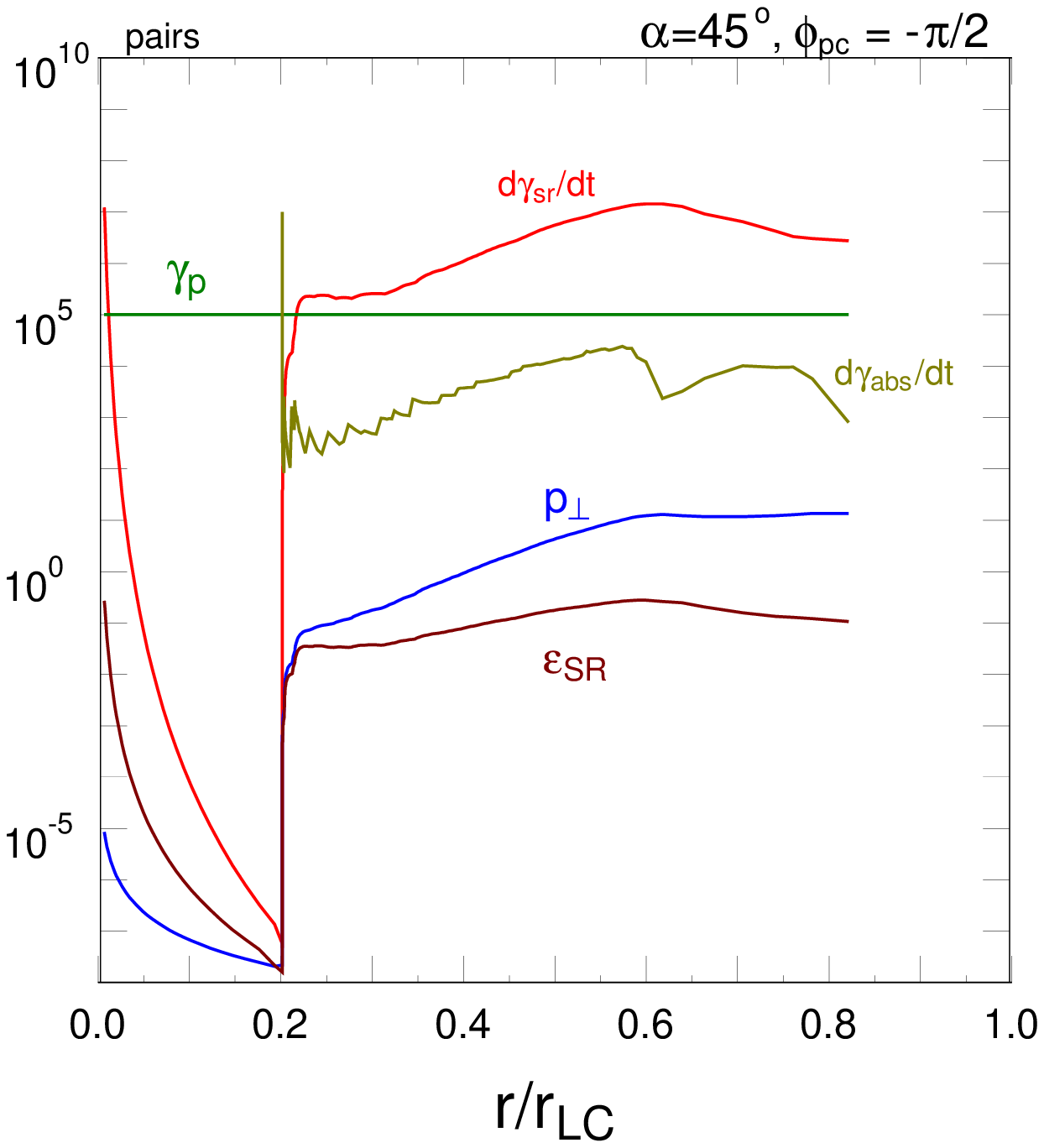}

\figureout{f1.eps}{
Evolution of the dynamics and radiation of particles as a function of distance (in units if light cylinder radius) 
along leading ($\phi_{\rm pc} = \pi/2$) and trailing ($\phi_{\rm pc} = -\pi/2$) magnetic field lines.  
Quantities plotted are particle Lorentz factor, $\gamma$, 
perpendicular momentum, $p_{\perp}$ (in units of $mc$), acceleration gain rate, $d\gamma_{\rm acc}/dt$ ($s^{-1}$), 
curvature loss rate, $d\gamma_{\rm cr}/dt$ ($s^{-1}$), synchrotron loss rate, $d\gamma_{\rm sr}/dt$ ($s^{-1}$), 
cyclotron absorption rate, $d\gamma_{\rm abs}/dt$ ($s^{-1}$), and critical synchrotron energy $\varepsilon_{_{\rm SR}}$ 
(in units of $mc^2$).}

\newpage 
~
\vskip -1.6cm\includegraphics[width=6.8cm]{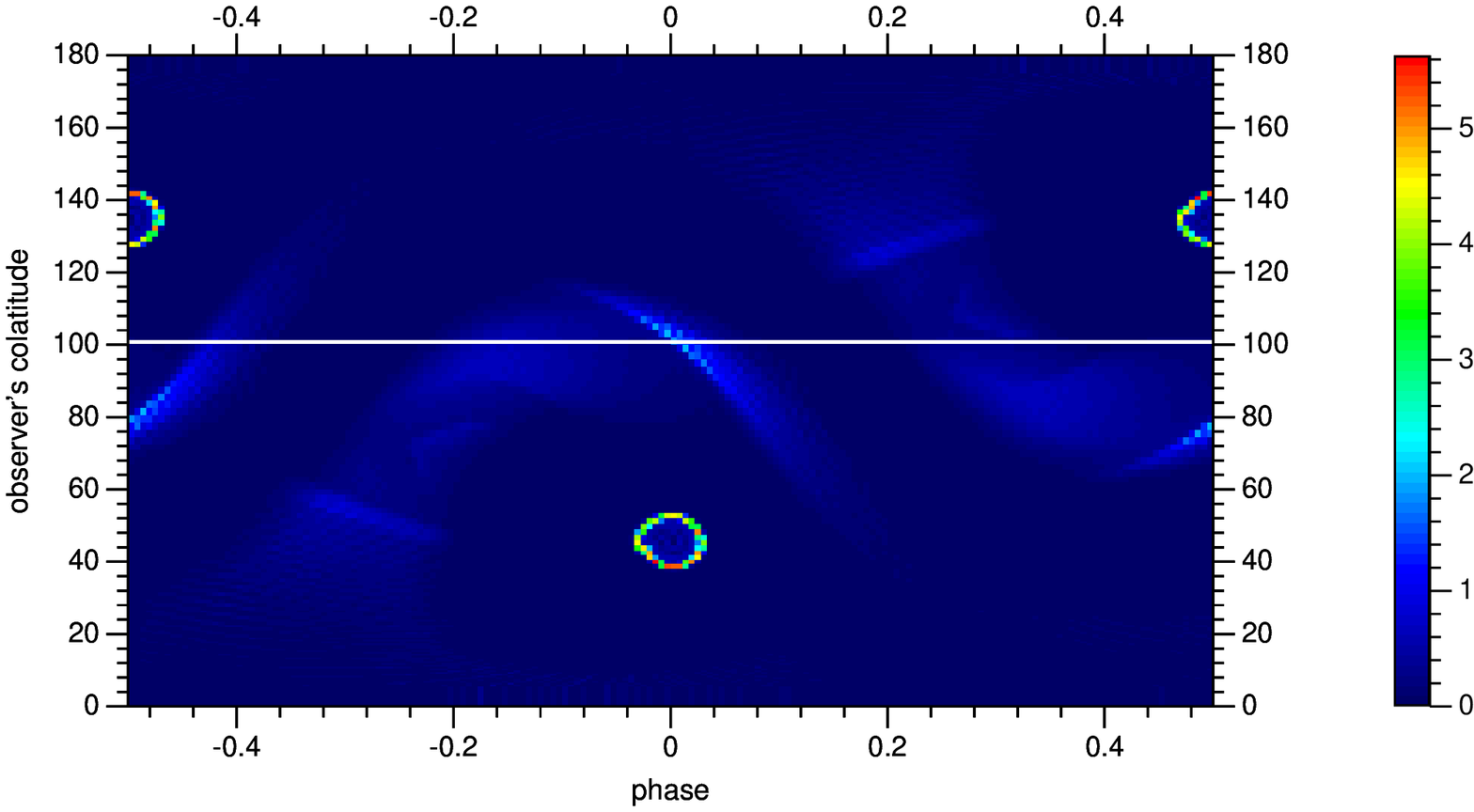}\hskip 1.8cm\includegraphics[width=8.6cm]{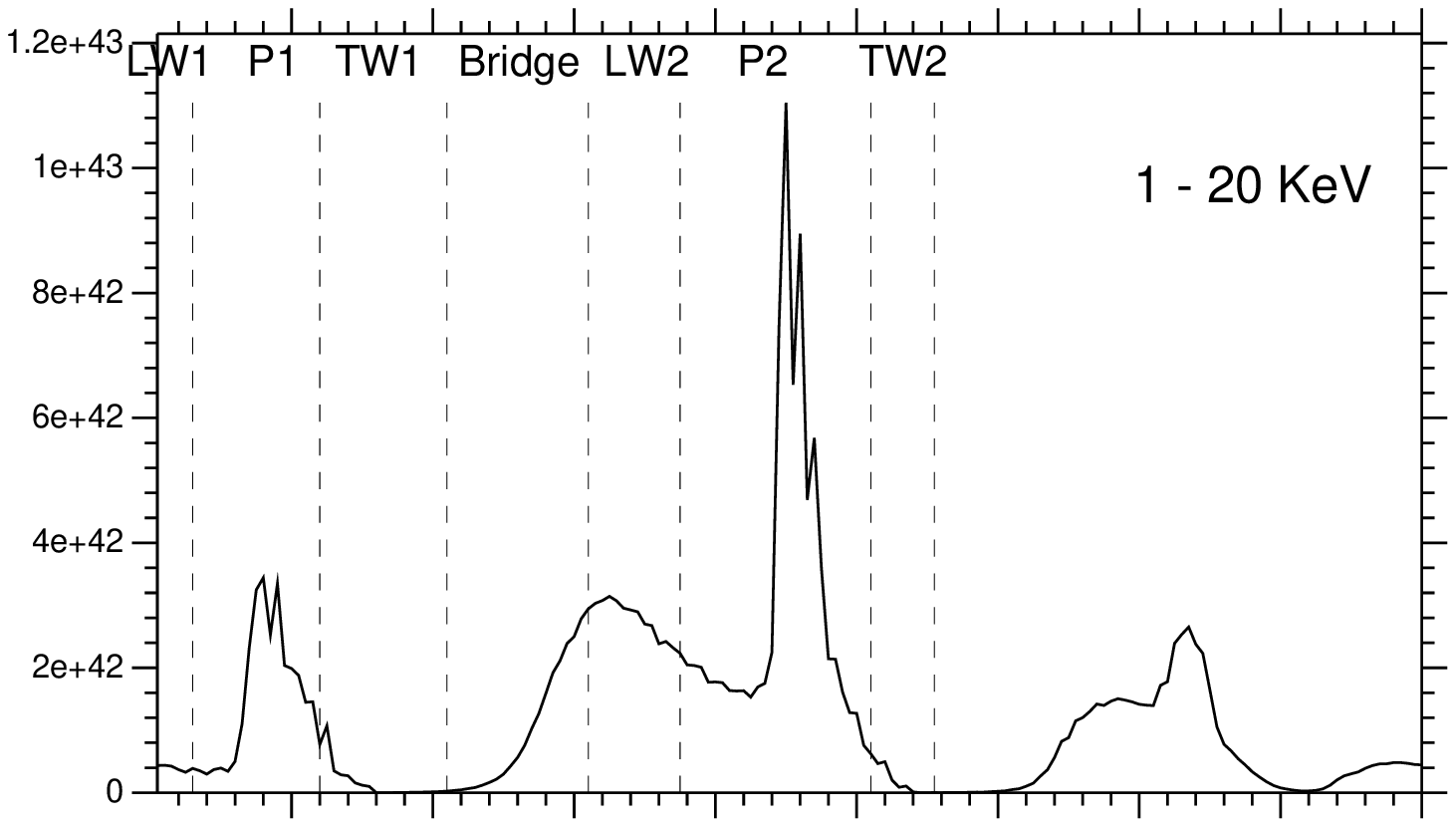}\hskip 0cm

\vskip -0.2cm\includegraphics[width=6.8cm]{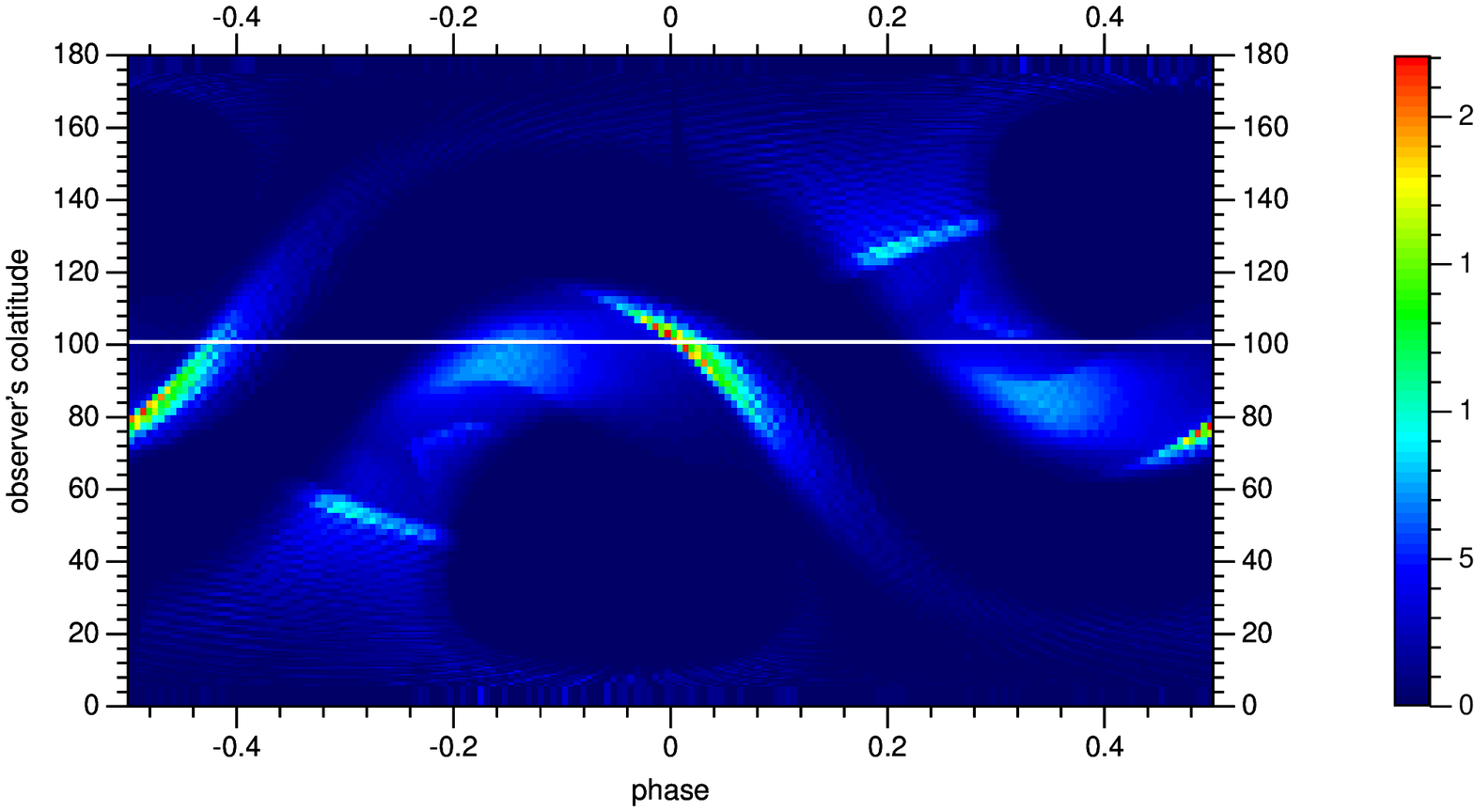}\hskip 1.8cm\includegraphics[width=8.6cm]{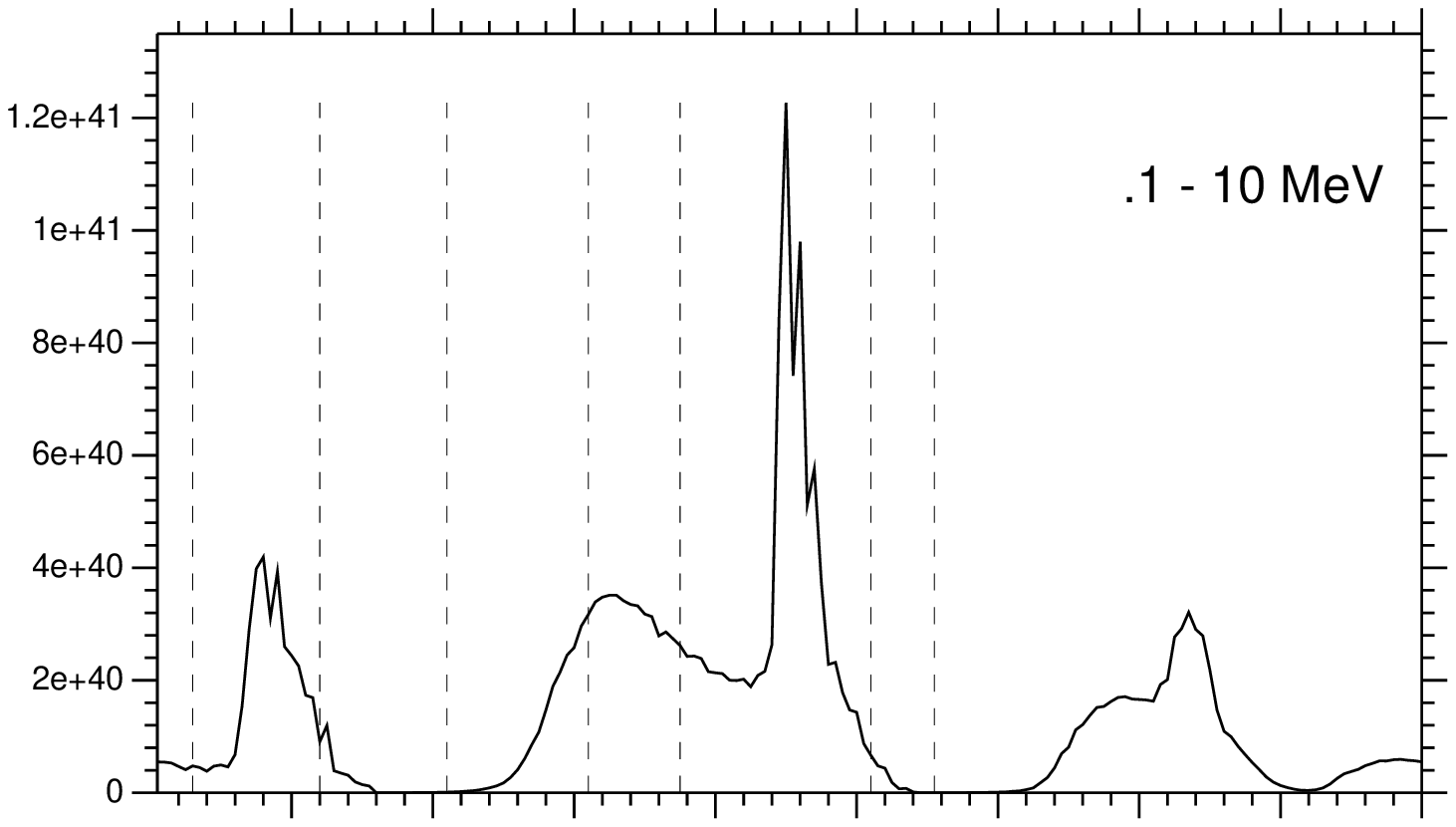}

\vskip -0.2cm\hskip 0cm\includegraphics[width=6.8cm]{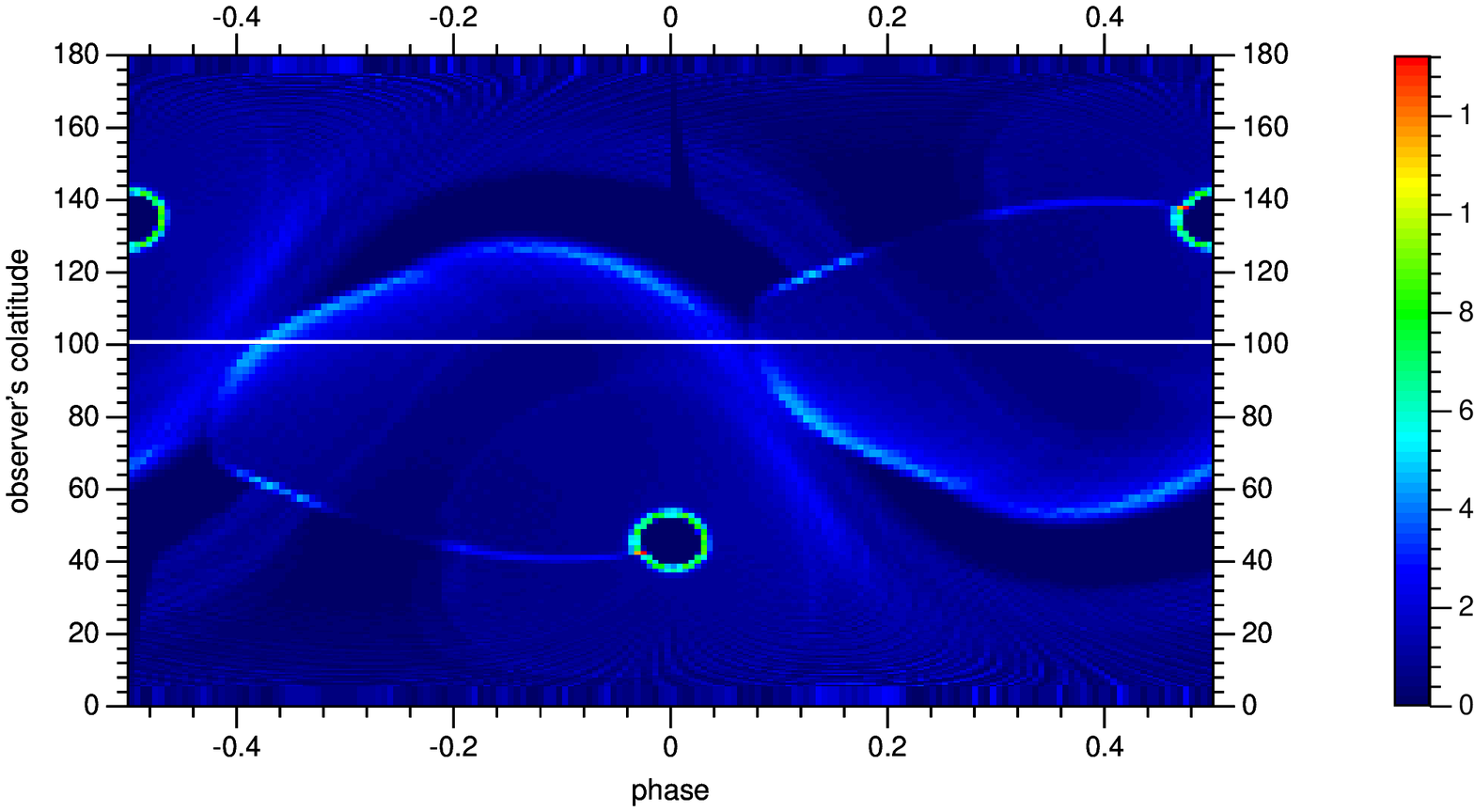}

\vskip -1.3cm\hskip 1.8cm\includegraphics[width=7.5cm,height=7.5cm,angle=90]{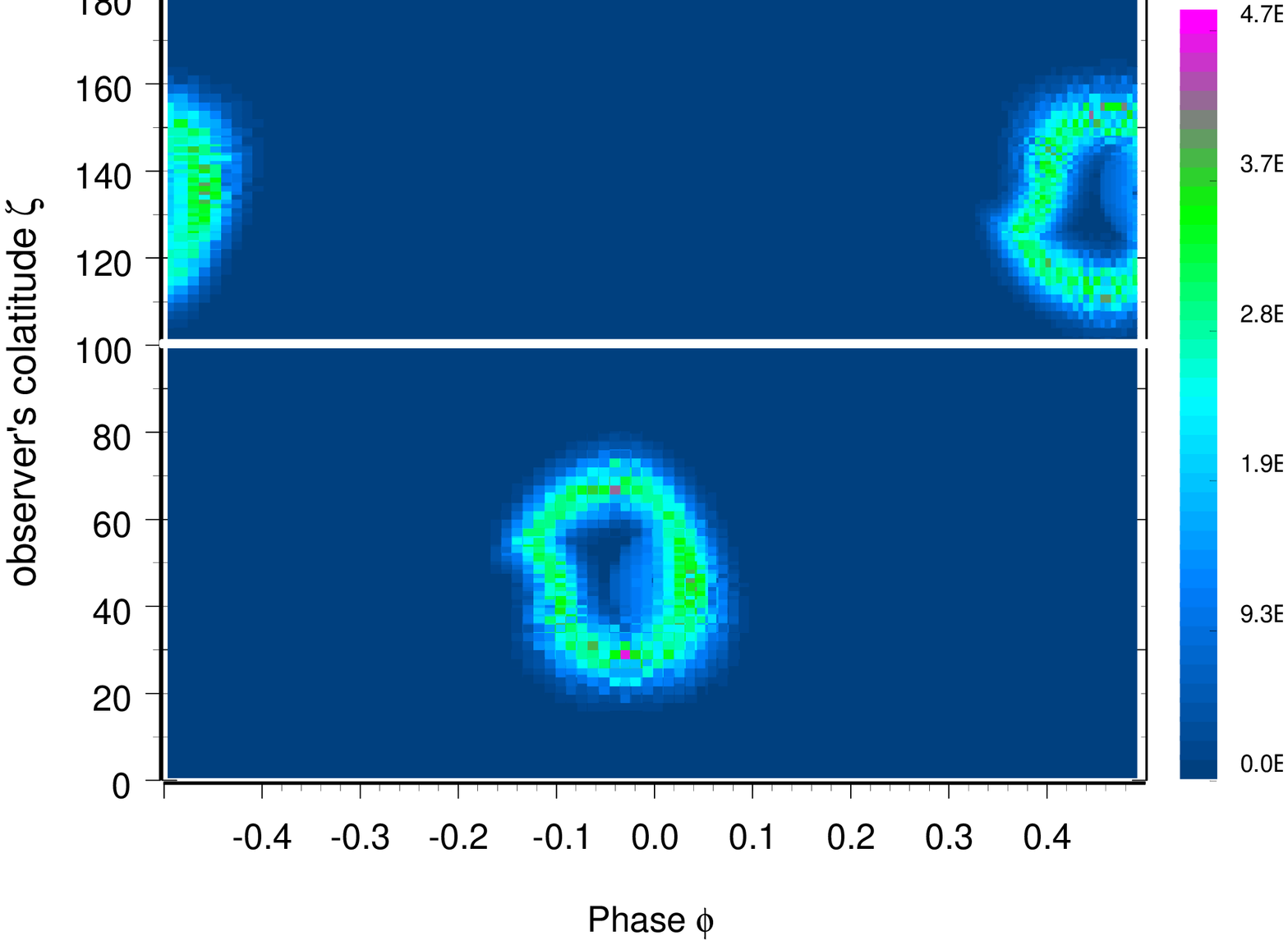}

\vskip -11.5cm\hskip 8.6cm\includegraphics[width=8.6cm]{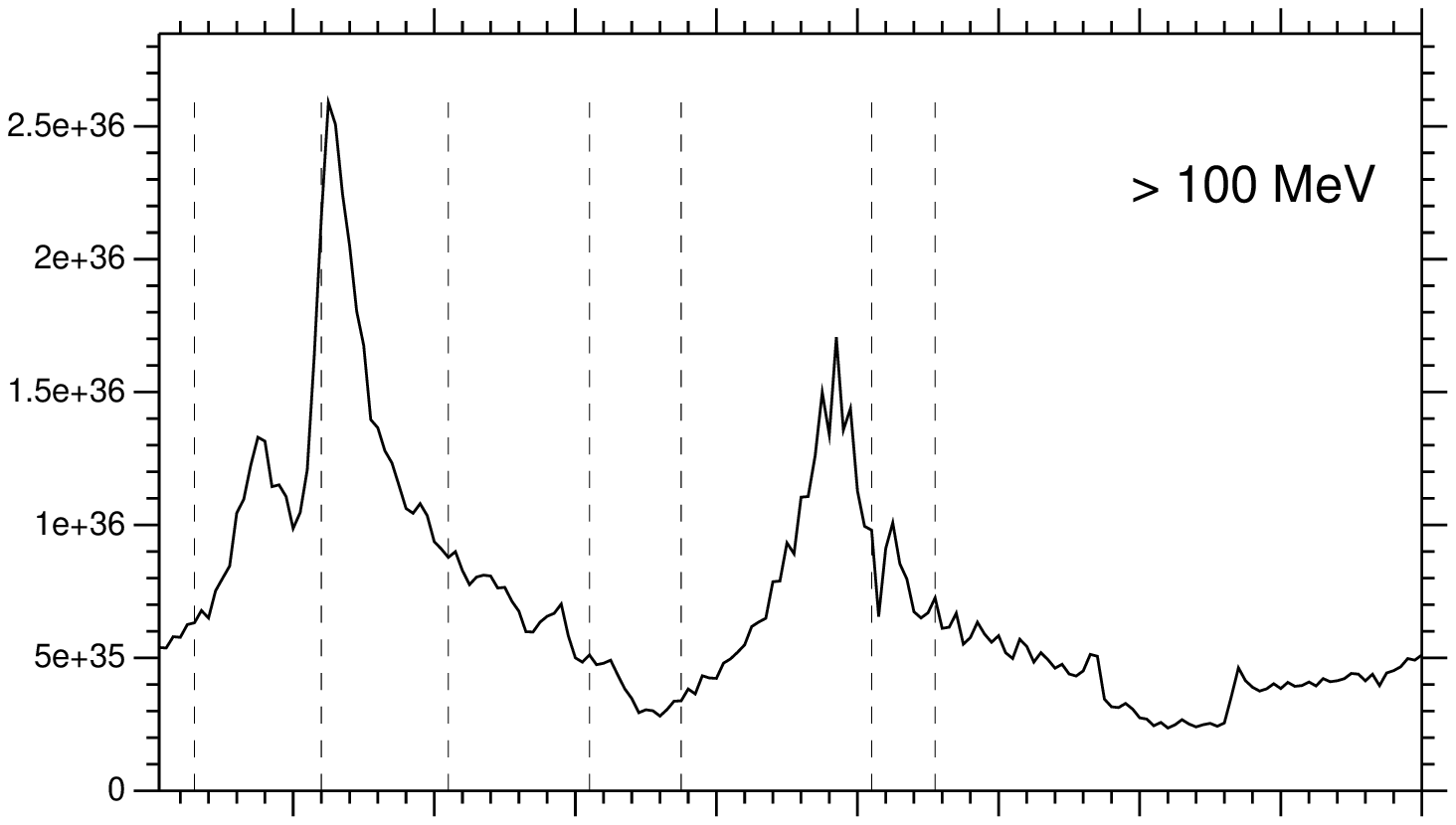}

\vskip -1.5cm\hskip 7.7cm\includegraphics[width=8.4cm]{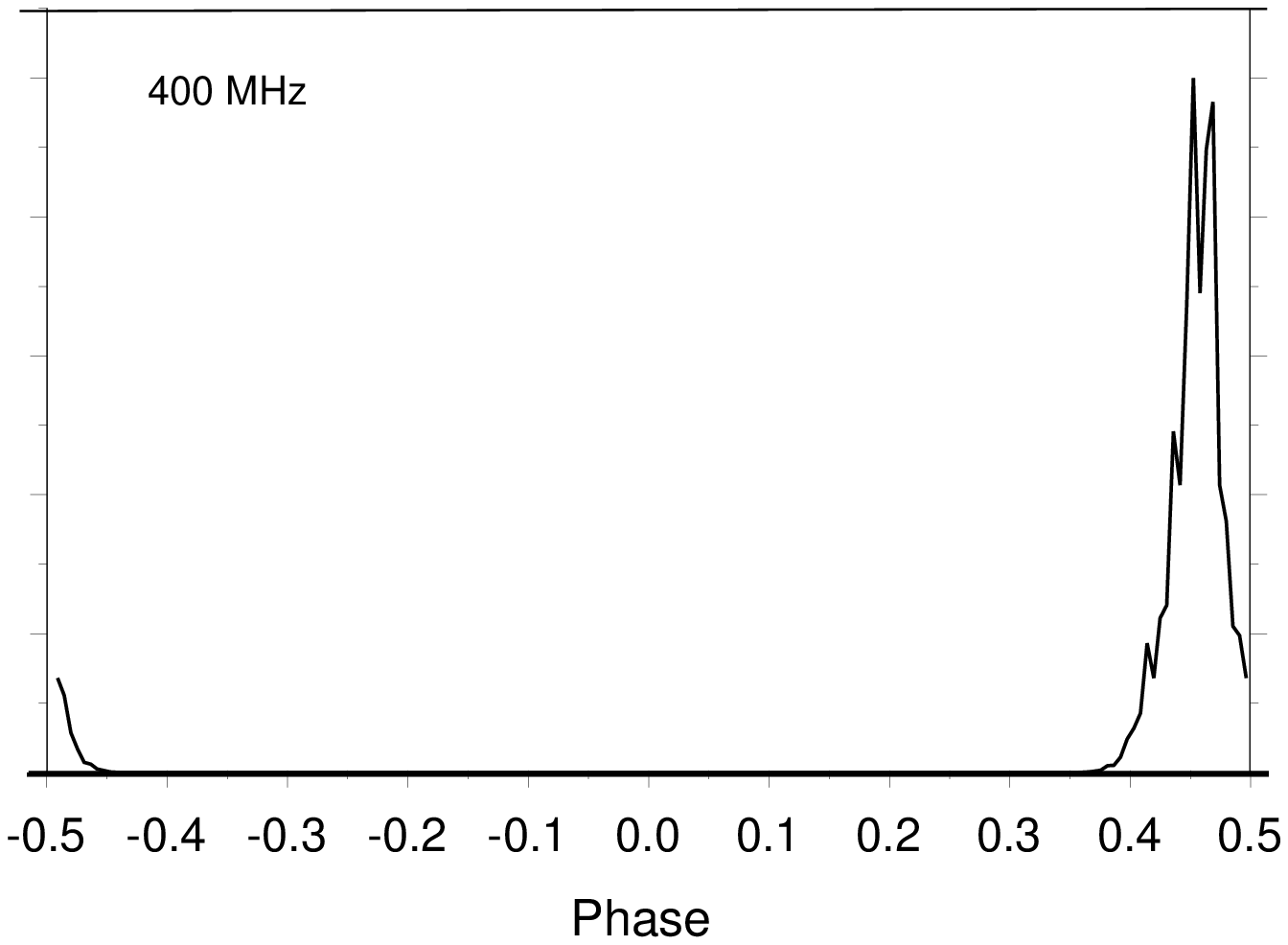}\vskip -0.7cm

\figureout{f2.eps}{
Model pulse profiles and intensity maps (observer angle $\zeta$ vs. rotation phase $\phi$)
in different frequency ranges as labeled for the case of standard radio beam model and inclination 
angle of $\alpha = 45^{\circ}$.  Profiles are shown for observer angle $\zeta = 100^{\circ}$.  The scale
of the high energy intensity maps is in units of $\rm ph \,s^{-1}/ster$ and the scale of the profiles is
in units of $\rm ph \,s^{-1}/ster/N_{\phi}$, where $N_{\phi} = 180$ are the number of phase bins.  The scale 
of the radio map is in $\rm mJy-kpc^2/ster$.
}

\newpage
\includegraphics[width=180mm]{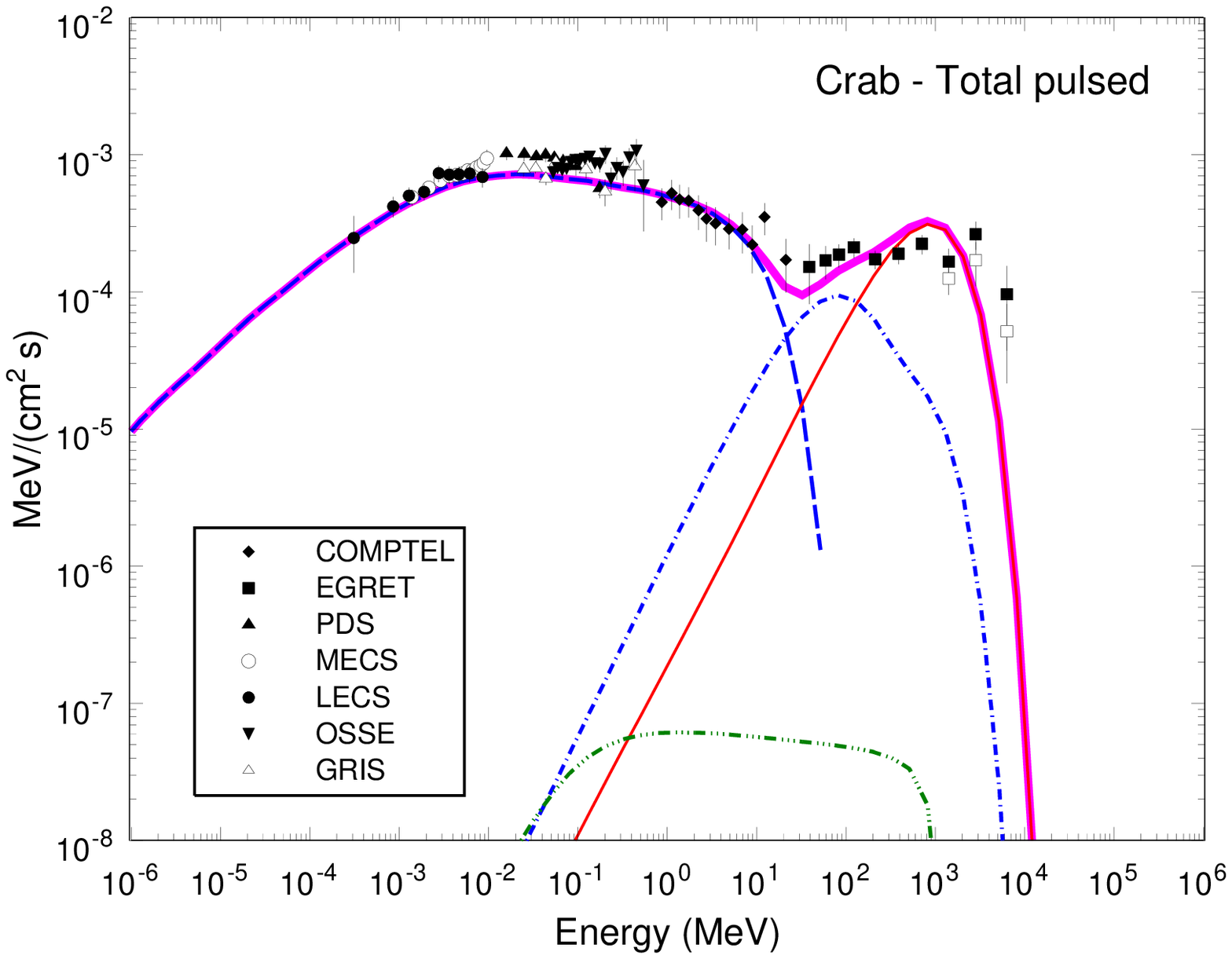}
\figureout{f3.eps}{
Model spectrum of phase-averaged total pulsed emission (heavy solid line) that is the sum of emission components
from curvature (light solid line), synchrotron  (dashed-dot line) and inverse Compton 
(dashed-dot-dot line) radiation of primary electrons in the slot gap and synchrotron radiation 
from pairs (dashed line) inside the slot gap.  Data points are from Kuiper et al. (2001) [http://www.sron.nl/divisions/hea/kuiper/data.html].  The open squares are corrected EGRET values above
1 GeV from Stecker et al. (2007).}

\newpage
\hskip -2.0cm
\includegraphics[width=7cm]{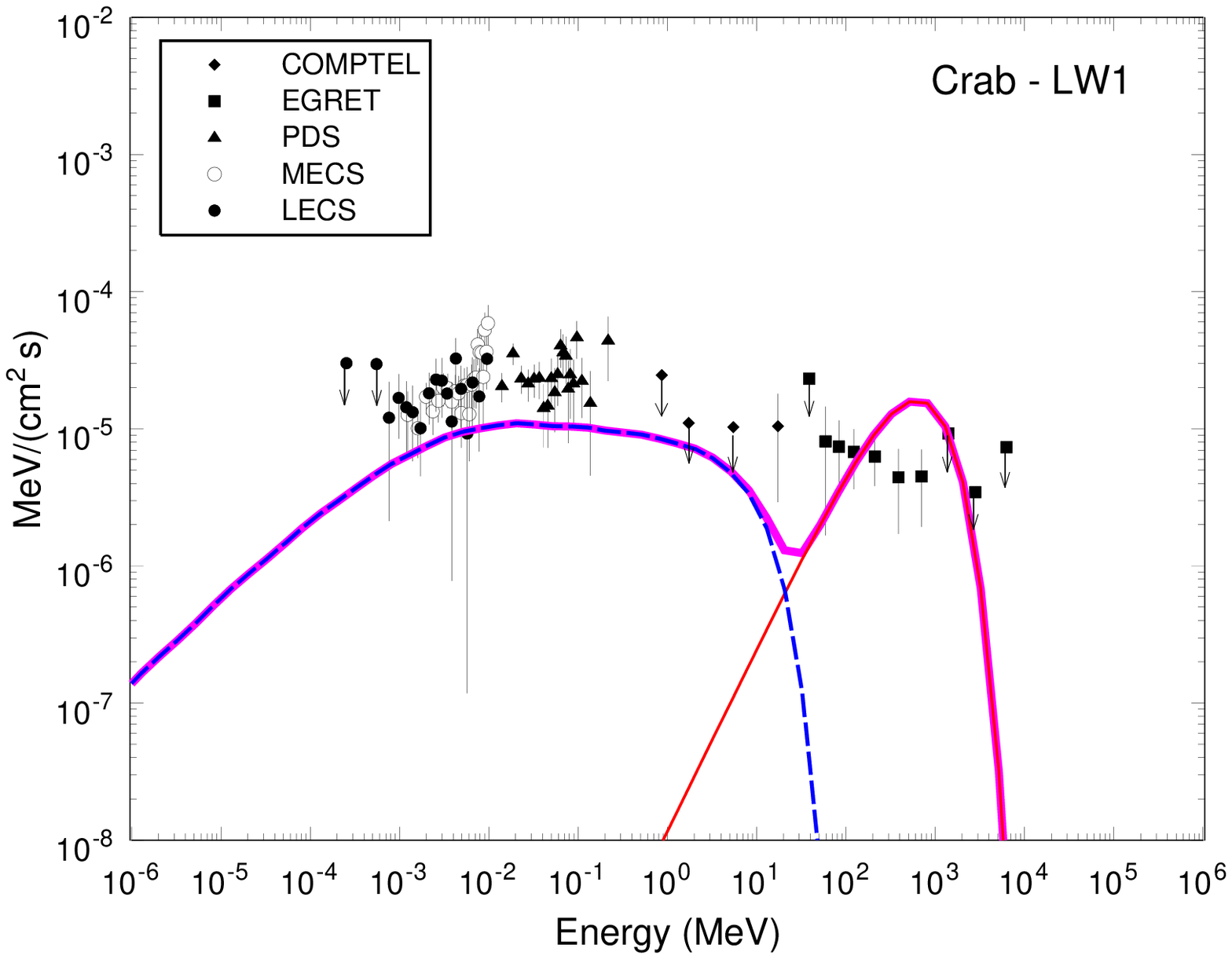}\hskip -0.5cm\includegraphics[width=7cm]{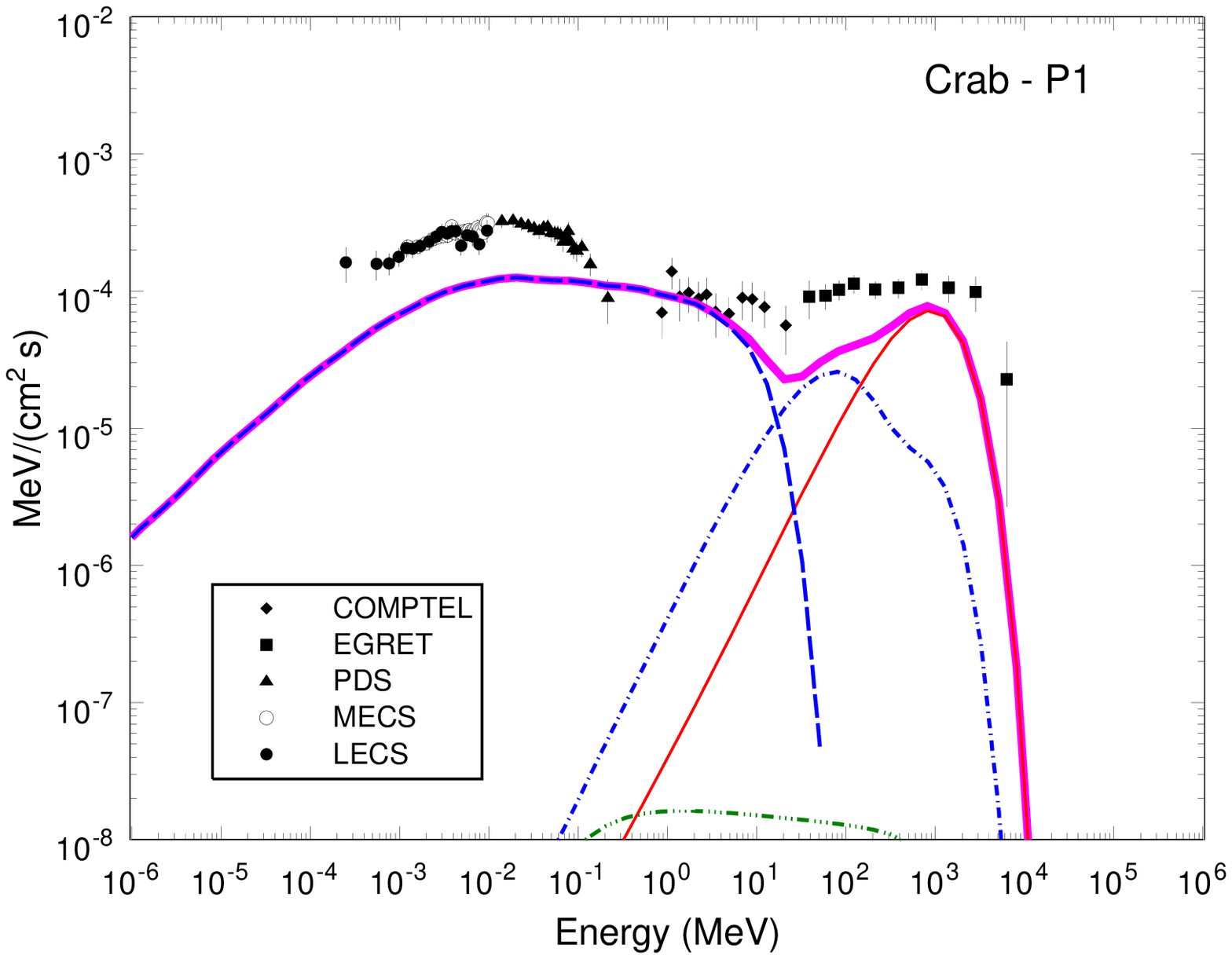}

\hskip -2.0cm
\includegraphics[width=7cm]{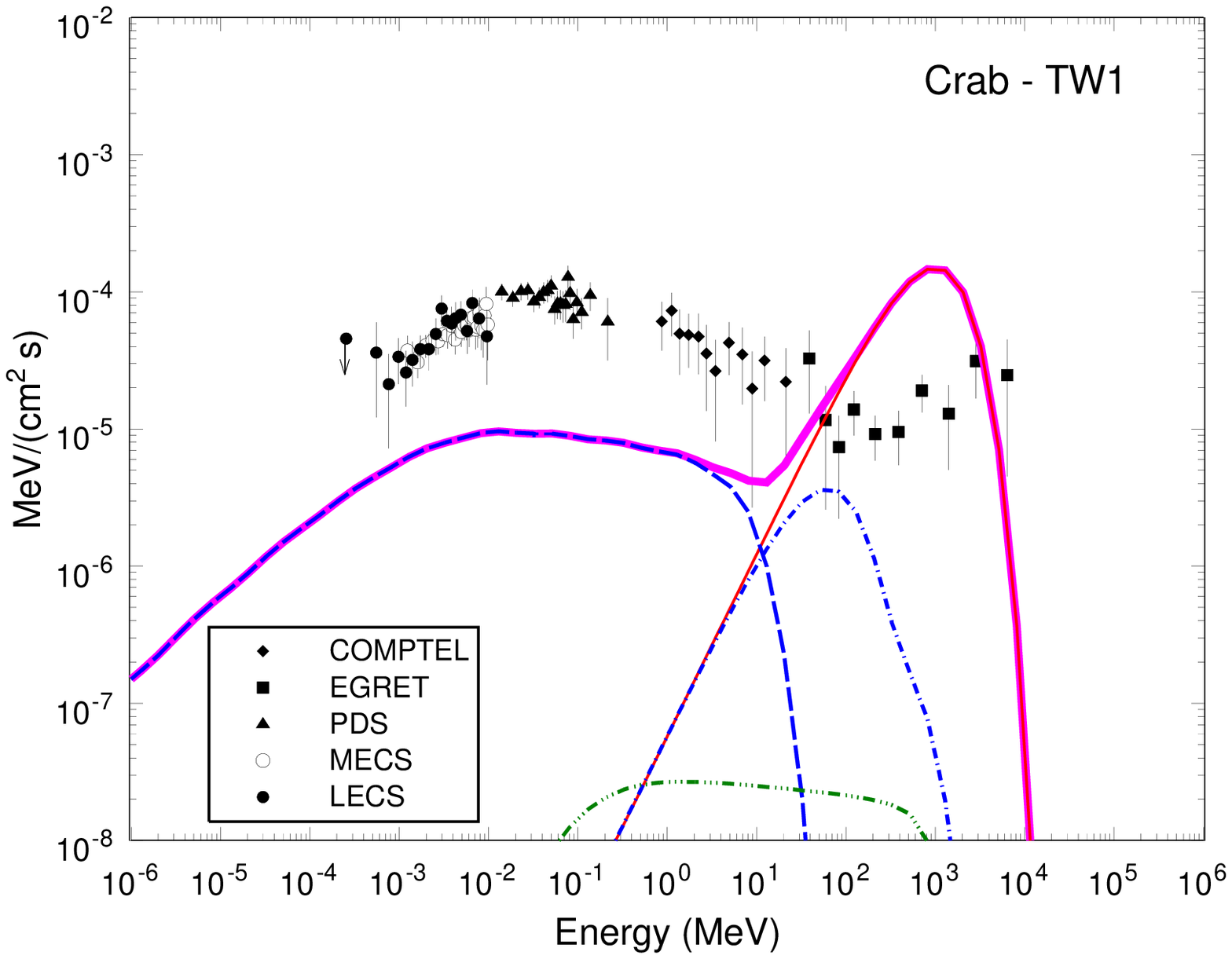}\hskip -0.5cm
\includegraphics[width=7cm]{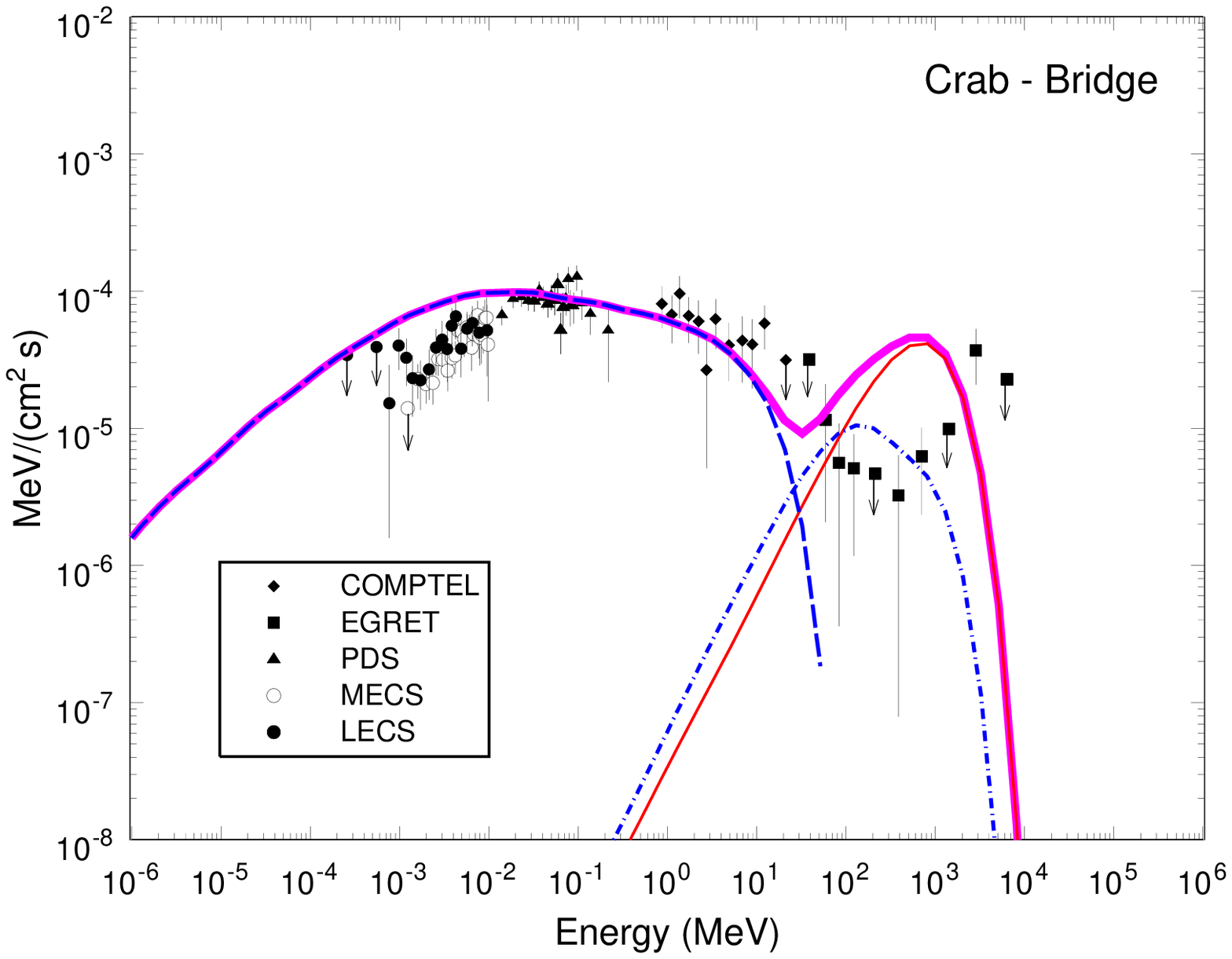}\hskip -0.5cm\includegraphics[width=7cm]{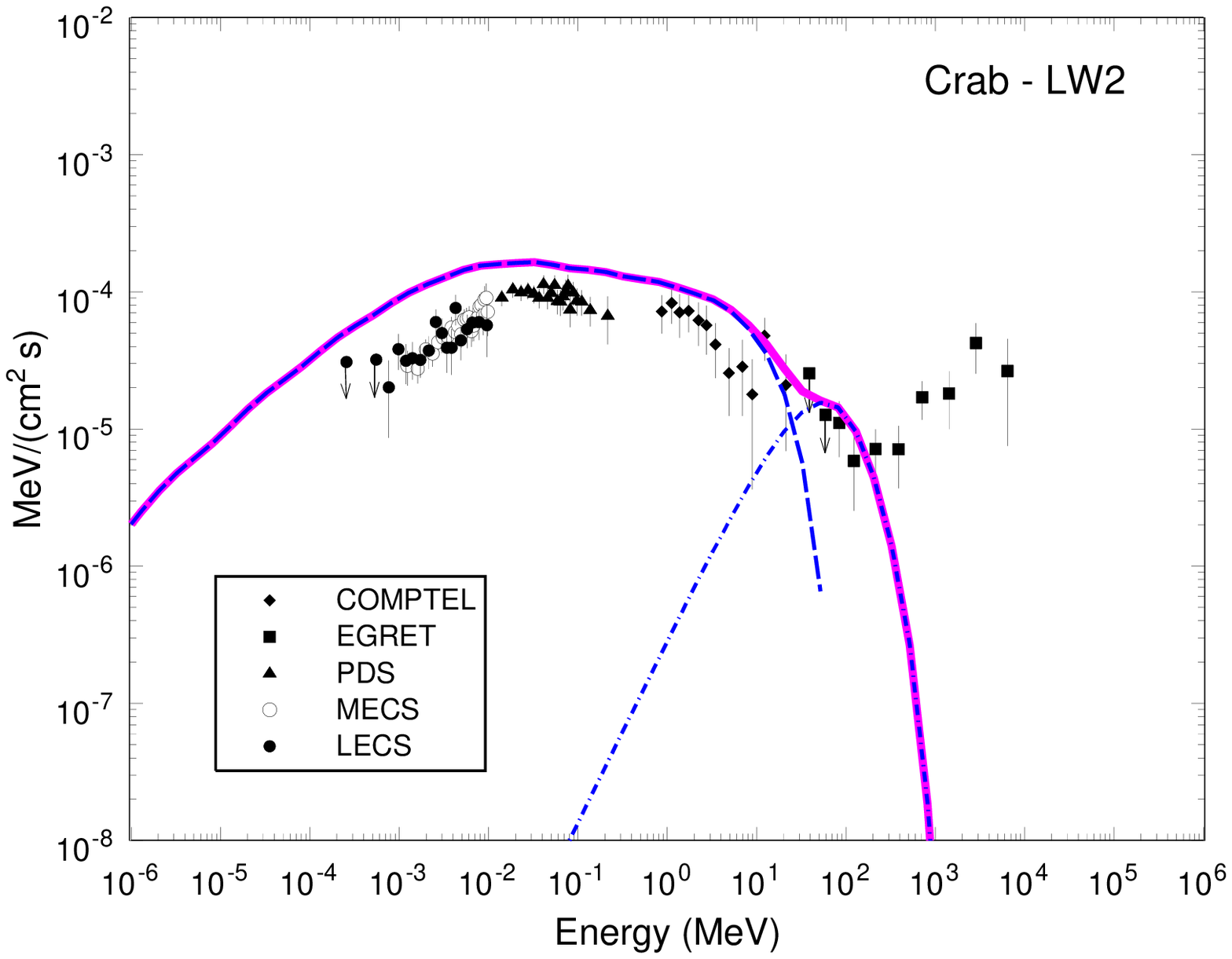}

\hskip -2.0cm
\includegraphics[width=7cm]{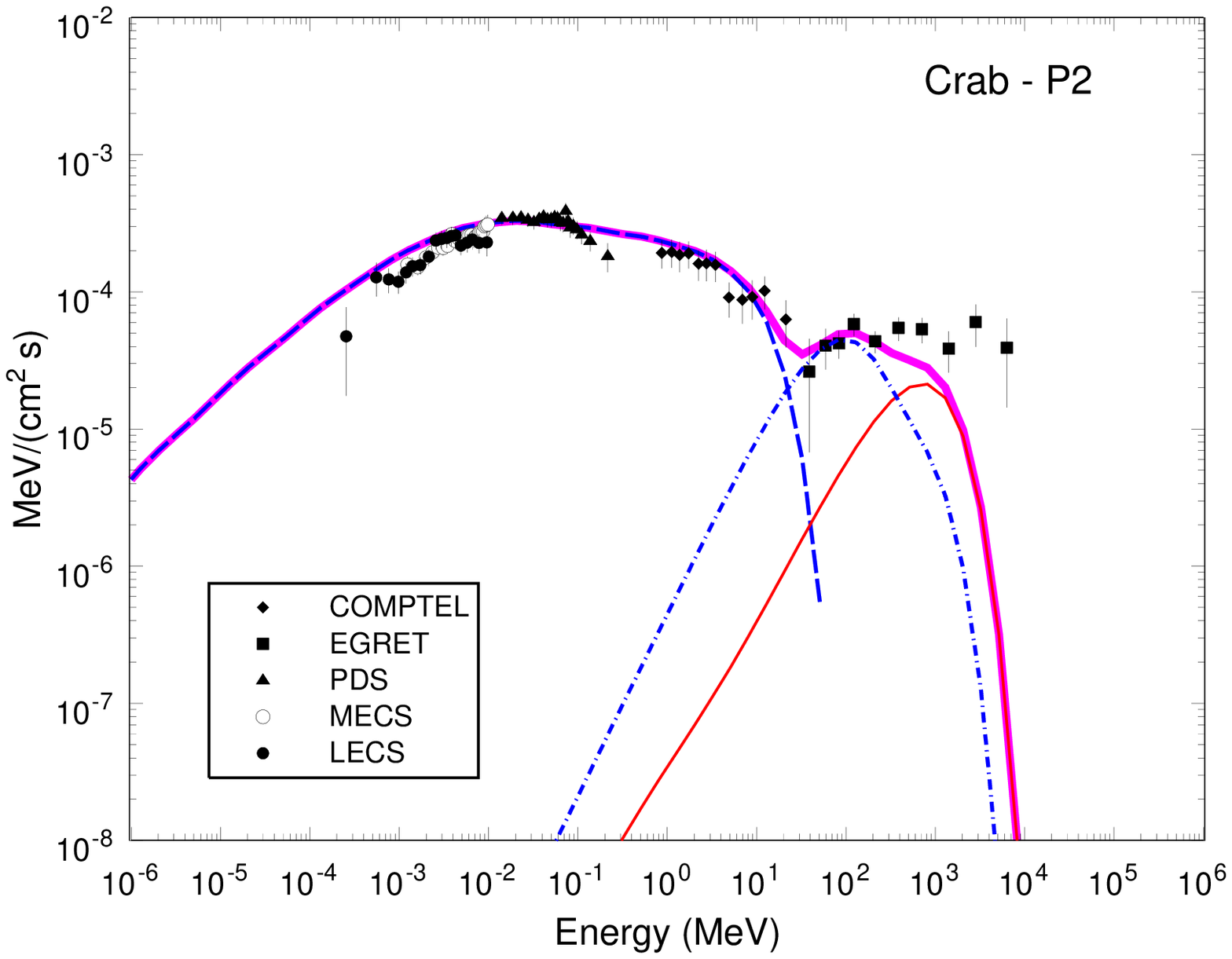}\hskip -0.5cm\includegraphics[width=7cm]{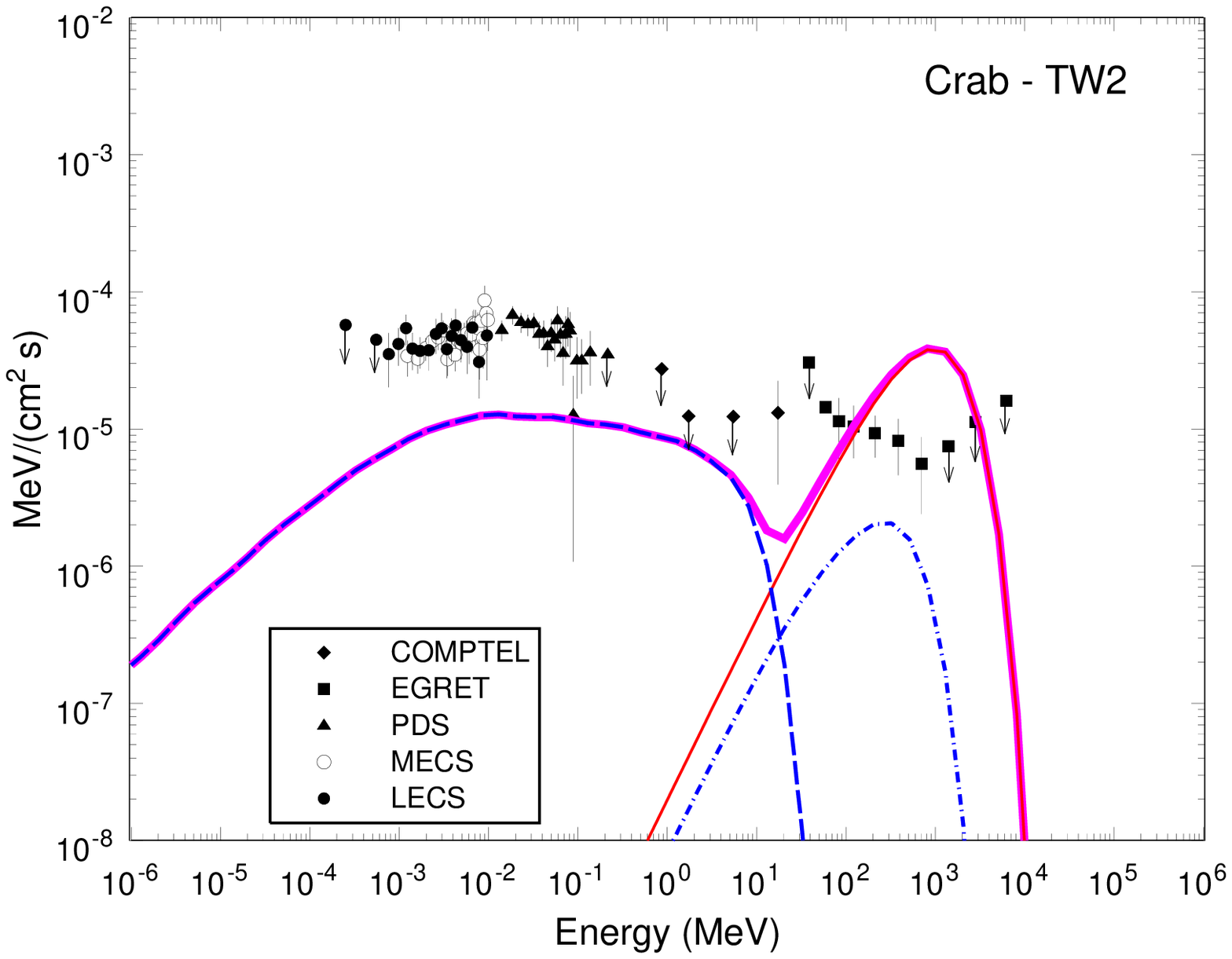}

\figureout{f3\4.eps}{
Model spectrum of total pulsed emission in different phase intervals (heavy solid line) 
that is the sum of emission components
from curvature (light solid line), synchrotron  (dashed-dot line) and inverse Compton 
(dashed-dashed-dot line) radiation of primary electrons in the slot gap and synchrotron radiation 
from pairs (dashed line) inside the slot gap.}
\newpage
~
\vskip -1cm\includegraphics[width=6.8cm]{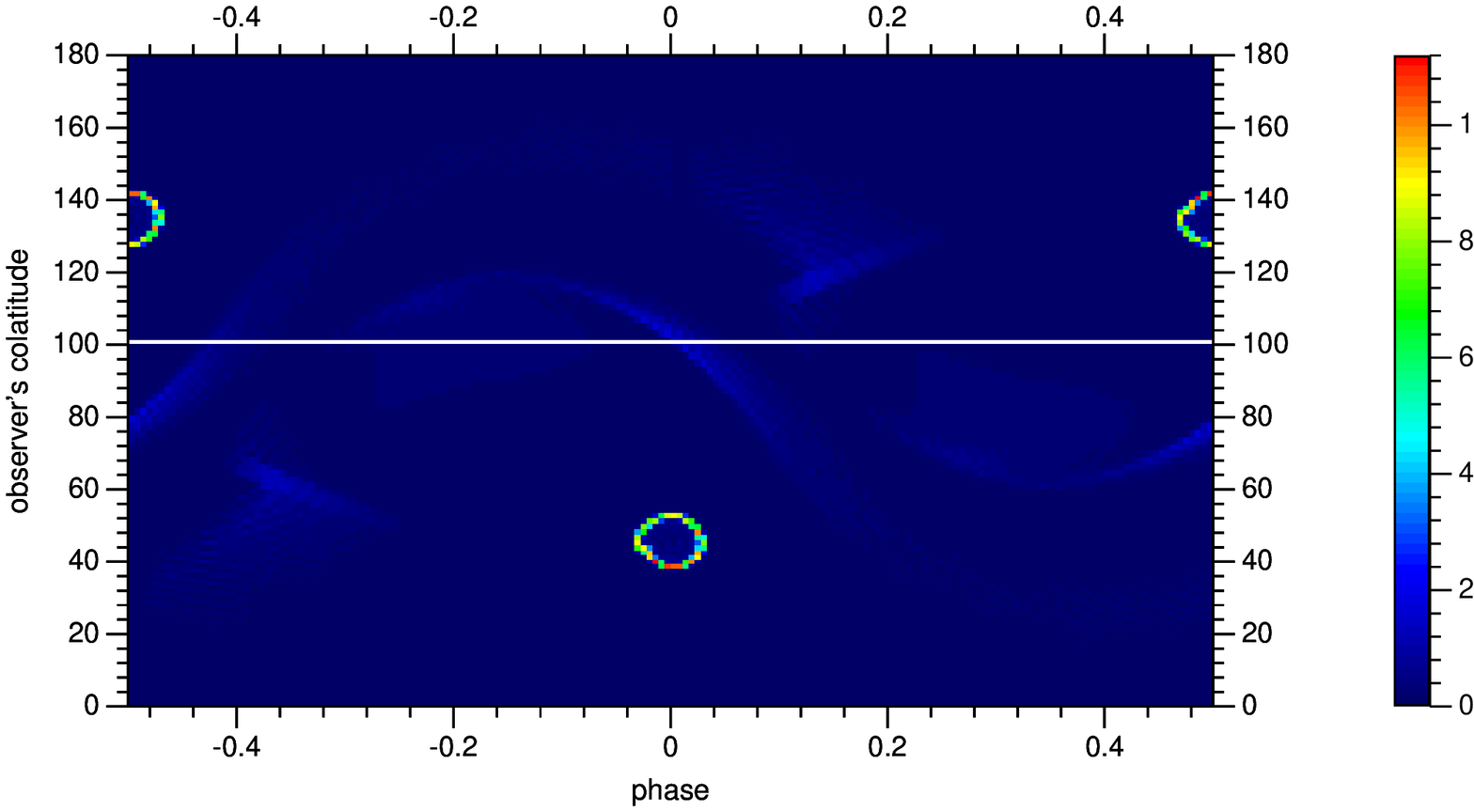}\hskip 1.8cm\includegraphics[width=8.6cm]{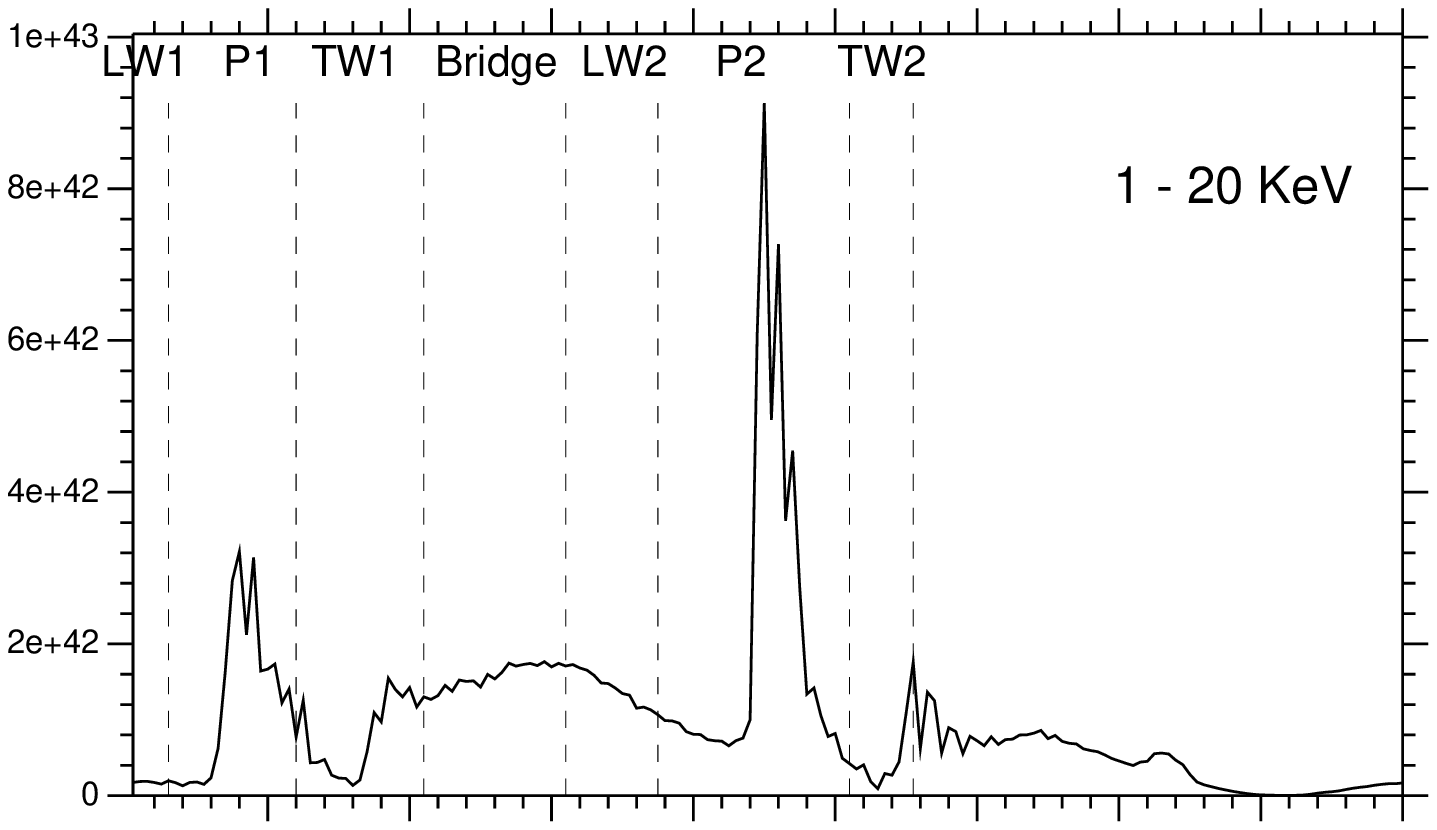}\hskip 0cm

\vskip -0.2cm\includegraphics[width=6.8cm]{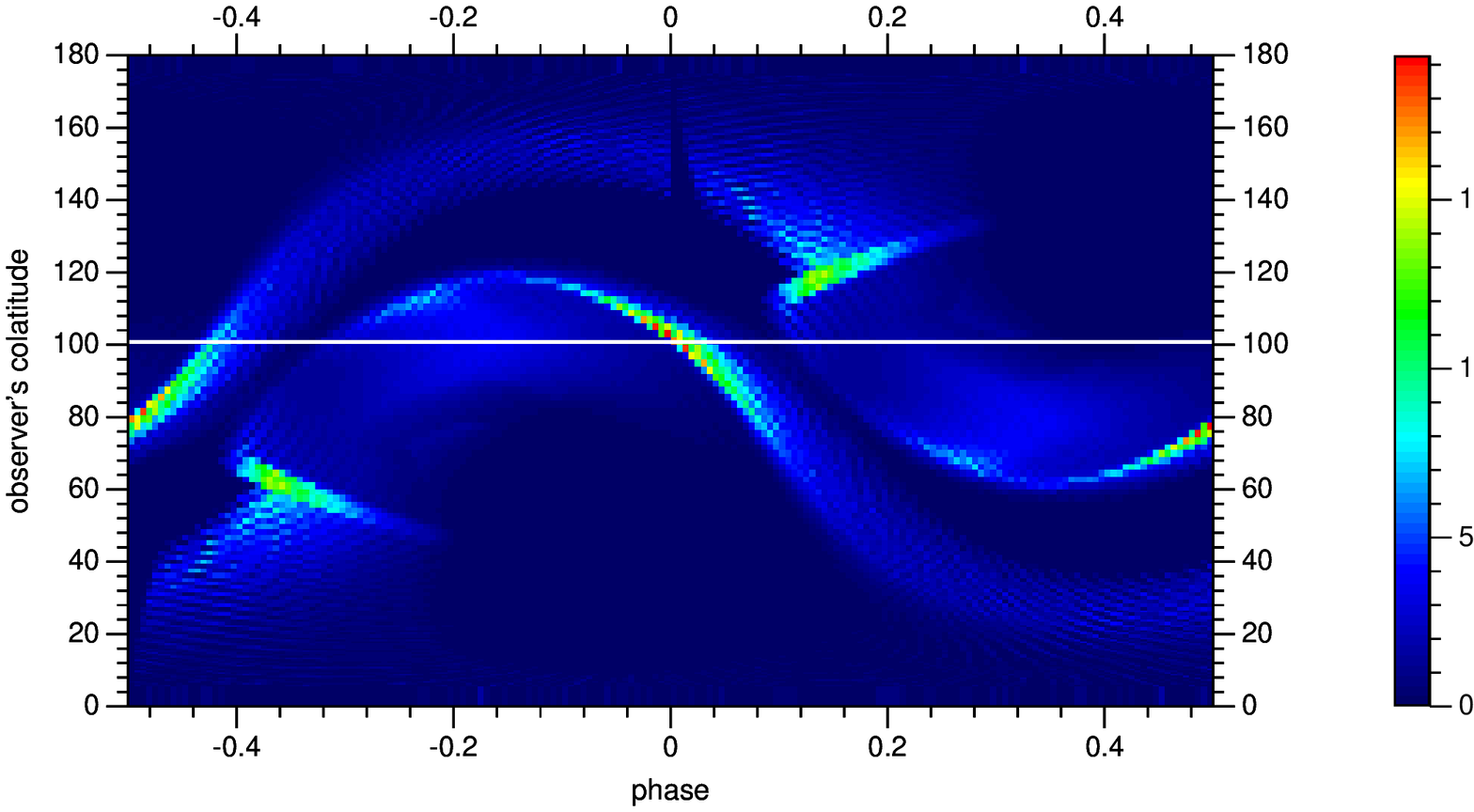}\hskip 1.8cm\includegraphics[width=8.6cm]{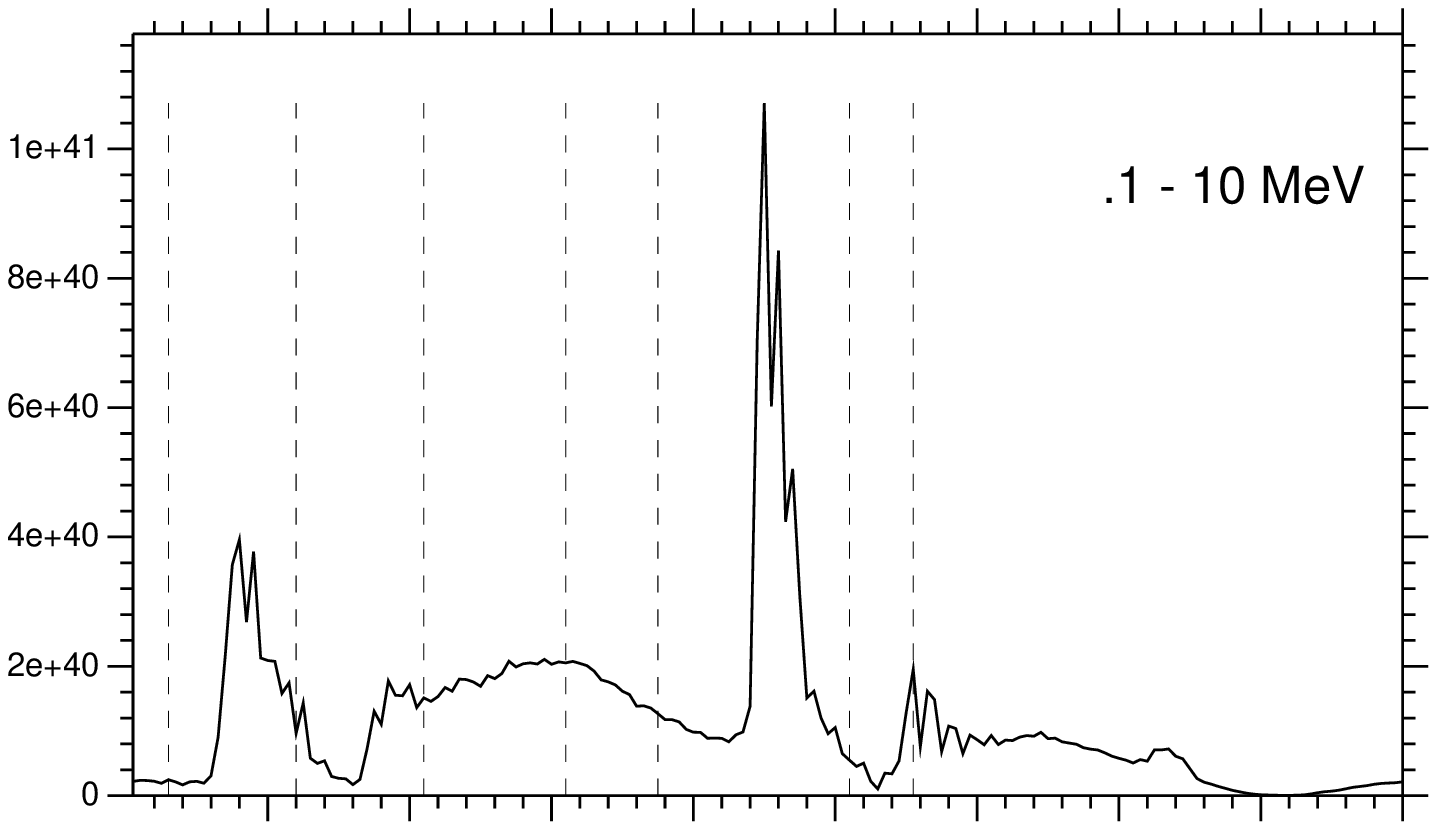}

\vskip -0.2cm\hskip 0cm\includegraphics[width=6.8cm]{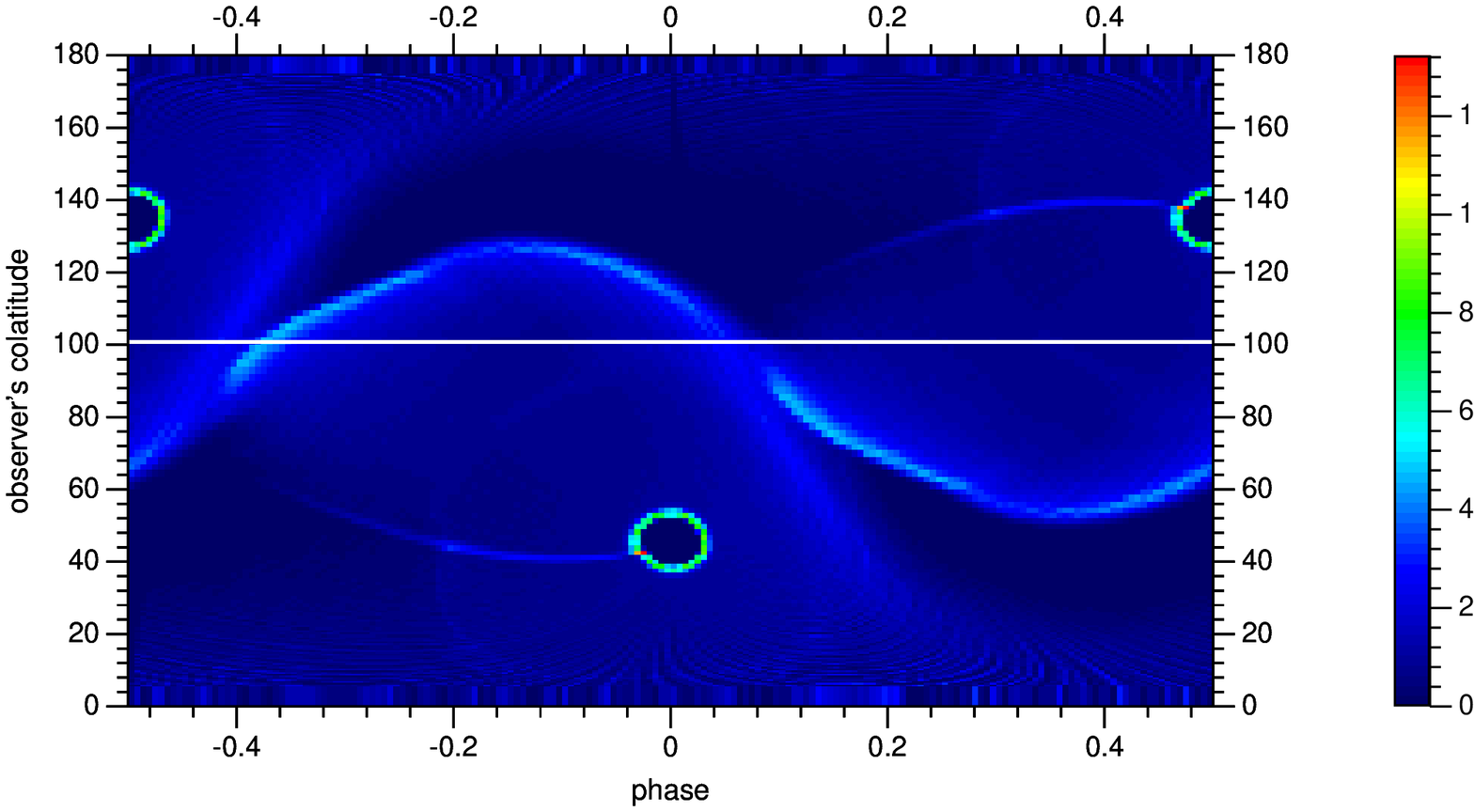}

\vskip -1.2cm\hskip -0.55cm\includegraphics[width=7.7cm,height=7.7cm,angle=90]{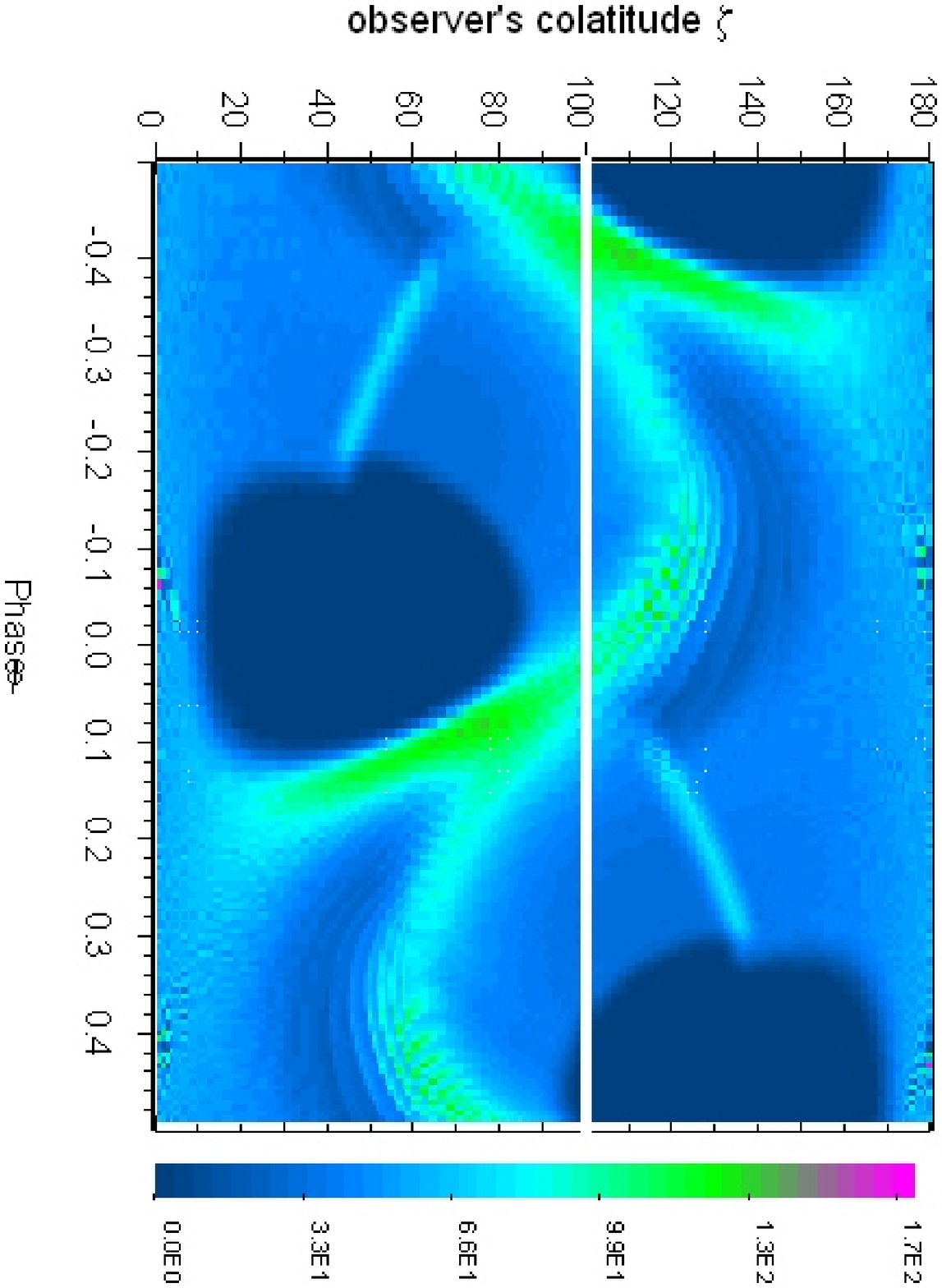}

\vskip -11.5cm\hskip 8.6cm\includegraphics[width=8.6cm]{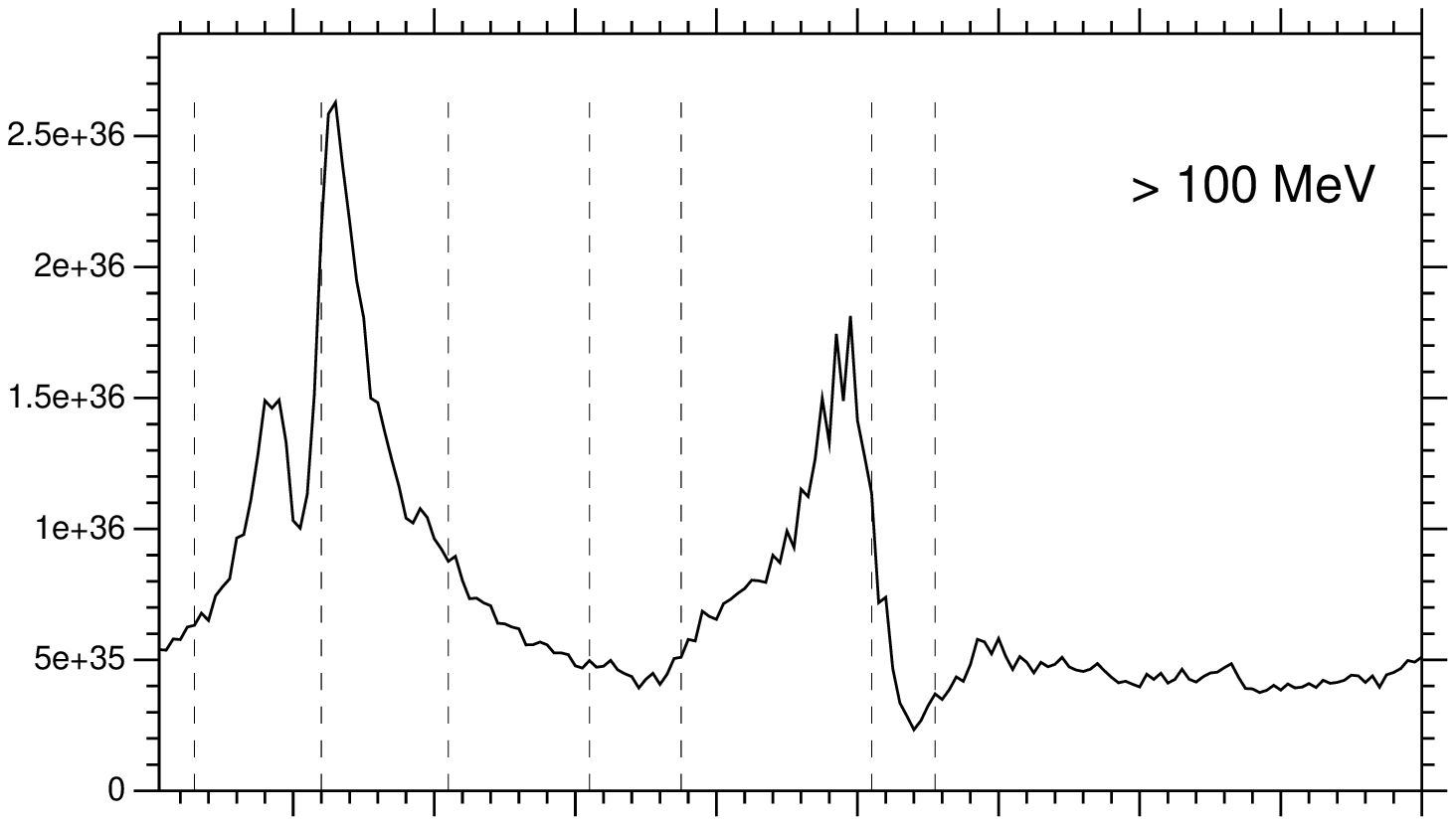}

\vskip -1.5cm\hskip 7.6cm\includegraphics[width=8.6cm,height=6.5cm]{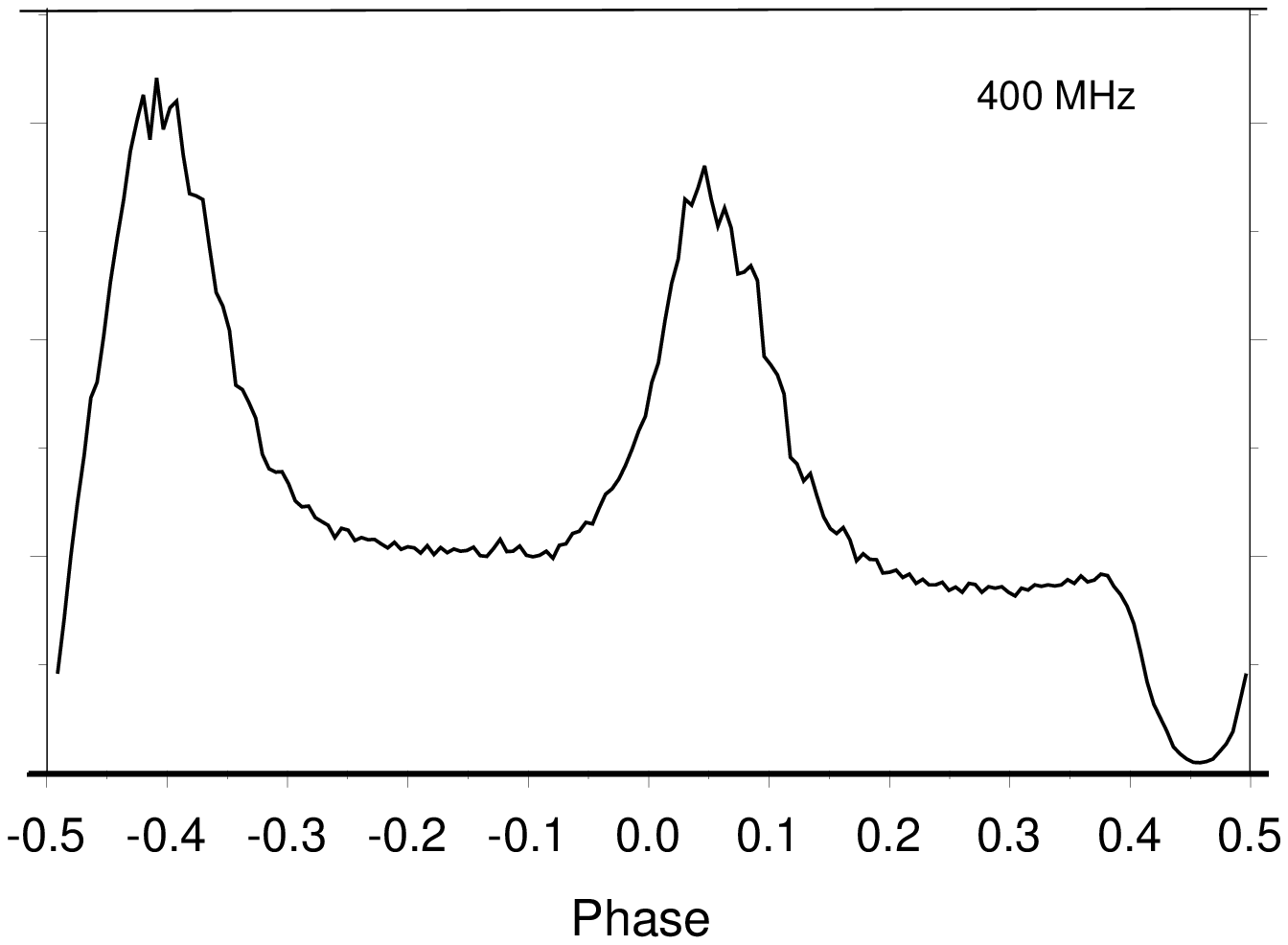}

\figureout{f5.eps}{
Model pulse profiles and intensity maps (observer angle $\zeta$ vs. rotation phase $\phi$)
in different frequency ranges as labeled for the case of an extended 
radio cone model and inclination angle of $\alpha = 45^{\circ}$.  Profiles are shown for observer angle
$\zeta = 100^{\circ}$. 
}
\newpage

\includegraphics[width=180mm]{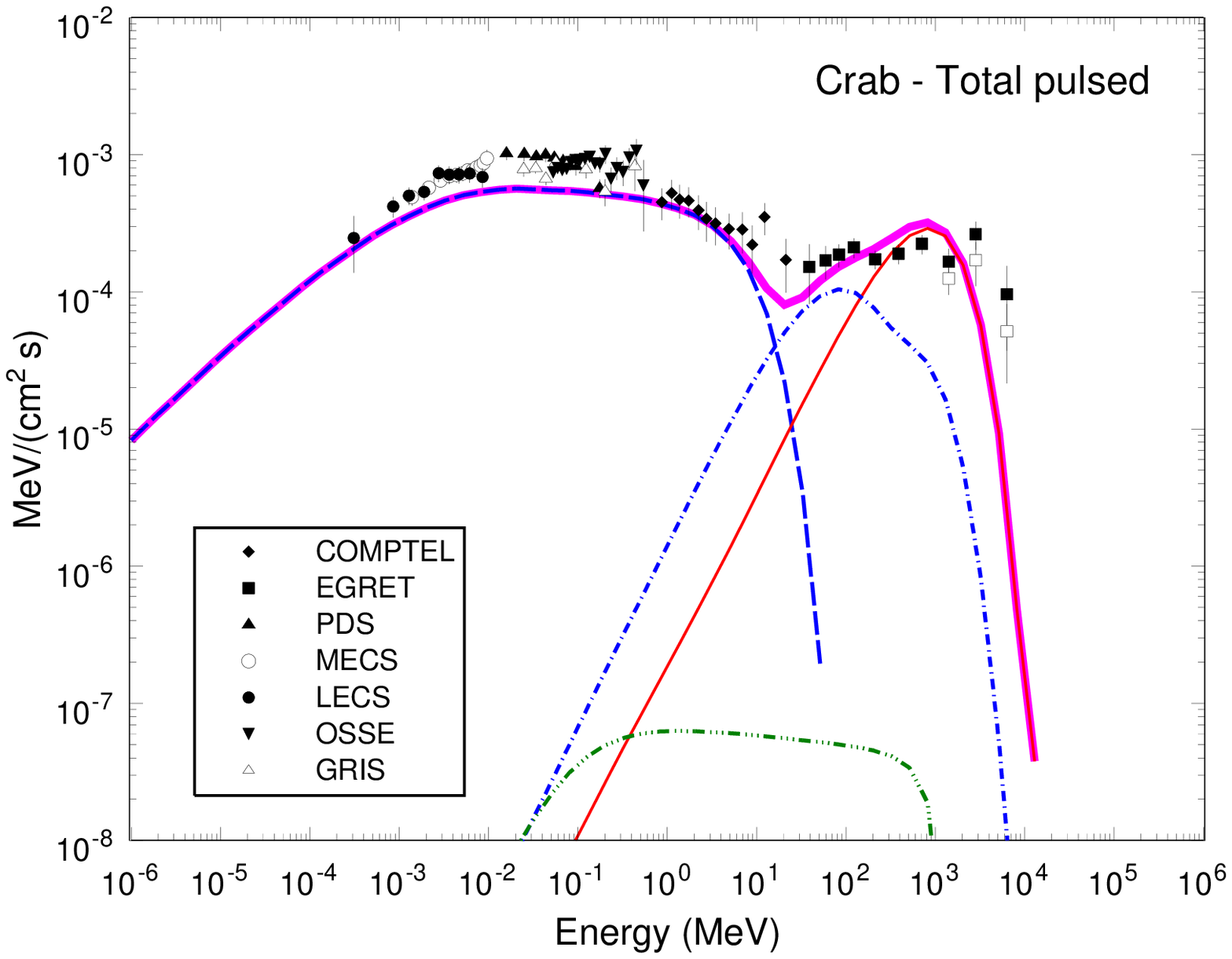}
\figureout{f6.eps}{
Same as Figure 3, but for the case of an extended radio cone model.}

\newpage
\hskip -2.0cm
\includegraphics[width=7cm]{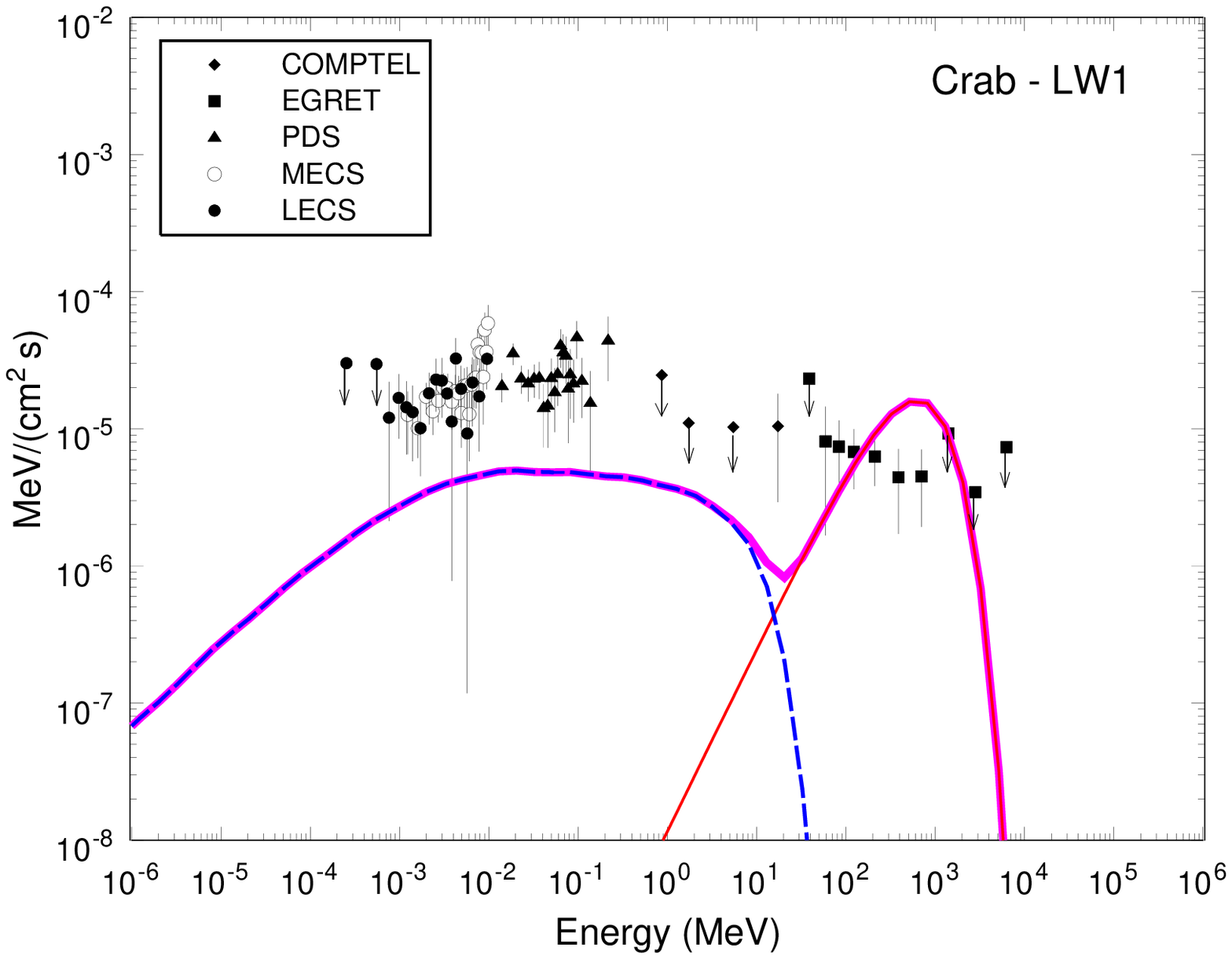}\hskip -0.5cm\includegraphics[width=7cm]{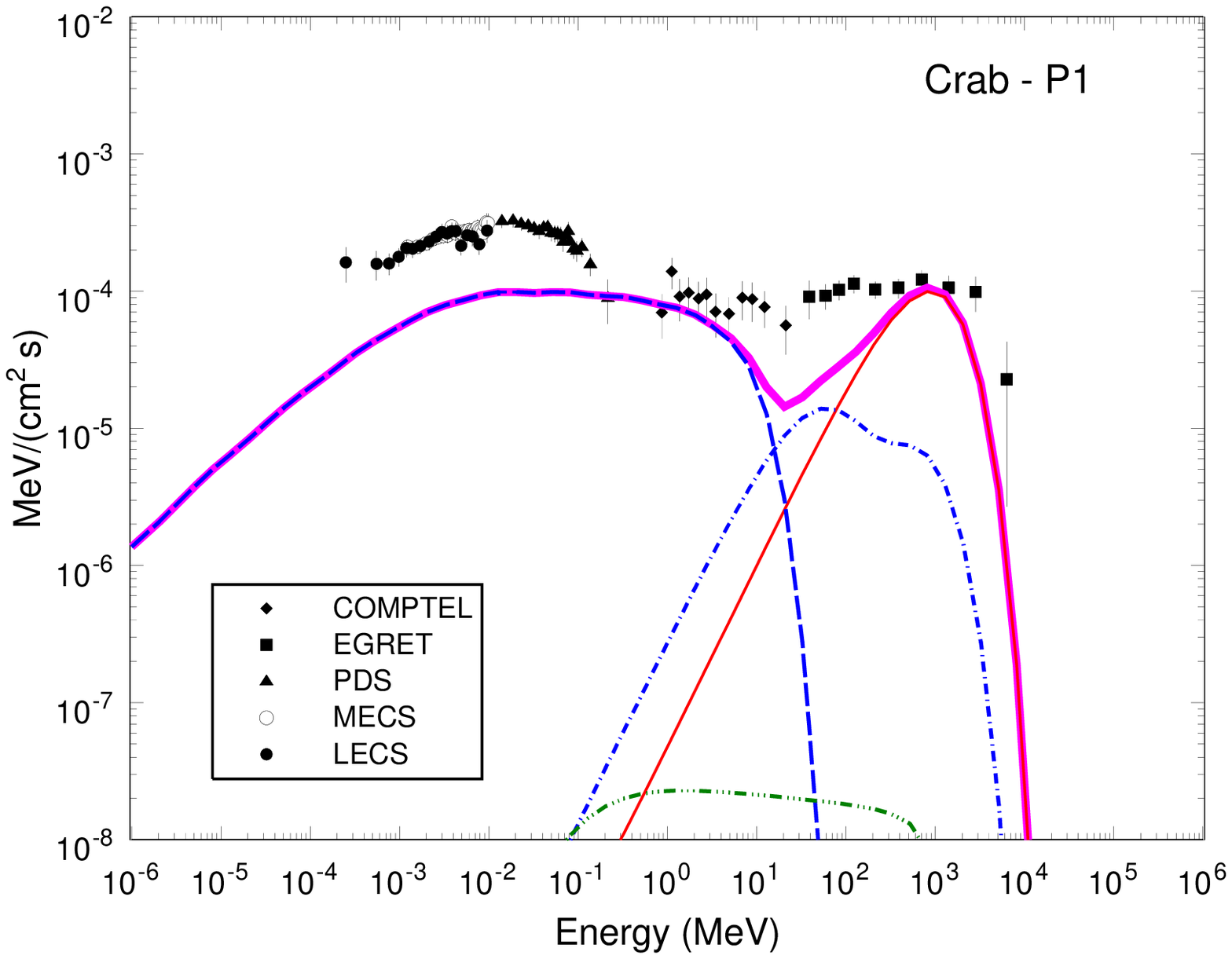}

\hskip -2.0cm
\includegraphics[width=7cm]{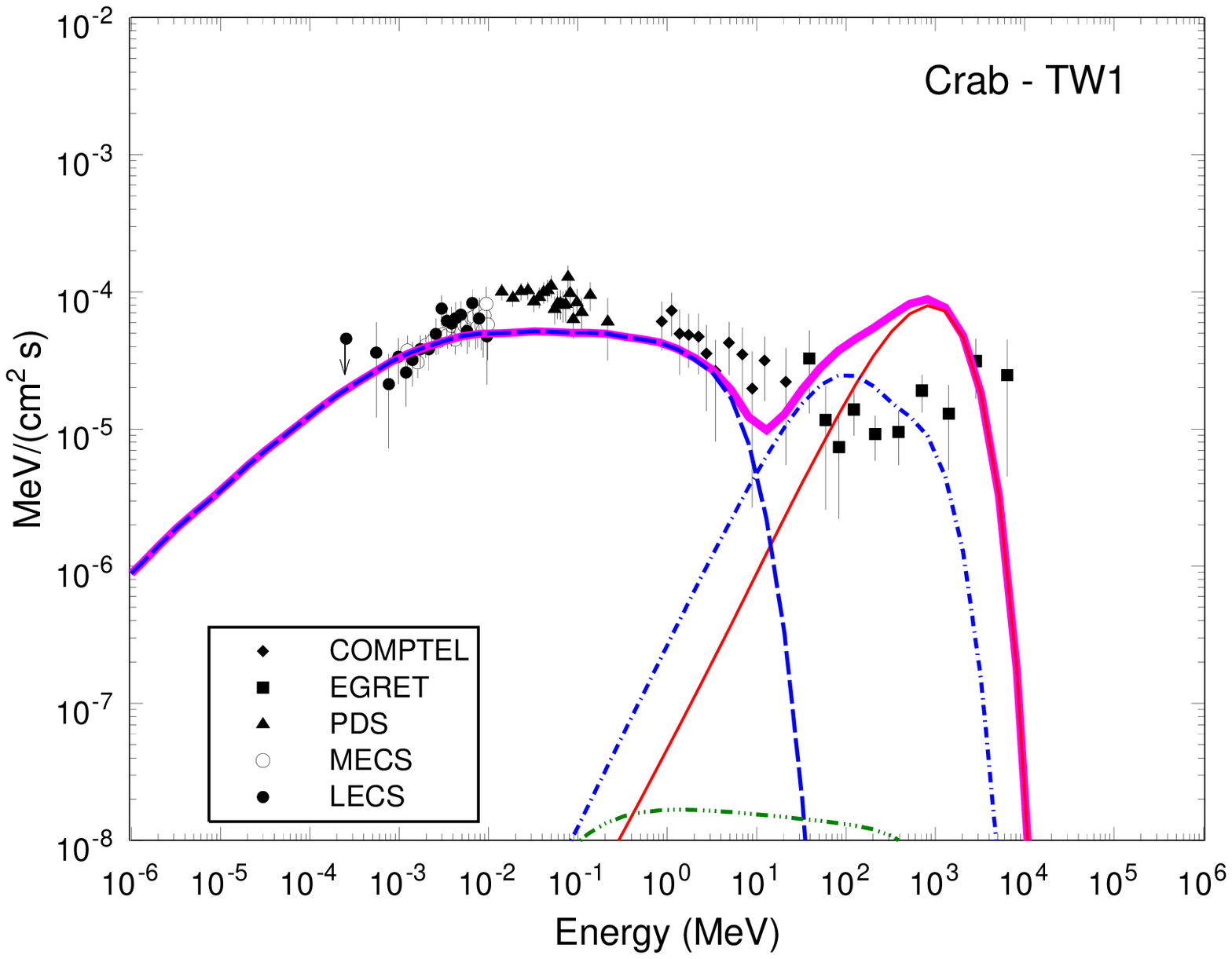}\hskip -0.5cm
\includegraphics[width=7cm]{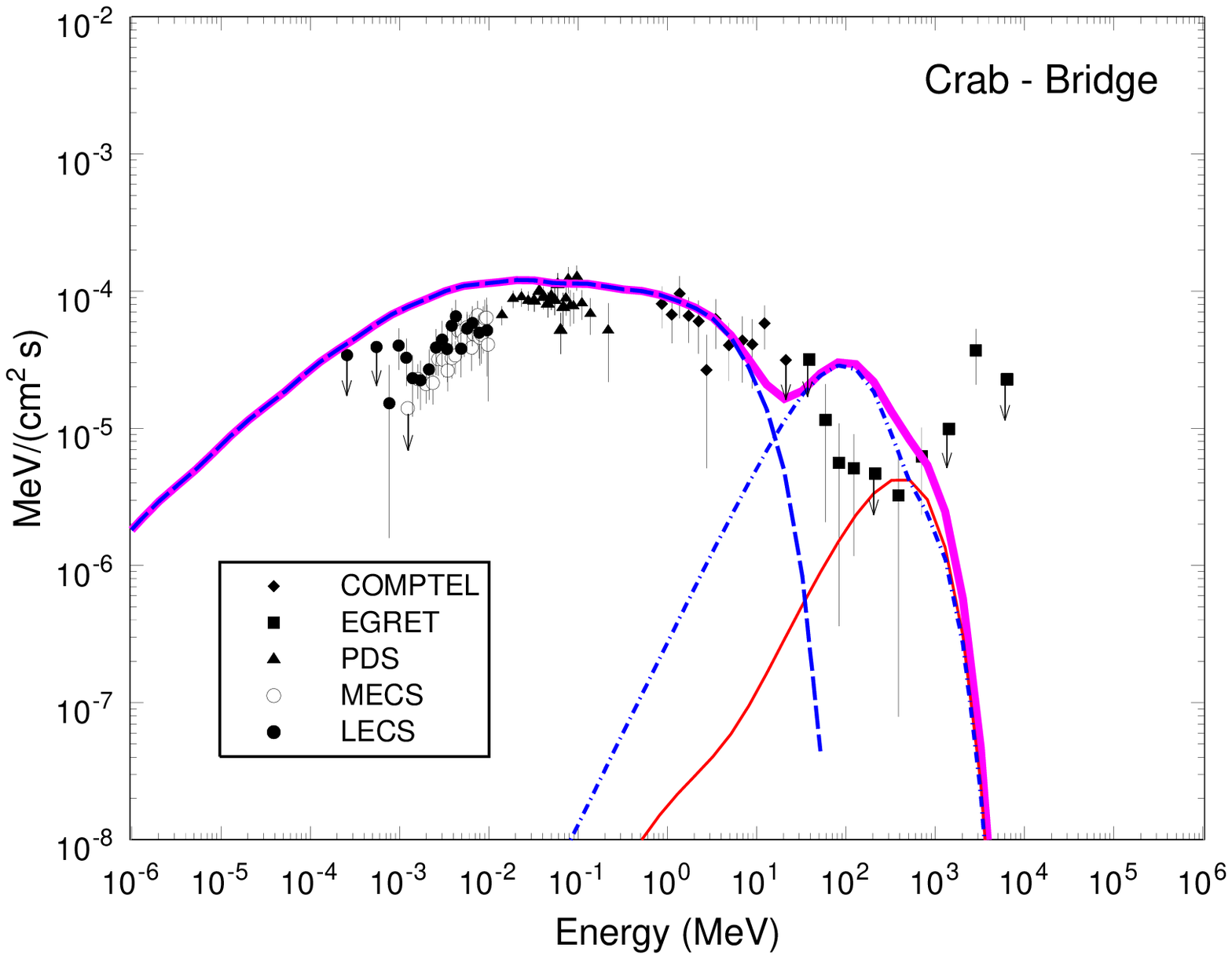}\hskip -0.5cm\includegraphics[width=7cm]{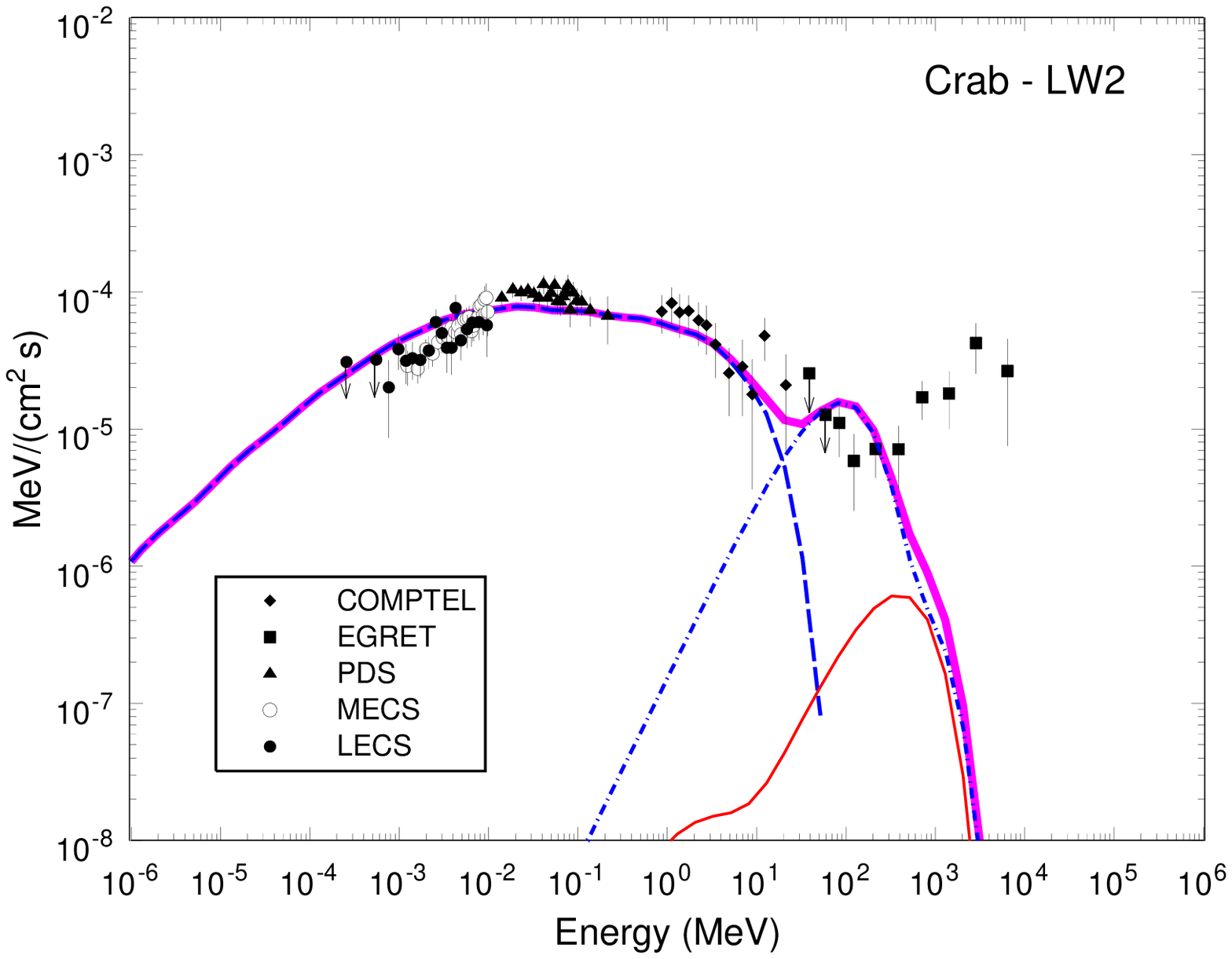}

\hskip -2.0cm
\includegraphics[width=7cm]{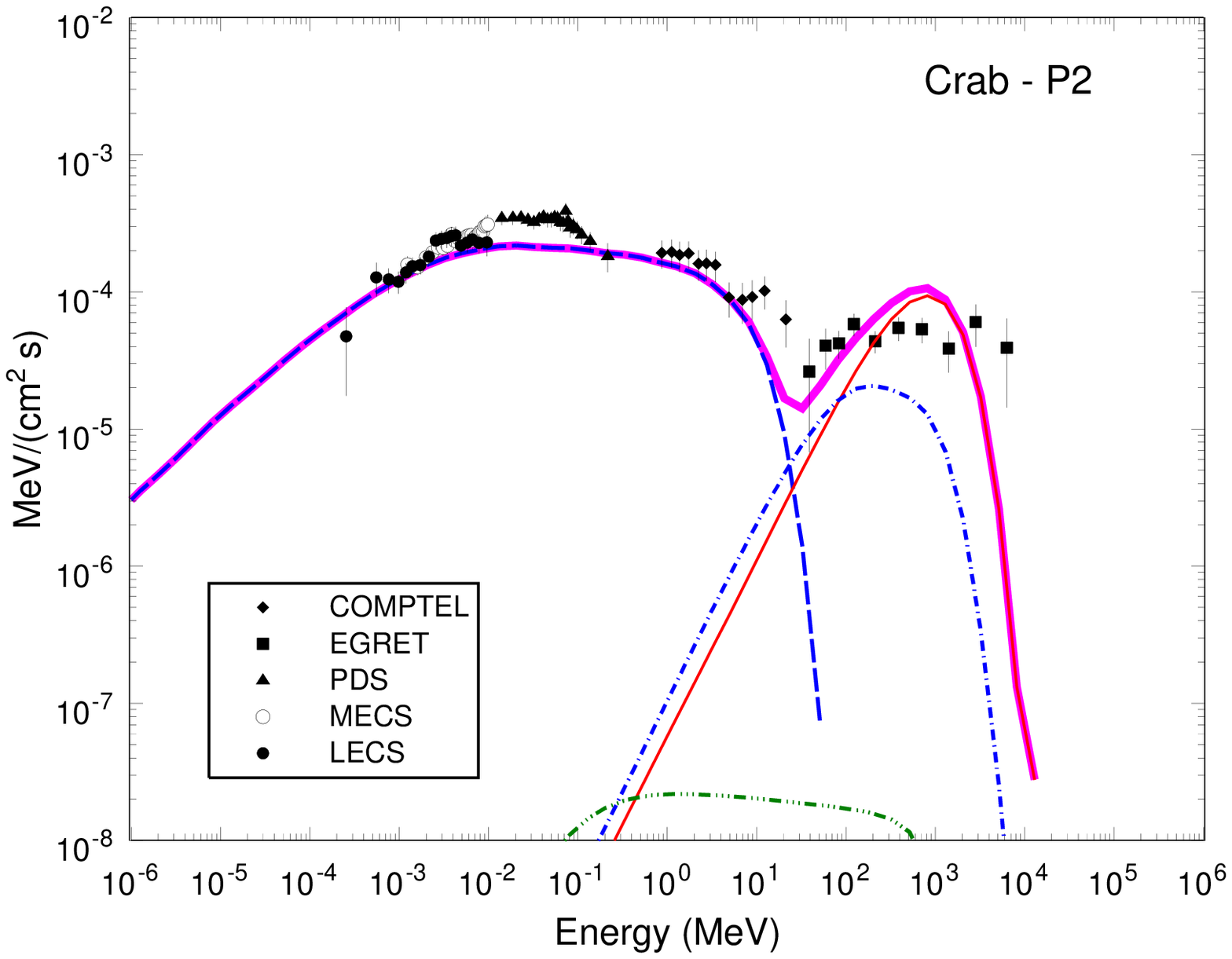}\hskip -0.5cm\includegraphics[width=7cm]{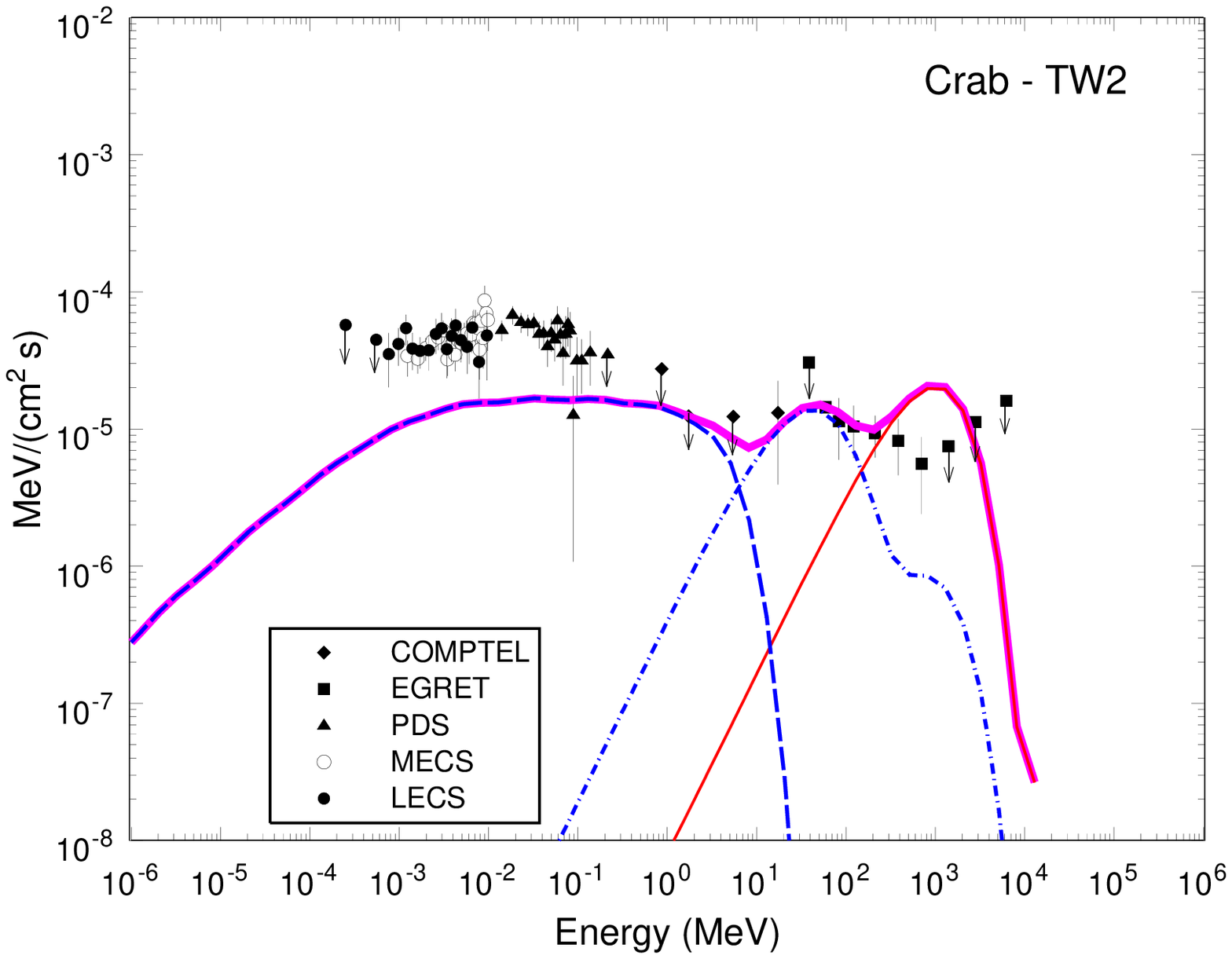}

\figureout{f7.eps}{
Same as Figure 4, but for the case of an extended radio cone model.}


\begin{references}
\reference{}
Arendt, P. N. \& Eilek, J. A. 1998, ApJ, submitted, (astro-ph/98011257).
\reference{}
Arons, J. 1983, ApJ, 266, 215.
\reference{}
Arons, J. \& Scharlemann, E. T. 1979, ApJ, 231, 854.
\reference{}
Arzoumanian, Z., Chernoff, D.F. \& Cordes, J.M. 2002, ApJ, 568, 289.
\reference{}
Blandford, R.~D., \& Scharlemann, E.~T. 1976, MNRAS, 174, 59.
\reference{}
Cheng, K.~S., Ho, C., \& Ruderman, M.~A. 1986, ApJ, 300, 500.
\reference{}
Cheng, K. S., Ruderman, M. A. \& Zhang, L. 2000, ApJ, 537, 964.
\reference{}
Daugherty, J. K. \& Harding, A. K., 1982, ApJ, 252, 337.
\reference{}
Daugherty, J. K. \& Harding, A. K., 1996, ApJ, 458, 278.
\reference{}
Dermer, C. D. 1990, ApJ, 360, 197.
\reference{}
Dyks, J. \& Harding, A. K.  2004, ApJ, 614, 869.
\reference{}
Dyks, J. \& B. Rudak, 2003, ApJ, 598, 1201.
\reference{}
Dyks, J., A. K. Harding \& B. Rudak, 2004, ApJ, 606, 1125. 
\reference{}
Gonthier, P.L., Van Guilder, R. \& Harding, A.K. 2004, ApJ, 604, 775.
\reference{}
Gonthier, P.L., Story, S.A., Clow, B. D. \& Harding, A.K. 2007, Ap \& SS, 309, 245.
\reference{}
Hankins, T. H. \& Eilek, J. A., 2007, astro-ph/0708.2505
\reference{}
Harding, A. K. 2005, in Proc. of 22nd Texas Symp. on Rel. Astrophys.,
ed. P.Chen et al., econf C041213 (astro-ph/0503300).
\reference{}
Harding, A.~K., \& Muslimov, A.~G. 1998, ApJ, 508, 328 (HM98).
\reference{}
Harding, A.~K., \& Muslimov, A.~G. 2001, ApJ, 556, 987 (HM01).
\reference{}
Harding, A.~K., \& Muslimov, A.~G. 2002, ApJ, 568, 862 (HM02).
\reference{}
Harding, A.~K., Usov, V. V. \& Muslimov, A.~G. 2005, ApJ, 622, 531.
\reference{}
Hirotani, K. \& Shibata, S. 2001, MNRAS, 325, 1228.
\reference{}
Jackson, P. D. 1965, Classical Electrodynamics, (Wiley:New York)
\reference{}
Kanbach G., Slowikowska, A., Kellner, S. \& Steinle, H.  2005, AIP Conference Series, Volume 801, pp. 306-311. 
\reference{}
Kijak, J. \& Gil, J. 2003, A \& A, 397, 969. 	
\reference{}
Kuiper, L. et al. 2001, A\&A, 378, 918.
\reference{}
Lommen, A., Donovan, J.; Gwinn, C.; Arzoumanian, Z.; Harding, A.; Strickman, M.; Dodson, R.; McCulloch, P.; Moffett, D.  2007, ApJ, 657, 436.
\reference{}
Lundgren, S. C. et al. 1995, ApJ, 453, 433.
\reference{}
Lyne, A. G. \& Manchester, R. N. MNRAS, 234, 477 (1988)
\reference{}
Lyubarskii, Y.~E., \& Petrova, S.~A. 1998, A \& A, 337, 433.
\reference{}
Morini, M. 1983, MNRAS, 303, 495.
\reference{}
Muslimov, A.~G., \& Harding, A.~K. 1997, ApJ, 485, 735.
\reference{}
Muslimov, A.~G., \& Harding, A.~K. 2003, ApJ, 588, 430.
\reference{}
Muslimov, A. G. \& Harding, A. K. 2004, ApJ, 606, 1143. 
\reference{}
Petrova, S.A. 2002, MNRAS, 336, 774.
\reference{}
Petrova, S.~A. 2003, A\&A, 408, 1057.
\reference{}
Radhakrishnan \& Cooke, ApL, 3, 225, 1969.
\reference{}
Rankin, J. M. 1993, ApJ, 405, 285. 
\reference{}
Romani, R.~W. 1996, ApJ, 470, 469.
\reference{}
Romani, R. W. \& Yadigaroglu, I.-A. 1995, ApJ, 438, 314.
\reference{}
Shearer, A. et al. 2003, Science, 301, 493.
\reference{}
Stecker, F. W., S. D. Hunter \& Kniffen, D. A. 2007, AstroPh, in press (astro-ph/0705.4311)
\reference{}
Shklovsky, I.~S. 1970, ApJ, 159, L77.
\reference{}
Spitkovsky, A. 2006, ApJ, 648, L51.
\reference{}
Sturner, S.~J., \& Dermer, C.~D. 1994, ApJ, 420, L79.
\reference{}
Tademaru, E. 1973, ApJ, 183, 625.
\reference{}
Tang, A. P. S., TAkata, J., Jia, J. J. \& Cheng, K. S. 2007. [astro-ph/0711.2719]
\reference{}
Takata, J. \& Chang, H. K. 2007, ApJ, 670, 677.
\reference{}
Timokhin, A.~N. 2006, MNRAS, 36, 1055.
\reference{} 
Thompson, D.~J. 2001, in High-Energy Gamma-Ray Astronomy, ed. F. A. Aharonian \& H. J. 
Volk, (AIP, New York), p. 103.
\end{references}
\end{document}